%% file: SED_paper_a2.tex
\documentclass{aa}

\usepackage{graphicx}
\usepackage[labelfont=bf,font=small,tableposition=top]{caption}
\usepackage{subcaption}
\usepackage{amsmath}
\usepackage{txfonts}

\usepackage{fix2col}
\usepackage{ulem}

\usepackage{color}
\usepackage{colortbl} 
\definecolor{pinegreen}{RGB}{1, 121, 111}
\definecolor{salmon}{RGB}{255,160,122}
\hypersetup{breaklinks=true,colorlinks,citecolor=pinegreen}

\usepackage[hyphenbreaks]{breakurl}
\usepackage{natbib}

\definecolor{c1}{RGB}{91, 44, 111}
\definecolor{c2}{RGB}{13, 71, 161}
\definecolor{c3}{RGB}{14, 102, 85}

\definecolor{ylvacolour}{rgb}{.55,.4,.75}
\definecolor{YC}{RGB}{115,80,185}

\newcommand{\Msun}{\ensuremath{\,M_\odot}\xspace}

\newcommand{\Lsun}{\ensuremath{\,L_\odot}\xspace}
\newcommand{\kms}{\ensuremath{\,\rm{km}\,\rm{s}^{-1}}\xspace}
\newcommand{\Msunyr}{\ensuremath{\,M_\odot\,\rm{yr}^{-1}}\xspace}
\newcommand{\ergHz}{\ensuremath{\,\rm{erg}^{-1}\,\rm{Hz}}}

\newcommand{\Qz}{\ensuremath{\,Q_{\rm{0, pop}}}\xspace}
\newcommand{\Qo}{\ensuremath{\,Q_{\rm{1, pop}}}\xspace}
\newcommand{\Qt}{\ensuremath{\,Q_{\rm{2, pop}}}\xspace}

\newcommand{\xii}{\ensuremath{\,\xi_{\rm{ion, 0}}}\xspace}
\newcommand{\xiio}{\ensuremath{\,\xi_{\rm{ion, 1}}}\xspace}
\newcommand{\xiit}{\ensuremath{\,\xi_{\rm{ion, 2}}}\xspace}

\newcommand{\HI}{\ensuremath{\,\ion{H}{I}}\xspace}
\newcommand{\HII}{\ensuremath{\,\ion{H}{II}}\xspace}
\newcommand{\HeI}{\ensuremath{\,\ion{He}{I}}\xspace}
\newcommand{\HeII}{\ensuremath{\,\ion{He}{II}}\xspace}

\newcommand{\OII}{\ensuremath{\,\ion{O}{II}}\xspace}
\newcommand{\OIII}{\ensuremath{\,\ion{O}{III}}\xspace}

\newcommand{\CIV}{\ensuremath{\,\ion{C}{IV}}\xspace}

\newcommand{\Op}{\ensuremath{\, \mathrm{O}^{+}}\xspace}
\newcommand{\Otwop}{\ensuremath{\, \mathrm{O}^{2+}}\xspace}
\newcommand{\Ctwop}{\ensuremath{\, \mathrm{C}^{2+}}\xspace}
\newcommand{\Cthreep}{\ensuremath{\, \mathrm{C}^{3+}}\xspace}

\newcommand{\code}[1]{{\textsc{#1}}}

\definecolor{midblue}{RGB}{44,133,255}
\definecolor{editcolor}{RGB}{54,105,224}

\definecolor{darkcyan}{rgb}{0.0, 0.55, 0.55}
\definecolor{amethyst}{rgb}{0.6, 0.4, 0.8}

\usepackage{refs}   
\usepackage{booktabs}    

\defcitealias{2017A&A...608A..11G}{Paper~I}
\defcitealias{2018A&A...615A..78G}{Paper~II}

\usepackage{morefloats}

\bibpunct{(}{)}{;}{a}{}{,}

\begin{document} 

     \title{The impact of stars stripped in binaries\\ on the integrated spectra of stellar populations\thanks{Our models are available in electronic form at the CDS via anonymous ftp to \url{cdsarc.u-strasbg.fr} (130.79.128.5) or via \url{http://cdsweb.u-strasbg.fr/cgi-bin/qcat?J/A+A/}}}
 \titlerunning{Stripped stars in stellar populations}

   \author{Y.~G\"{o}tberg$^{1,2}$, S.~E.~de~Mink$^{1,3}$, J.~H.~Groh$^{4}$, C.~Leitherer$^{5}$, \and C.~Norman$^{6}$}  
   \authorrunning{G\"{o}tberg et al.}

   \institute{
            Anton Pannekoek Institute for Astronomy, University of Amsterdam, 1090 GE Amsterdam, The Netherlands
             \and
             The Observatories of the Carnegie Institution for Science, 813 Santa Barbara St., Pasadena, CA 91101, USA \\
             \email{ygoetberg@carnegiescience.edu}
             \and
             GRAPPA, GRavitation and AstroParticle Physics Amsterdam, University of Amsterdam, 1090 GE Amsterdam, The Netherlands 
             \and
             School of Physics, Trinity College Dublin, The University of Dublin, Dublin 2, Ireland
             \and
             Space Telescope Science Institute, 3700 San Martin Drive, Baltimore, MD 21218, USA
             \and
             Dept. of Physics \& Astronomy, Johns Hopkins University, Baltimore, MD 21218, USA}
 
  \abstract
   {
Stars stripped of their envelopes from interaction with a binary companion emit a significant fraction of their radiation as ionizing photons. They are potentially important stellar sources of ionizing radiation, however, they are still often neglected in spectral synthesis simulations or simulations of stellar feedback. 
In anticipating the large datasets of galaxy spectra from the upcoming James Webb Space Telescope, we modeled the radiative contribution from stripped stars by using detailed evolutionary and spectral models. We estimated their impact on the integrated spectra and specifically on the emission rates of \HI-, \HeI-, and \HeII-ionizing photons from stellar populations. 

We find that stripped stars have the largest impact on the ionizing spectrum of a population in which star formation halted several Myr ago. In such stellar populations, stripped stars dominate the emission of ionizing photons, mimicking a younger stellar population in which massive stars are still present. Our models also suggest that stripped stars have harder ionizing spectra than massive stars.

The additional ionizing radiation, with which stripped stars contribute affects observable properties that are related to the emission of ionizing photons from stellar populations. In co-eval stellar populations, the ionizing radiation from stripped stars increases the ionization parameter and the production efficiency of hydrogen ionizing photons. They also cause high values for these parameters for about ten times longer than what is predicted for massive stars. The effect on properties related to non-ionizing wavelengths is less pronounced, such as on the ultraviolet continuum slope or stellar contribution to emission lines. However, the hard ionizing radiation from stripped stars likely introduces a characteristic ionization structure of the nebula, which leads to the emission of highly ionized elements such as \Otwop and \Cthreep. We, therefore, expect that the presence of stripped stars affects the location in the BPT diagram and the diagnostic ratio of \OIII to \OII nebular emission lines.
Our models are publicly available through CDS database and on the \code{Starburst99} website. 
 } 

   \keywords{ultraviolet: galaxies -- binaries: close -- stars: atmospheres -- galaxies: starburst -- galaxies: stellar content}

   \maketitle
%

\section{Introduction}

Spectra of stellar populations provide us with powerful tools to study stars and their host galaxies across cosmic time. 
Existing surveys and those anticipated with future facilities, such as the James Webb Space Telescope \citep[JWST,][]{2006SSRv..123..485G}, are expected to deliver a wealth of observational data that can potentially revolutionize our understanding. Translating these data into measurements of the physical quantities of interest, such as star formation rates, requires the use of theoretical or semi-empirical models for the spectra of stellar populations. Accurate models for the spectra of stellar populations, and in particular the ionizing radiation, are therefore indispensable \citep{2013ARA&A..51..393C}.  

Ionizing photons are of primary interest for two reasons. Ionizing photons from stellar sources can be reprocessed by nearby gas and dust, giving rise to infrared excess and various prominent emission lines \citep{2001MNRAS.323..887C}. These include lines that are used as diagnostics to infer star formation rates, metallicities, as well as possible variations in the initial stellar mass function and to infer the presence or absence of active galactic nuclei (AGN). The reliability of these measurements depends on the accuracy of the underlying models. Ionizing photons from stellar populations also play a crucial role as a source of stellar feedback. For example, it is thought to be important in regulating the efficiency of star formation \citep{2006ApJ...653..361K, 2013MNRAS.431.1062D}. On a larger scale, ionizing radiation from stellar populations that escapes the host galaxies can ionize gas in the intergalactic medium (IGM). These galaxies are generally held responsible for the reionization of the Universe \citep{2001PhR...349..125B, 2010Natur.468...49R, 2017ApJ...835...37L}. What type of sources that caused cosmic reionization is uncertain. However, recently, the role of interacting binary stars as sources of ionizing photons has been shown to potentially be important during the reionization \citep{2016MNRAS.459.3614M, 2016MNRAS.458L...6W}.
 
Traditionally, massive single stars are considered to be the main producers of ionizing photons in stellar populations. They are rare and short-lived, but they have high luminosities of $\sim 10^4-10^6\Lsun$. Furthermore, with temperatures higher than $\sim 25\,000$~K, they emit most of their radiation at energies above the threshold for hydrogen ionization \citep[e.g.,][]{2002MNRAS.337.1309S, 2005A&A...436.1049M, 2012A&A...537A.146E, 2015A&A...573A..71K, 2018A&A...618A..73S}. 
	The most massive stars can eventually lose their hydrogen-rich envelopes as a result of strong stellar winds or eruptions creating Wolf-Rayet (WR) stars \citep{2005A&A...429..581M}, which can be so hot that they emit photons sufficiently energetic to ionize even helium \citep{2007ARA&A..45..177C}. 

Recent studies of nearby resolved stellar populations show that massive and intermediate mass stars are often accompanied by a companion star that orbits so close that interaction between the two stars is inevitable as the stars evolve and swell up \citep{2012Sci...337..444S, 2017ApJS..230...15M}. Interaction in binary systems can completely change the future evolution of the stars in the system, leading to mass accretion, rejuvenation, rapid rotation, mass loss, and possibly even coalescence (e.g., \citealt{1979A&A....73...19V, 1992ApJ...391..246P, 1999A&A...350..148W, 2008MNRAS.384.1109E, 2014ApJ...782....7D}; \mbox{\citealt{2015ApJ...805...20S, 2018MNRAS.479...75S}}). Including, or improving the treatment of, the effects of binary interaction in models for the spectra of stellar populations is therefore warranted.

In this study, we focus on stars that have been stripped of their envelope as a result of interaction with a binary companion \citep{1967ZA.....65..251K, 1967AcA....17..355P}. This is expected to be a very common outcome of binary interaction \citep[e.g.,][]{2012Sci...337..444S}, resulting in very hot and compact stars \citep{1987A&A...178..170V, 2010ApJ...725..940Y,2017ApJ...840...10Y}. Due to their high temperatures, they are expected to emit the majority of their radiation at ionizing wavelengths \citep[][hereafter \citetalias{2017A&A...608A..11G}]{2017A&A...608A..11G}.  This makes them potentially important, but still often neglected, contributors to the budget of ionizing photons produced by stellar populations, as argued, for example, in \citet{2003A&A...400...63V} and \citet[][see also \citealt{2003A&A...400..429B} and \citealt{2016MNRAS.456..485S}]{2007ApJ...662L.107V}.

One of the challenges to properly account for the impact of these stripped stars on the ionizing spectra of stellar populations, was, until recently, the scarcity of appropriate atmosphere models. Direct observations of stars stripped in binaries are still scarce, likely because they are typically outshone by their companion \citep[see, however,][]{1998ApJ...493..440G, 2008A&A...485..245G, 2008ApJ...686.1280P, 2013ApJ...765....2P, 2017ApJ...843...60W, 2018ApJ...853..156W, 2018ApJ...865...76C}. As a result, there had been little motivation to calculate atmosphere models for stripped stars. The earliest estimates of the integrated spectra of stellar populations including stripped stars, therefore, relied on black-body approximations or rescaling of atmosphere models that were originally computed for more luminous WR stars. Modern simulations make use of spectral libraries, but these typically do not cover the full parameter space of interest for stripped stars nor consider the effect of metallicity. 

 This motivated us to undertake the effort of computing a library of evolutionary and spectral models custom-made for stripped stars that result from binary interaction for a range of masses and metallicities \citep[][hereafter \citetalias{2018A&A...615A..78G}]{2018A&A...615A..78G}.  
In \citetalias{2017A&A...608A..11G} we showed how metallicity affects the stripping process. At lower metallicity, the progenitor star remains more compact leading to an incomplete stripping of the envelope. The resulting star is therefore partially stripped and more luminous, but also cooler. In \citetalias{2018A&A...615A..78G} we presented model grids that cover a range of masses and metallicities. We showed how these stars span a continuous range of spectral types, from WR-like spectra characterized by emission lines formed in the winds of the more massive and metal-rich stripped stars, to subdwarf-like spectra dominated by absorption features resulting from the photosphere of stripped stars with transparent outflows. We further predicted the existence of a hybrid intermediate class of spectra showing a combination of absorption and emission lines, very similar to those observed for the recently discovered new stellar class labeled WN3/O3 \citep{2014ApJ...788...83M, 2015ApJ...807...81M,2017ApJ...837..122M, 2017ApJ...841...20N, 2018ApJ...863..181N, 2018MNRAS.475..772S}. 

The aim of this work is to estimate the radiative contribution from stripped stars to the spectral energy distribution of stellar populations and measure the additional emission rate of ionizing photons that stripped stars produce. For this purpose, we performed a population synthesis to estimate the number and type of stripped stars that are present in a population at any given time. We used the custom-made grid of spectral models for individual stripped stars that we published in \citetalias{2018A&A...615A..78G}.  We use this to investigate the impact on the integrated spectra. We discuss the integrated spectra, the emission rate of \HI-, \mbox{\HeI-,} and \HeII-ionizing photons, and further quantities that can be derived from observations, namely the production efficiency of ionizing photons, the ionization parameter, the UV luminosity and continuum slope, and stellar spectral features.

The first theoretical and semi-empirical studies of the integrated spectra of stellar populations primarily focussed on the effects of single stars. These include codes based on single star models, such as \code{Starburst99} \citep{1999ApJS..123....3L, 2010ApJS..189..309L, 2014ApJS..212...14L} \code{GALAXEV} \citep{2003MNRAS.344.1000B}, and \code{PEGASE} \citep{1997A&A...326..950F, 1999astro.ph.12179F, 2004A&A...425..881L}. However, more recently, several groups have considered the effects of binary interaction. These include the Brussels code \citep{1999NewA....4..173V, 2003A&A...400..429B, 2007ApJ...662L.107V}, the Yunnan simulations \citep{2004A&A...415..117Z,2010Ap&SS.329...41H, 2010Ap&SS.329..277C, 2012MNRAS.421..743Z, 2012MNRAS.424..874L, 2015MNRAS.447L..21Z}, and the \code{BPASS} code \citep{2009MNRAS.400.1019E, 2012MNRAS.419..479E, 2017PASA...34...58E}. 


The diversity of codes has helped improve our understanding of the importance of binary interactions throughout the years. However, the outcome of binary interactions and their spectra is still uncertain \citep[see e.g.,][]{2017PASA...34....1D} and detailed studies of each binary evolutionary channel is required. One important part for spectral synthesis models that has not been investigated in detail is how the spectra of stars stripped in binaries look like and what their contribution is to the spectrum of the full stellar population. 

In this study, we use our spectral models custom-made for stars stripped in binaries to make predictions for their contributions to the spectrum of a full stellar population. We choose to focus on this particular binary evolutionary channel, both so that we can carefully examine the evolution and spectra of these objects and so that we can understand in detail the effects of stripped stars in particular. We compare our results with the results from \code{Starburst99} to understand the effect of envelope-stripping compared to a population containing only single stars. The code \code{BPASS} takes a sophisticated approach, combining very large grids of binary evolution models with spectral libraries with the aim to account for all binary evolutionary channels at once. The developers of \code{BPASS} have provided testable predictions for a large range of observable phenomena (\citet{2012MNRAS.419..479E, 2016MNRAS.462.3302E, 2016MNRAS.461L.117E, 2019MNRAS.482..870E, 2014MNRAS.444.3466S, 2016MNRAS.456..485S, 2018MNRAS.479...75S, 2018MNRAS.477..904X, 2019MNRAS.482..384X}). By comparing our results to those from \code{BPASS}, we can identify effects that come from stripped stars. This is not only valuable for understanding the effects of stripped stars and how various uncertain binary parameters would alter the results for stripped stars, but this approach will also enable the identification of parts that require careful modeling and which effects that are bluntly unreliable because of uncertainties in the models for stripped stars.

Our hope is that the predictions provided in this work will be of use for the interpretation of the data that is resulting from several recent and ongoing surveys that are probing the ionizing properties of stellar populations. This includes the anticipated James Webb Space Telescope \citep[][see for example \citealt{2016MNRAS.456.3354F}, \citealt{2016MNRAS.462.1757G}, \citealt{2018MNRAS.477..904X}]{2006SSRv..123..485G}, but our simulations are also interesting for various surveys currently conducted from the ground. For example, the MOSDEF survey \citep{2015ApJS..218...15K}, the KBSS \citep{2012ApJ...750...67R, 2014ApJ...795..165S}, the KLCS \citep{2018ApJ...869..123S}, the GLASS \citep{2015ApJ...812..114T}, the VUDS \citep{2015A&A...576A..79L}, and the ZFIRE \citep{2016ApJ...828...21N}. We will, therefore, make our predictions available online in electronic format. They can be retrieved from the CDS database\footnote{\url{http://cdsweb.u-strasbg.fr/cgi-bin/qcat?J/A+A/}} and they will also be available through the \code{Starburst99} web-portal.

The structure of this paper is as follows. \secref{sec:modeling} describes the models for the evolution and spectra of stripped stars that we presented previously in \citetalias{2018A&A...615A..78G} and how we use these to construct a model for the integrated spectrum of stripped stars in a stellar population. In \secref{sec:nbr_stripped}, we present the model predictions for how many stripped stars are present in stellar populations. In \secref{sec:SED_time}, we show how the presence of stripped stars affects the total spectral energy distribution of a stellar population. We quantify the contribution from stripped stars to the emission rate ionizing photons in \secref{sec:ionizing_time}. In \secref{sec:observables}, we discuss the impact of stripped stars on observable quantities. In \secref{sec:summary_conclusions} we summarize our findings and conclusions. The current paper is the third in a series with \citetalias{2017A&A...608A..11G}  and \citetalias{2018A&A...615A..78G}, but it can be read independently.


\section{Accounting for the radiative emission from stripped stars in stellar populations}\label{sec:modeling}

We created an estimate of the radiative contribution from the stripped stars in stellar populations. We first describe the models for the evolution of individual stripped stars and their spectra in \secref{sec:individual}. We then describe the assumptions that we make to model stripped stars in stellar populations in \secref{sec:modeling_population} and how we represent the emission from a full stellar population in \secref{sec:model_pop_stripped_SB99}. 

\subsection{Models of individual stripped stars}\label{sec:individual}

\subsubsection{Binary evolutionary models}

We used the models presented in \citetalias{2018A&A...615A..78G} to describe the evolution of stars that lose their envelope through interaction with a binary companion \citepalias[see also][for an in-depth discussion]{2017A&A...608A..11G}. These are models of binary systems in which stable mass transfer is initiated early during the Hertzsprung gap, after which the H-rich envelope is stripped off \citep[commonly referred to as Case~B type mass transfer,][]{1967ZA.....65..251K}. Stripped stars can also result from stable mass transfer initiated during the main-sequence evolution of the most massive star in the system \citep[Case~A type mass transfer,][]{1967ZA.....65..251K}, or from unstable mass transfer and a subsequent successful ejection of the common envelope \citep{paczynski_1976, 2011ApJ...730...76I}. The contribution from the different formation channels vary depending on the progenitor mass, but we expect that Case~B type mass transfer is responsible for the majority of the stripped stars. Despite the variety of evolutionary histories, the properties of the stripped stars are remarkably similar. These properties are primarily dependent on the mass of the stripped stars alone. This assumption works well for most systems, see however \citet{2011A&A...528A.131C, 2017ApJ...840...10Y, 2018A&A...614A..99S, 2018arXiv180807580S} for evolution that leads to a larger fraction of the H-rich envelope is left, which is the case for systems with long initial periods or very low metallicity. We used our models of stripped stars created through Case~B type stable mass transfer as an approximation for stripped stars formed via any evolutionary channel. This approximation is sufficient for our current purposes. 

The evolutionary models were computed using the binary stellar evolution code \code{MESA} \citep{2011ApJS..192....3P,2013ApJS..208....4P, 2015ApJS..220...15P, 2018ApJS..234...34P,2019ApJS..243...10P}. The models have the metallicities $Z = 0.014$, 0.006, 0.002, and 0.0002, which are representative of the metallicity of the Sun, the Large and Small Magellanic Clouds, and an environment with very low metallicity that may be representative of early stellar populations. Each metallicity grid consists of 23 models, which are different from each other by the initial mass of the donor star, $M_{\text{1, init}}$. The initial donor star masses were chosen between $M_{\text{1, init}} = 2$\Msun and 20\Msun with equal spacing in the logarithm of the mass. Since, in most cases, the properties of the stripped star are not sensitive to the formation channel or initial orbital period and initial mass ratio, one binary model was computed for each initial mass of the donor star. We chose the initial mass of the accretor star by setting the initial mass ratio to $q \equiv M_{\text{2, init}}/M_{\text{1, init}} = 0.8$. We chose the initial periods such that mass transfer was initiated early during the Hertzsprung gap. This corresponded to initial periods between $P_{\text{init}} = 3$~days and 35~days depending on the initial mass of the donor star (see Table~1 of \citetalias{2018A&A...615A..78G}). The resulting stripped stars have masses between 0.35\Msun and 7.9\Msun. 
The wind mass-loss from stripped stars is an uncertain parameter because few stripped stars have been observed. The models, therefore, employ extrapolations of the wind mass-loss algorithm for hot and subluminous stars of \citet{2016A&A...593A.101K} from the low mass end and of the empirical WR mass-loss recipe of \citet{2000A&A...360..227N} from the high mass end. The switch between the two wind regimes is chosen to occur for stripped stars with progenitor masses of $6 \Msun$. Low-mass stripped stars are affected by diffusion processes that impact the surface composition \citep{2016PASP..128h2001H}. An algorithm accounting for the effect of gravitational settling \citep[see][also \citealt{1994ApJ...421..828T} and \citealt{1986ApJS...61..177P}]{2011ApJS..192....3P} is included when modeling the evolution of stripped stars. It has strong effects for stripped stars with masses below $0.7\Msun$ \citepalias[see][]{2018A&A...615A..78G}.

Our evolutionary model grids cover the evolution of stripped stars from low to high mass. They stretch from stripped stars at the lower mass limit of central helium burning \citep[$\sim0.35\Msun$,][]{2002MNRAS.336..449H} up to close to the mass limit where massive stars lose their envelope via their own wind \citep[e.g.,][]{1986ARA&A..24..329C}. It is likely that stars of higher mass than what we consider experience envelope-stripping in binaries, for example, through mass-transfer initiated on the main sequence evolution of the donor star \citep[see e.g.,][]{2010ApJ...725..940Y}. However, these stars would primarily contribute at early stages because the progenitor stars are more massive and, therefore, live for a shorter time. Here, we focus on the contribution from stripped stars that can not have been created by strong wind mass-loss and, therefore, they have lower masses than most WR stars. For this mass range, we consider that our models are appropriate to use as a representation of the stellar evolution of stripped stars given the careful choices for the wind mass-loss rates and the treatment of diffusion processes on the stellar surfaces.
We use the properties of the stripped stars that were computed in the evolutionary models as input for the spectral models described below. We also adopt the time of stripping and the duration of the stripped phase computed in the evolutionary models in our simulations of the integrated spectra of stellar populations, see \secref{sec:modeling_population}.

\subsubsection{Spectral models}

The spectral models for individual stripped stars that we used in this work were computed with the non-LTE radiative transfer code \code{CMFGEN} \citep{1990A&A...231..116H, 1998ApJ...496..407H} and were custom-made for stripped stars by employing the surface parameters given by the evolutionary models as input at the base of the atmosphere. The evolutionary models were used at the time when the stripped star had reached half-way through central helium burning ($X_{\text{He, c}} = 0.5$).
We can use these models as a good approximation for the spectral properties throughout most of the stripped star phases. This is because the luminosity and the effective temperature do not change significantly during most of the time the stars are stripped. We show this in \appref{app:stripped_phase}. The spectral models for individual stripped stars are publicly available at the CDS database\footnote{\url{http://vizier.cfa.harvard.edu/viz-bin/VizieR?-source=J/A+A/615/A78}}.

The shape of the spectral energy distribution and the emission rates of ionizing photons \citepalias[see Table~1 of][]{2018A&A...615A..78G} depend on the assumed wind mass-loss rates, wind speeds, and wind clumping. These parameters are uncertain. Theoretical predictions are now available \citep[e.g.,][]{2016A&A...593A.101K, 2017A&A...607L...8V}, but they have not yet been thoroughly tested against observations, because only very few stripped stars with sufficiently strong wind mass-loss have been identified and studied in detail so far \citep[e.g.,][]{2008A&A...485..245G}. In \citetalias{2017A&A...608A..11G}, we showed that variations in wind mass-loss rate primarily affect the predicted emission rate of \HeII-ionizing photons, while the emission rates of \HI- and \HeI-ionizing photons are not significantly affected. The mass-loss rates assumed in our models were chosen to smoothly connect the mass-loss rates of subdwarfs \citep{2016A&A...593A.101K} with the observed mass-loss rates of WR stars \citep{2000A&A...360..227N}. Our assumed mass-loss rates also match well with the observed mass-loss rate of the stripped star in the binary system HD~45166 \citep[][]{2008A&A...485..245G}. The recent theoretical predictions by \citet{2017A&A...607L...8V} suggest that the mass-loss rates of stripped stars may be ten times lower than what we assume in this paper. The winds of stripped stars are likely not reaching close to the Eddington limit, in contrary to massive main-sequence and WR stars \citep[cf.][]{2014A&A...570A..38B}. This suggests that the wind mass-loss rate from stripped stars is lower than that from WR stars and thus not well-described by the recipe for WR stars of \citet{2000A&A...360..227N}. To establish which are the wind mass-loss rates from stripped stars, observations of a sample of stripped stars are necessary. 
If, as suggested by \citet{2017A&A...607L...8V}, the mass-loss rates from stripped stars indeed are lower than what the recipe from \citet{2000A&A...360..227N} predicts, it would likely imply an increase of the emission rates of \HeII-ionizing photons presented in this work. The emission rates of \HI- and \HeI-ionizing photons are robust against wind uncertainties.

The wind parameters also affect the stellar spectral features. Higher wind mass-loss rate, slower winds, or more clumping result in stronger emission features as the stellar wind becomes denser. The wind speed could be slower than our assumptions. We assumed terminal wind speeds of 1.5 times the escape speed of the surface of the stripped star, which resulted in values of $\sim1500 - 2500$~\kms. The observed stripped star HD~45166 has an anisotropic wind that partially is slow, which gives rise to the strong emission lines the star exhibits \citep[][]{2008A&A...485..245G}. For a better understanding of the spectral features from stripped stars, an observed sample is necessary.

\subsection{Modeling the contribution of stripped stars to a stellar population}\label{sec:modeling_population}

We modeled the number and type of stripped stars that are present in a population as a function of time by taking a Monte Carlo approach. We first created a sample of stars by randomly drawing initial masses, $M_{\text{init}}$, from the initial mass function of \citet[][]{2001MNRAS.322..231K}, $dN/dM_{\text{init}} \propto M_{\text{init}}^{\alpha}$, where $\alpha = -1.3$ for $M_{\text{init}} < 0.5\Msun$ and $-2.3$ for $M_{\text{init}} > 0.5\Msun$ until we reached a total stellar mass of $10^{6} \Msun$. We assumed mass limits of 0.1\Msun and 100\Msun. We then chose which stars that have a companion star using the mass-dependent binary fraction of \citet{2017ApJS..230...15M} that follows closely the linear function $f_{\text{bin}} = 0.09 + 0.63\log_{10} (M_{\text{init}}/\Msun)$ (we fit this polynomial to the data presented in Fig.~42 of \citealt{2017ApJS..230...15M}, for comparison purposes also see \citealt{2013A&A...552A..69V}). The mass of the companion stars were randomly drawn, such that the mass ratio, $q = M_{\text{2, init}}/M_{\text{1, init}}$, followed a flat distribution sampled between 0.1 and 1 (consistent with the observations of \citealt{2012ApJ...751....4K}, \citealt{2012Sci...337..444S}, and \citealt{2017ApJS..230...15M}, for early-type stars). After assigning companions and their masses, the total stellar mass increased and we therefore randomly removed single or binary stars until we reached a total stellar mass of $10^{6}\Msun$ again. The initial orbital periods of the binary systems were randomly drawn from a distribution that is flat in the logarithm of the period \citep[e.g.,][]{1924PTarO..25f...1O, 2007A&A...474...77K, 2017ApJS..230...15M}. For systems where the most massive star of the system has a mass $M_{\text{1, init}} \geq 15\Msun$, we used the distribution by \citet{2012Sci...337..444S}, which favors short-period systems. As a lower limit for the initial period, we chose the shortest period that allows both stars to fit inside their Roche-lobes at zero-age main-sequence. For the upper limit of the initial period, we followed \citet{2017ApJS..230...15M} and set $10^{3.7}$~days. \citet{2017ApJS..230...15M} also presented period distributions for binaries, but with lower resolution in the period than \citet{2012Sci...337..444S} do for the massive stars. We, therefore, combined the period distribution for massive stars presented by \citet{2012Sci...337..444S} with the flat distribution that is consistent with the data presented by \citet{2017ApJS..230...15M} for lower mass stars.

We used evolutionary models of single stars to follow the radius evolution of the donor star and to determine when it will start to interact with its companion star. We created these models with \code{MESA}, using the same physical assumptions as we adopted for the models of binary stars \citepalias[see][]{2018A&A...615A..78G}. The moment mass transfer starts can then be determined by comparing the radius evolution of the most massive star with its Roche radius \citep{1983ApJ...268..368E}. We predicted the further evolution from the initial mass ratio of each binary system and whether the donor star had developed a deep convective envelope at the time of interaction. We assumed that stable mass transfer occurs in systems with an initial mass ratio larger than a critical value \citep[$q_{\text{crit, MS}} = 0.65$ and $q_{\text{crit, HG}} = 0.4$ for interaction initiated on the main-sequence and Hertzsprung gap following][and \citealt{2002MNRAS.329..897H}, respectively]{2007A&A...467.1181D}. For systems with a smaller initial mass ratio than the critical value and systems which have donor stars that have a deep convective envelope, we assumed that a common envelope developed. Since stars do not have well-defined cores during the main sequence, we assumed that stripped stars are not formed through common envelope evolution initiated during the main sequence evolution of the donor star. We also did not include stripped stars formed after or during central helium burning because the product is likely short-lived.

We used the classical $\alpha$-prescription to determine whether the common envelope is successfully ejected or not \citep{1984ApJ...277..355W}. For this, we employed the standard value of one for the efficiency of the ejection of the common envelope, $\alpha_{\text{CE}}$ \citep[see e.g.,][]{2002MNRAS.329..897H}. We also assumed that the parameter $\lambda_{\text{CE}}$, which describes how strongly the envelope is bound to the core of the donor star \citep{2000A&A...360.1043D,2001A&A...369..170T}, is $0.5$, which is the average for Hertzsprung gap stars \citep[see Appendix~E of][]{2004PhDT........45I}. \citet{2000A&A...360.1043D} showed that $\lambda_{\text{CE}}$ is dependent on the stellar mass and evolutionary stage. They found that $\lambda_{\text{CE}}$ decreases with increasing mass and that $\lambda_{\text{CE}}$ typically is between 0.1 and 1 for Hertzsprung gap stars of initial masses between 3\Msun and 10\Msun. The mass dependence of $\lambda_{\text{CE}}$ suggests that lower mass donors easier eject a common envelope. However, since $\alpha_{\text{CE}}$ is considered to be a very uncertain parameter, we assumed the simple, mass-independent value for $\lambda_{\text{CE}}$.
 
We assumed that the stars that shared a common envelope coalesced if either the core of the donor star or the companion star filled their Roche-lobe during the in-spiral inside the common envelope. To control whether this occurs, we assumed that the core of the donor star had the same mass and radius as a stripped star that originated from a progenitor star with the same initial mass as the donor \citepalias[see Table~1 of][]{2018A&A...615A..78G}. For the companions, we assumed that they have the same mass and radius as on the zero-age main sequence. We used our models for stripped stars and for single stars and interpolated over the mass to find the radii of the stars in each system. In cases when the companion star had lower mass than our lowest mass model, we extrapolated to smaller radii, however, this never led to negative radii. Once mass transfer initiates within a common envelope, it likely leads to coalescence since the stars spiral closer together due to friction from the surrounding material. If coalescence occurred, we did not consider the further evolution of the star. 
The $\alpha$-prescription likely oversimplifies the evolution of stars in a common envelope \citep[see e.g.,][]{2013A&ARv..21...59I}. We therefore consider our approach approximate. To better constrain the formation of stripped stars through common envelope evolution, a large sample of observed stripped stars is needed.

Summarizing the above descriptions, we considered that stripped stars are formed through three main evolutionary channels: stable mass transfer initiated on the main-sequence evolution of the donor star, stable mass transfer initiated on the Hertzsprung gap evolution of the donor star, and un-stable mass transfer initiated on the Hertzsprung gap evolution of the donor star followed by a successful ejection of the common envelope.
We assigned the masses of the stripped stars by interpolating between the masses of the progenitor stars and assuming that the stripped star masses are constant throughout the stripped phases, taken at $X_{\text{He, c}} = 0.5$ from the evolutionary models (see \secref{sec:individual}). We also interpolated over the progenitor masses in the evolutionary grid to determine the time of envelope-stripping and the duration of the stripped phases. This approach neglects that Case~A type mass transfer can result in somewhat lower mass stripped stars \citep[see e.g.,][]{1994A&A...290..119P}. However, this evolutionary channel is responsible for less than a fourth of the stripped stars of intermediate or high mass. We therefore expect that the total effect is small for the ionizing emission. 
We interpolated the spectral models for stripped stars over the mass of the stripped star to find the best-fitting radiative representation for each stripped star. With the described approach, we assumed that the spectrum of a stripped star remains stationary throughout the stripped phase. This approximation is good for most of the time since stripped stars have roughly the same effective temperature and luminosity throughout the long lasting central helium burning phase \citepalias[see \appref{app:stripped_phase} and ][]{2018A&A...615A..78G}. Our approach also assumes that stripped stars have the same properties as long as they are formed from a progenitor with the same initial mass and independent on the formation channel and initial orbital period and mass ratio. This assumption is also valid for most cases. The exception is for systems that interact on the main sequence, that have very low metallicity, or that have initially very long orbital periods \citep[cf.][]{2011A&A...528A.131C, 2017ApJ...840...10Y, 2018arXiv180807580S}.

We considered two different star formation histories. The first is an instantaneous starburst with initially $10^6 \Msun$ of mass in stars. The second is a population in which stars form at a constant rate of 1\Msunyr. For the case of continuous star formation, we used the predictions for the co-eval stellar population and convolved the number of stripped stars, the spectral energy distribution and emission rates of ionizing photons over time. We performed the convolution every 1~Myr, which produces stochastic effects expected for a constant star formation rate of 1\Msunyr.
In \appref{app:SFH} we consider three additional, more realistic, star formation histories and we discuss how much the stripped stars in those populations affect the emission rates of ionizing photons.

\subsection{Including the contribution from stripped stars to the model of a full stellar population}\label{sec:model_pop_stripped_SB99}

In the previous subsection, we described how we model the radiative contribution from stripped stars. To model the radiation from a realistic population, we also need to model the contribution of the remainder of the population, which includes primarily single stars and stars in binary systems that have not yet interacted.

\code{Starburst99} provides well-established models for the integrated spectra of stellar populations, including models of main-sequence stars, giant stars, and stars in more evolved stages of the stellar life \citep{1999ApJS..123....3L, 2010ApJS..189..309L}. Most of the stars in a stellar population are main-sequence stars that have not yet interacted. Binary interaction primarily occurs at later evolutionary stages as the stars expand significantly after central hydrogen exhaustion and only mildly during the main-sequence. \code{Starburst99} thus constitutes a fair approximation for stars that have not interacted with a binary companion. However, including stripped stars implies that there should be fewer giant stars as a fraction of them have become stripped. Moreover, the companions to stripped stars are expected to have accreted material and thus become more massive and somewhat rejuvenated. Similar is expected for binary stars that have merged. This leads to a slight increase in the radiation in optical and UV wavelengths since the mass-gaining star, in most cases, is a main-sequence star \citep[cf.][]{1999NewA....4..173V}. We expect the total effect from mass-gainers, mergers, and the lack of giant stars on the emission in the optical and UV wavelengths to be small, compared to the total emission from the full stellar population.

We used the combination of models from \code{Starburst99} and our model for the contribution from stripped stars to represent the radiation of a full stellar population in which stripped stars are formed. 
We made our models publicly available on the \code{Starburst99} online interface\footnote{\url{www.stsci.edu/science/starburst99/}}, providing the addition from stripped stars to the spectral energy distribution, the emission rates of \HI-, \HeI-, and \HeII-ionizing photons, and the high-resolution UV and optical spectra. 
In this study, we compare the contribution from stripped stars to the version of \code{Starburst99} that uses the initial mass function from \citet{2001MNRAS.322..231K} with mass limits of $0.1\Msun$ and $100\Msun$, that is, the same as what we employed for the stripped stars in the population. We chose to compare to the \code{Starburst99} models that use non-rotating stellar evolutionary models from the Geneva grids \citep{2012ApJ...751...67L} and spectral models from \code{WM-Basic} for OB-stars \citep{2001A&A...375..161P, 2010ApJS..189..309L}, \code{CMFGEN} for WR stars \citep{1998ApJ...496..407H, 2002MNRAS.337.1309S}, and \code{BaSeL} v3.1 for later type main sequence stars, cooler stars, and red supergiants \citep{1997A&AS..125..229L}. 
When comparing our models of various metallicity with \code{Starburst99}, we used the \code{Starburst99} models of $Z = 0.014$, 0.008, 0.002, and 0.001 together with our models with metallicity of $Z = 0.014$, 0.006, 0.002, and 0.0002, respectively (see also \tabref{tab:Z_combination}). We consider that the difference in metallicity between the two model sets is small and expect that the difference when using models with exactly the same metallicity is also small. 
We consider that the combination of the stripped star models and \code{Starburst99} is a good assumption for radiation with wavelengths shorter than $\sim 5000$~\AA. For the model to be accurate at longer wavelengths, we would need to decrease the radiation from giant stars to compensate for stars that we assume have become stripped. Giant stars emit their radiation primarily at wavelengths longer than $\sim 5000$~\AA. We expect the decrement of radiation at these long wavelengths to be at maximum about 30\% \citep[as also suggested by][see e.g., their Figs.~5 and 15]{2017PASA...34...58E}. This is approximately the fraction of massive stars that get stripped \citep{2012Sci...337..444S} and thus the fraction of giant stars that should be missing. This topic is beyond the scope of this paper, but we hope to address it more in detail at a later stage.

In several cases in this study, we compare our results to those from the code \code{BPASS} \citep[version 2.2.1, or \textit{Tuatara}, see][but also \citealt{2009MNRAS.400.1019E, 2012MNRAS.419..479E, 2017PASA...34...58E}]{2018MNRAS.479...75S}. We used the version of the code that accounts for binary interactions and employs a similar initial mass fraction as we do \citep[$\alpha = -2.35$ for $M_{\text{init}} > 0.5\Msun$ and $\alpha = -1.3$ for $M_{\text{init}} < 0.5\Msun$,][]{1993MNRAS.262..545K} and the same mass limits as we do. In \textit{Tuatara}, the binary fraction, period distribution, and mass ratio distribution have been adapted following the data presented in \citet{2017ApJS..230...15M}. 
The \code{BPASS} code uses stellar evolutionary models computed with a modified version of the \code{STARS} code that accounts for binary interaction, and the spectral models that are used are from \code{BaSeL} v3.1 and v2.2 for most part of the stellar evolution, \code{WM-Basic} for O-stars, and \code{PoWR} models for WR stars defined as when $X_{\text{H,s}} < 0.4$ and $\log_{10} T_{\star} \geq 4.45$. We therefore expect that the emission from stripped stars of masses above $\sim 0.5\Msun$ is represented using models from \code{PoWR} in \code{BPASS}. We present a comparison between our models and \code{BPASS} for individual stripped stars in \appref{app:BPASS_comparison}, where we also discuss differences in the modeling of stripped stars in our simulations and in \code{BPASS}. 
We compare our models with metallicity $Z = 0.014$, 0.006, 0.002, and 0.0002 with the models of \code{BPASS} with metallicity $Z = 0.014$, 0.006, 0.002, and 0.0001, respectively (see \tabref{tab:Z_combination}).


\section{The number of stripped stars in stellar populations}\label{sec:nbr_stripped}

The model for a population of stripped stars gives estimates for the number of stripped stars that are present in stellar populations, what are their properties and how they were formed. The number of stripped stars in a population is expected to be much smaller than the number of main sequence stars since the helium-core burning phase is about ten times shorter than the main sequence evolution and only a fraction of stars loose their envelopes through binary interaction.

\begin{figure*}
\centering
\includegraphics[width=.45\textwidth]{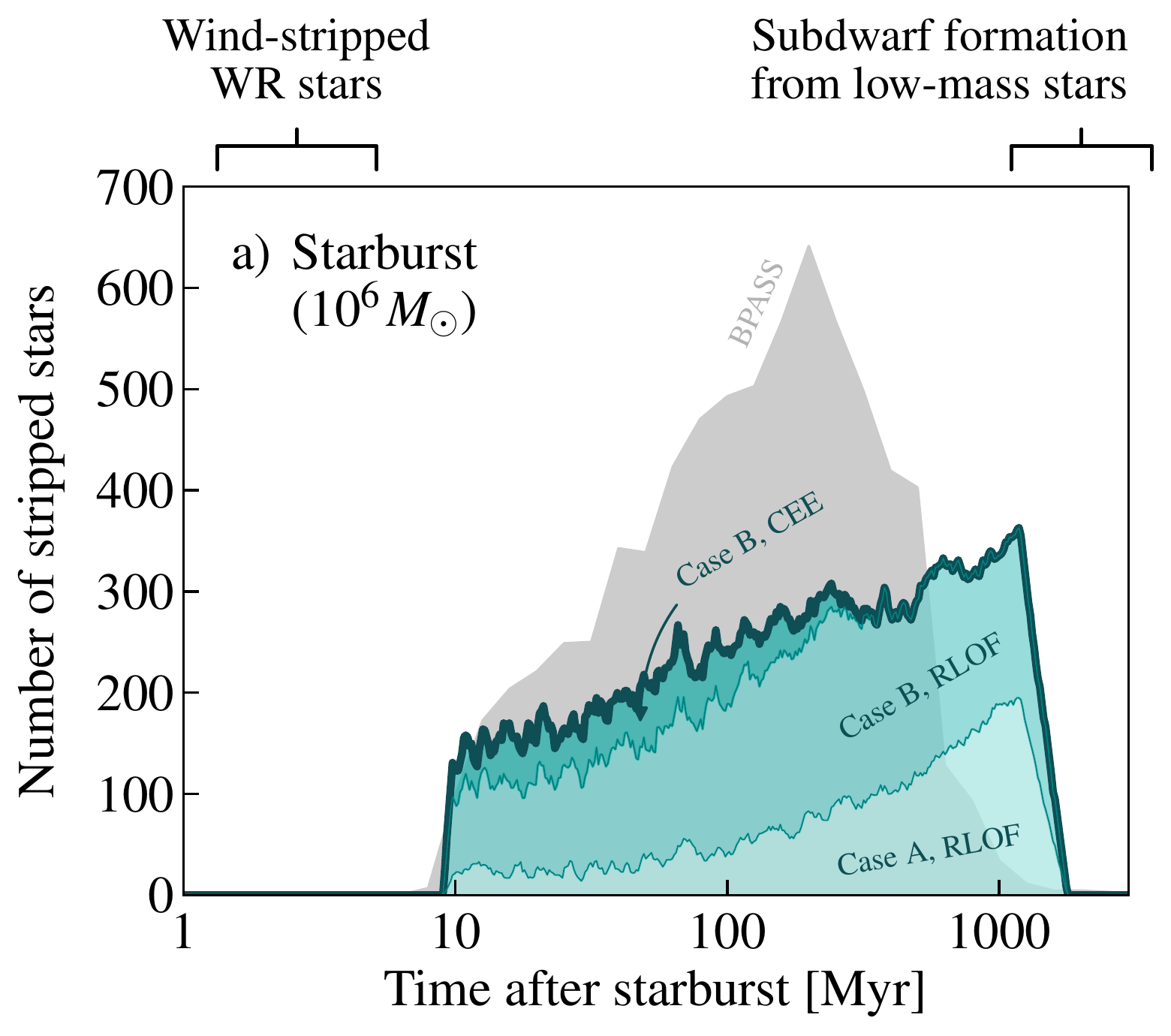}
\hspace{5mm}
\includegraphics[width=.44\textwidth]{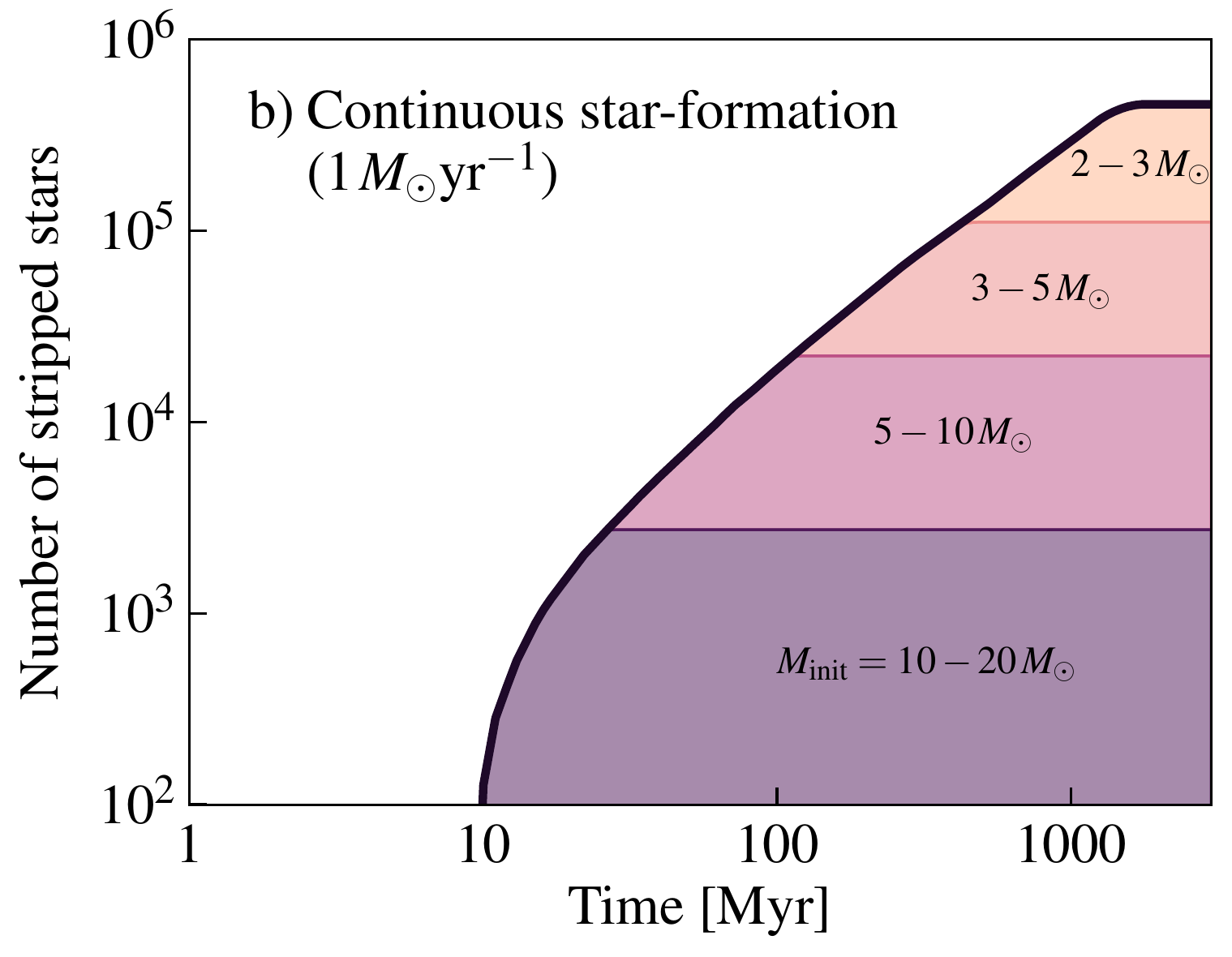}
\caption{Number of stripped stars shown as function of time for a population of solar metallicity. 
\textit{Panel a:} The numbers for a co-eval stellar population with initially $10^{6}\,\Msun$ in stars. We highlight the three formation channels that we consider to lead to the formation of stripped stars: stable mass transfer initiated during the main sequence evolution of the donor star (Case~A, RLOF in light green), stable mass transfer initiated when the donor star passes over the Hertzsprung gap (Case~B, RLOF in medium green), and successful ejection of a common envelope that developed during interaction initiated during the donor's Hertzsprung gap evolution (Case~B, CEE). For comparison, we show the number of low-luminosity WR stars predicted by \code{BPASS} in light gray \citep{2017PASA...34...58E}. We indicate the expected time of WR star formation through solely stellar winds and the time of subdwarf formation through low-mass stellar evolution above the panel. 
\textit{Panel b:} The case of continuous star formation at a rate of 1\Msunyr. We show the parts of the population of stripped stars that are created from progenitor stars with initial masses between $10-20$\Msun in purple, $5-10$\Msun in pink, $3-5$\Msun in salmon, and $2-3$\Msun in peach.  
}
\label{fig:nbr_time}
\end{figure*}

We show the number of stripped stars that are formed as a function of time for the two star formation histories and for solar metallicity in \figref{fig:nbr_time}. 
For the case of the instantaneous starburst of $10^{6} \Msun$ (panel a), between $\sim150$ and $\sim 400$ stripped stars are present in the population between 10~Myr and 2~Gyr after the starburst. Because we only consider progenitors with masses between 2\Msun and 20\Msun, the formation of stripped stars start after about 10~Myr, when the 20\Msun stars evolve off of the main sequence and can become stripped. The number of stripped stars in the population increases with time because stars of initially lower mass, that are also more common, evolve off of the main sequence and can get stripped. 

Panel a of \figref{fig:nbr_time} shows the fraction of stripped stars that are formed through the three considered formation channels: stable mass transfer initiated during the main sequence evolution of the donor star (labeled Case~A, RLOF), stable mass transfer initiated during the Hertzsprung gap evolution of the donor star (labeled Case~B, RLOF), and unstable mass transfer initiated during the Hertzsprung gap evolution of the donor star (labeled Case~B, CEE). The figure shows that Case~B type stable mass transfer is the most common evolutionary channel for stripped stars. This is expected since massive stars swell up significantly when the hydrogen shell is ignited. The expansion during the Hertzsprung gap decreases with lower stellar mass. This can be seen in the figure since the relative importance of Case~A type mass transfer becomes a more important formation channel for stripped stars that form later and thus originate from lower mass stars. In our model, the contribution from common envelope evolution is relatively small and primarily occurring for stars more massive than $\sim 4 \Msun$. However, depending on uncertain details in the treatment of common envelope evolution (see \secref{sec:modeling_population}) it is likely that the contribution from common envelope evolution is slightly different from our predictions. It is possible that our prediction for the fraction of stripped stars that form via Case~A type mass transfer is slightly exaggerated. The reason is that we do not account for conservative mass transfer, which might lead the accreting star to expand sufficiently and form an overcontact binary that eventually leads to coalescence. However, this is expected to occur only for very short periods \citep[see e.g.,][for a massive example]{2015ApJ...812..102A}. 

For reference, we also show the number of low-luminosity WR stars produced in \code{BPASS} in panel a of \figref{fig:nbr_time}. These are defined as stars that have luminosities of $\log_{10} (L/\Lsun) < 4.9$, temperatures of $\log_{10} (T_{\star}/\text{K}) \geq 4.45$ and surface abundance of hydrogen of $X_{\text{H, surf}} \leq 0.4$ (see the entry for WNH in Table~3 of \citealt{2017PASA...34...58E}, we note that the WNH entry in this table includes all low-luminosity WR stars). 
Since no main sequence stars should fit with these constraints and WR stars formed from wind mass loss likely have higher luminosity, it is plausible that these are stripped stars. We expect that the subdwarfs with lower mass than $\sim 0.5\Msun$ are excluded since they have surface temperatures that are lower than $10^{4.45}$~K. Such subdwarfs are primarily formed after $\sim 400$~Myr, which could explain why the number of low-luminosity WR stars predicted by \code{BPASS} drops around that time. The predicted number of stripped stars is larger in \code{BPASS} than in our model for times before around 500~Myr. The reason could be that \code{BPASS} treats mass transfer and common envelope evolution different \citep[cf.][]{2017PASA...34...58E}. However, our predictions and the predictions from \code{BPASS} are of the same order with just a factor of about two difference at maximum. 

For the case of continuous star formation (panel~b of \figref{fig:nbr_time}), the number of stripped stars steadily increases with time as lower mass stars start to interact with binary companions. The figure shows that a stellar population that has formed stars at a constant rate of 1\Msunyr for 100~Myr contains roughly 10\,000 stripped stars. About 3\,000 of these originate from progenitor stars more massive than 10\Msun and are therefore likely to be hotter than 70\,kK and more luminous than $10^{4} \, \Lsun$ (following Table~1 of \citetalias{2018A&A...615A..78G}). The figure shows that no or few subdwarfs are predicted to be present in such a young stellar population, but that the population of stripped stars is dominated by subdwarfs at later times. After about 1~Gyr of constant star formation, there is about half a million stripped stars present in the population, with the majority of these being subdwarfs (over 90\%).

The effect of metallicity is small on the predicted number of stripped stars. We describe the results for lower metallicity environments in \appref{app:metallicity}.


\section{The impact of stripped stars on the spectral energy distribution}\label{sec:SED_time}

In this section, we describe the effect of stripped stars on the total spectral energy distribution of a stellar population. 
Although the contribution from stripped stars to the total bolometric luminosity is small, the hard, ionizing spectra of stripped stars do significantly change the shape of the spectral energy distribution. The differences mainly occur in the extreme ultraviolet. We discuss the effects for co-eval stellar populations in \secref{sec:SED_coeval} and for the case of continuous star formation in \secref{sec:SED_cont}.

\subsection{Predictions for co-eval stellar populations}\label{sec:SED_coeval}

\begin{figure}
\centering
\includegraphics[width=.85\hsize]{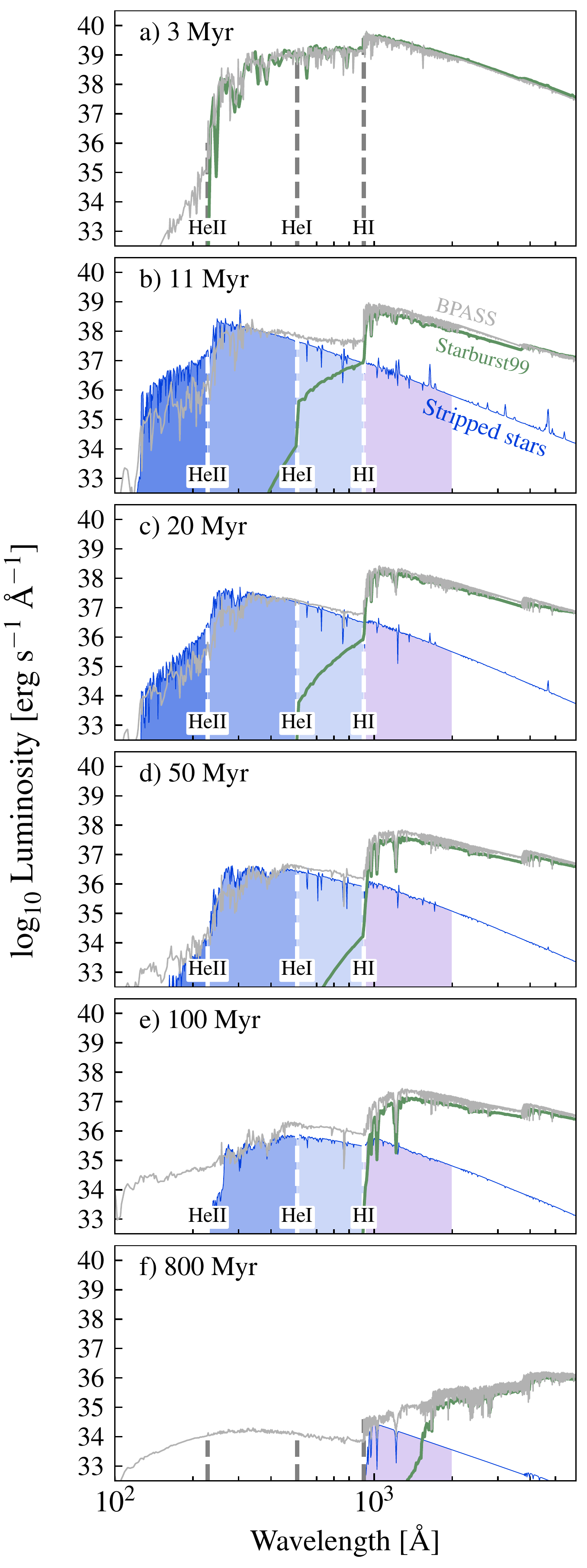}
\caption{Spectral energy distribution of a co-eval stellar population. The contribution from stripped stars is highlighted with a blue line. The parts of the spectra that are \HI-, \HeI-, and \HeII-ionizing are shaded in blue, while the UV is shaded in purple. For comparison, we show the spectral energy distribution of a population containing only single stars using \code{Starburst99} (green line), which can be interpreted as the contribution from the remaining stars in the stellar population. We also show the predictions from \code{BPASS} (gray line), where the effects of binary interaction are included. The panels correspond to different times after the instantaneous starburst of $10^{6}\Msun$. The model shown here assumes solar metallicity (see \appref{app:metallicity} for lower metallicity models).}
\label{fig:SED_population}
\end{figure}

\figref{fig:SED_population} shows how the spectral energy distribution is affected by the presence of stripped stars. The panels of the figure show snapshots taken 3, 11, 20, 50, 100, and 800~Myr after an instantaneous starburst of $10^6 \Msun$, assuming solar metallicity.

Stripped stars dominate the ionizing output from the stellar population after about 10~Myr and up to at least 100~Myr. 
At 3~Myr after starburst, shown in \figref{fig:SED_population}a, the ionizing radiation originates from massive main-sequence stars. Stars stripped in binaries are not yet present, because it takes time for them to form. Their progenitors, the donor stars, need time to evolve and swell up to fill their Roche-lobe. Stripped stars are therefore formed with a delay corresponding roughly to the main-sequence lifetime of the progenitor star. For a 20\Msun progenitor star, the main-sequence lifetime, and thus the delay with which stripped stars with such progenitors form, is about 10 Myr. Also, stars that initiate mass transfer during the main-sequence evolution result in stripped stars after a time delay corresponding to the main-sequence lifetime of the donor star. The reason for this is that the mass transfer rate is slow during the main-sequence evolution and it is not terminated until central hydrogen depletion is reached.

Stars stripped in binaries are created over an extended period of time. The main-sequence lifetime, and therefore the time delay with which stripped stars are created, varies with the mass of the progenitor star. The stars that form stripped stars 11, 20, 50, 100, and 800~Myr after starburst have initial masses of about 18, 12, 7, 5, and 2\Msun. The resulting stripped stars have masses of about 7, 4, 2, 1, and 0.5\Msun, respectively. The mass range of the stripped stars that are present at each point in time is small since the duration of the stripped phase is about 10~\% of the main-sequence lifetime of the progenitor star, which is the time delay with which the stripped stars are created. The temperature of stripped stars decreases with decreasing stellar mass as seen in Table~1 of \citetalias[][]{2018A&A...615A..78G}, which shows that a 7~\Msun WR star has a temperature of $100\,000$~K, while a 1~\Msun subdwarf has a temperature of $40\,000$~K. This decrease in temperature causes their contribution to the integrated spectrum to become softer with time as the mass of the stripped stars that are present decreases.

As seen in \secref{sec:nbr_stripped} and \figref{fig:nbr_time}, the number of stripped stars in a population increases as time proceeds. 
Despite the increase in their total numbers with time, we find that the total bolometric luminosity produced by stripped stars decreases. This is because the luminosity of individual stripped stars is a steep function of mass.

About 500~Myr after starburst, the stripped stars no longer significantly contribute with ionizing photons, according to our models. The reason for this is that the stripped stars that are still present at these late times are subdwarfs. These subdwarfs are affected by diffusion processes, which alter their surface composition and structure (for a discussion see Sect.~3 of \citetalias{2018A&A...615A..78G}). The result is an increase of the abundance of hydrogen at their surfaces, which creates a sharp cut-off of the spectral energy distribution at the Lyman limit. The integrated spectra are still significantly different from what is expected for a population of single stars, see panel f of \figref{fig:SED_population}. At these late times, we expect that white dwarfs contribute with ionizing radiation \citep{1984ApJ...287..315P}. However, more detailed modeling is needed to further understand the relative contributions of ionizing photons in late starbursts.

For comparison, we also show the spectral energy distributions predicted by the \code{BPASS} models in \figref{fig:SED_population}. The \code{BPASS} predictions for the shape of the ionizing part of the spectral energy distribution match well with our predictions for populations younger than about 50~Myr. It is interesting to note that despite our simpler approach, similar results are reproduced for starbursts of these ages. After this time, the \code{BPASS} models predict that the ionizing radiation is harder than what we find in our simulations. The reason is likely the adopted atmosphere models for central stars in planetary nebulae in \code{BPASS}, which are represented by hot WR star models \citep[see][for the models used]{2002A&A...387..244G} and the implementation of emission from X-ray binaries (priv.\ communication with JJ Eldridge). After about 300~Myr, the effects of diffusion on the stellar surface cause our stripped star models to appear slightly cooler, which could contribute to the difference between our models and the predictions of \code{BPASS} at late times. However, it is not likely that this is the full explanation, since the hard ionizing emission the \code{BPASS} predicts at 800~Myr shown in \figref{fig:SED_population}f requires very high temperatures, possibly similar to or higher than the massive WR stars. 

The effect of metallicity on the shape of the spectral energy distribution is relatively small. At lower metallicity, stripped stars are slightly cooler (see \citetalias{2017A&A...608A..11G} for a discussion), which means that the ionizing part of the spectral energy distribution is slightly softer. The first notable differences occur at very low metallicities, when $Z < 0.002$, see \appref{app:metallicity} and \figref{fig:SED_Z}.

\subsection{Predictions for continuous star formation}\label{sec:SED_cont}

\begin{figure}
\centering
\includegraphics[width=.95\hsize]{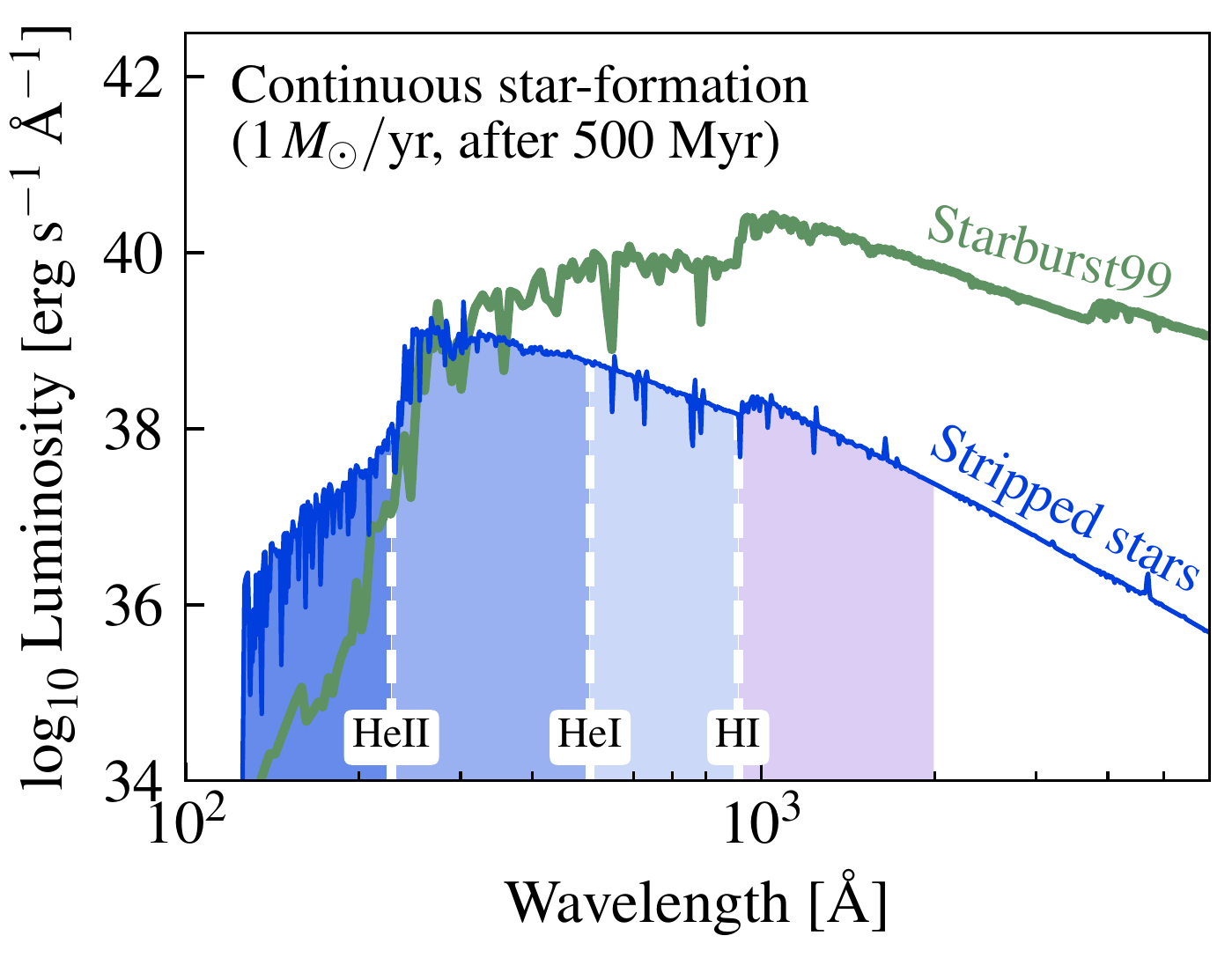}
\caption{Spectral energy distribution of a stellar population in which star formation has taken place at a constant rate of 1\Msunyr for 500~Myr. Otherwise similar to \figref{fig:SED_population}. We show models with solar metallicity.}
\label{fig:SED_cont_500}
\end{figure}

In \secref{sec:SED_coeval} we showed that, for co-eval stellar populations, stripped stars make a distinct contribution to the ionizing spectra at late times, while massive main-sequence stars dominate at early times. In \figref{fig:SED_cont_500}, we show the integrated spectrum for the idealized case of constant star formation. The spectrum is for a population in which stars have formed at a constant rate of 1\Msunyr, and for a prolonged period of time, here chosen to be 500~Myr. At this time, the ionizing spectrum has reached equilibrium. The figure shows that the spectrum is heavily dominated by single stars at almost the entire wavelength range. Also, the emission of \HI-ionizing photons is dominated by massive stars. The contribution from stripped stars to the total bolometric luminosity is negligible. They only dominate the emission of the hardest \HeII-ionizing photons. We note that our predictions for this part of the spectrum are uncertain and depend on the treatment of the stellar winds. 

Realistic stellar populations do not follow constant star formation, but are rather made up of a number of starbursts in star clusters that started to form stars at different times. The result is a star formation rate that varies with time. 
For such, more realistic, star formation histories, the relative contribution from stripped stars depends on the recent star formation activity. For populations that formed stars at a significant rate in the very recent past, $\lesssim 10$~Myr, massive stars likely dominate the output of ionizing photons. For populations that did not form stars very recently, we expect stripped stars to play a significant role. Since stripped stars emit ionizing photons for more than ten times longer than massive stars, it is likely that they will significantly contribute with the emission of ionizing photons in-between starbursts in stellar populations, see \appref{app:SFH} where we consider how the emission rate of ionizing photons is affected by the presence of stripped stars for three additional star formation histories.


\section{Impact on the budget of ionizing photons}\label{sec:ionizing_time}

In this section, we discuss the emission rates of ionizing photons: \Qz, \Qo, and \Qt for \HI-, \HeI-, and \HeII-ionizing photons, respectively. 
We use the common definition of the emission rate of ionizing photons as the number of emitted photons with wavelengths shorter than the ionization threshold of the considered atom or ion, which can be calculated from the emitted luminosity, $L_{\lambda}$, in the following way:
\begin{equation}
Q_i = \dfrac{1}{hc} \int _0^{\lambda_i} \lambda L_{\lambda}\, d\lambda ,
\end{equation}
where $h$ is the Planck's constant, $c$ is the speed of light, and the subscript $i$ refers to the considered ion. This method provides a good approximation for the emission rates of ionizing photons if the surrounding medium is sufficiently dense. The probability that an ionizing photon will lead to ionization once it encounters an atom or ion is decreasing with increasing photon energy \citep{2006agna.book.....O}. This decreasing ionization probability gives rise to a 100 times longer mean-free path for a photon with a wavelength of 228~\AA\ compared to one of 912~\AA\ in a medium containing only hydrogen (the mean-free path can be expressed as $\langle l \rangle = 1/(n \, \sigma)$, where $n$ is the number density of the surrounding medium and $\sigma$ is the ionization cross-section, which is wavelength dependent in the following way for hydrogen: $\sigma  = \sigma_0 (\lambda_0/\lambda)^3$). In a typical density for \HII regions of $10^2$~cm$^{-2}$, we calculate that these mean-free paths are of order $0.0005$~pc and $0.05$~pc, which are negligible length-scales in terms of the size of star-forming regions. We, therefore, consider the definition of the emission rates of ionizing photons as a realistic assumption (see however \citealt{2017ApJ...845..111M} for cases when low densities are considered).

\begin{figure*}
\centering
\includegraphics[width=.42\hsize]{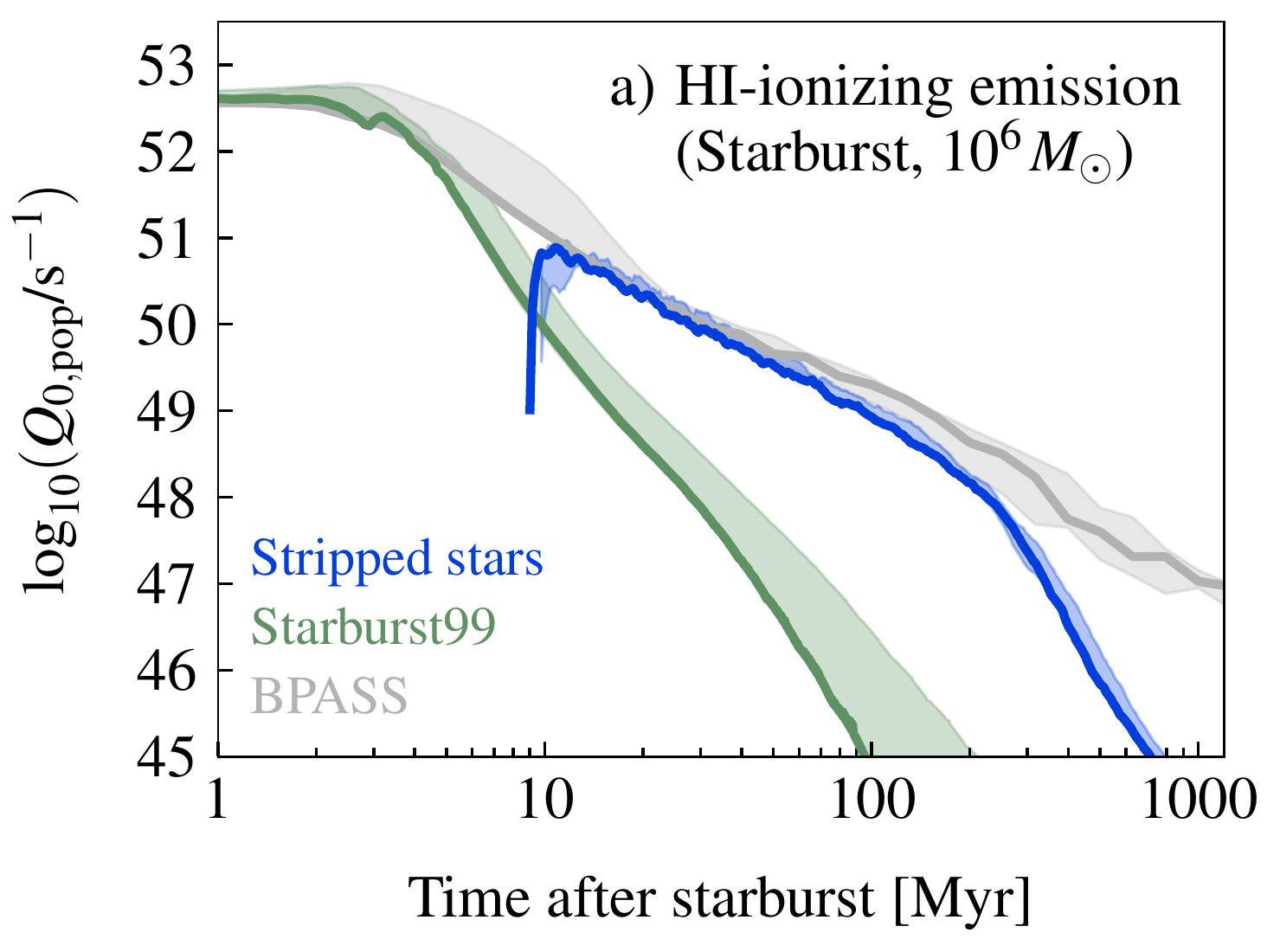}
\hspace{5ex}
\includegraphics[width=.42\hsize]{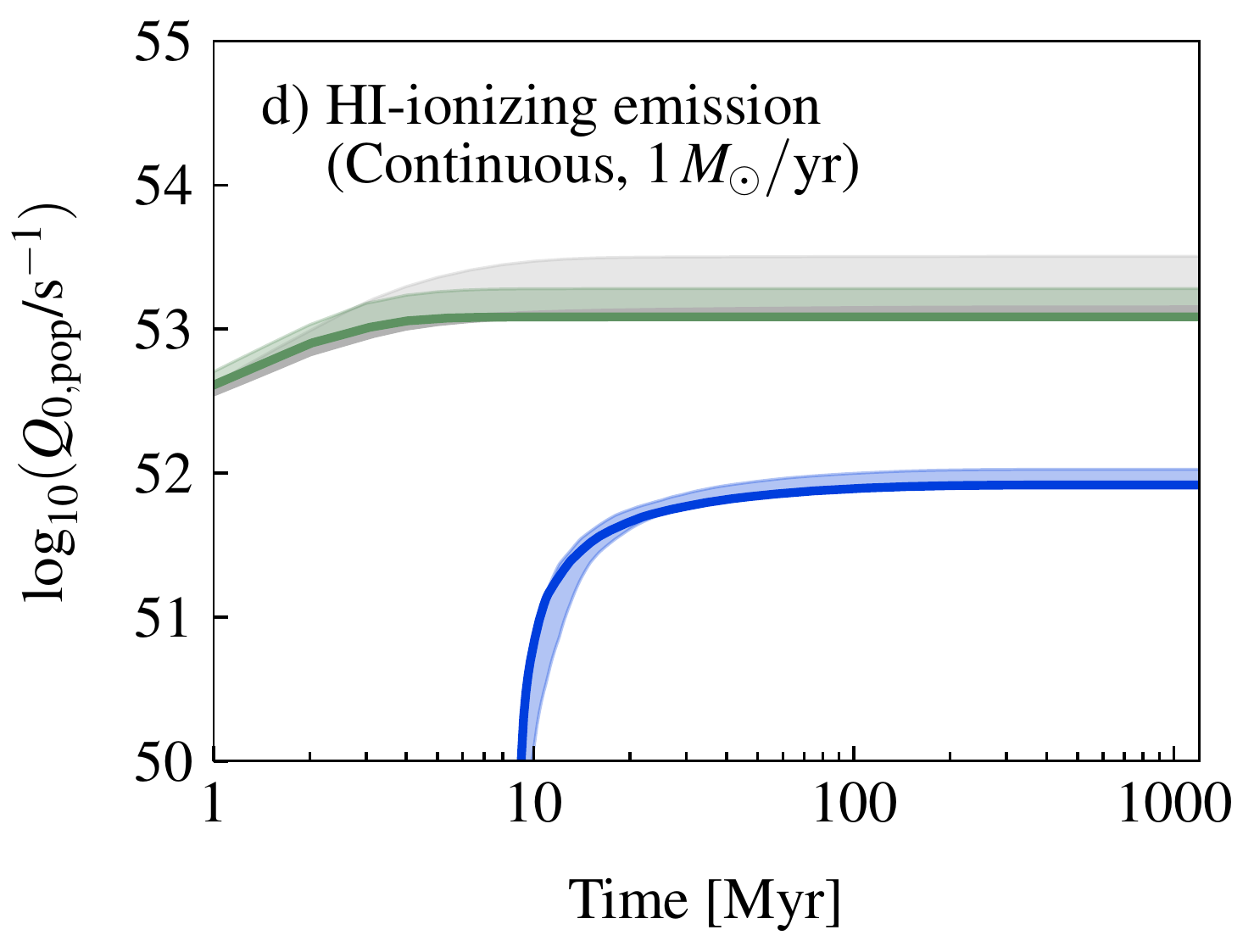}
\includegraphics[width=.42\hsize]{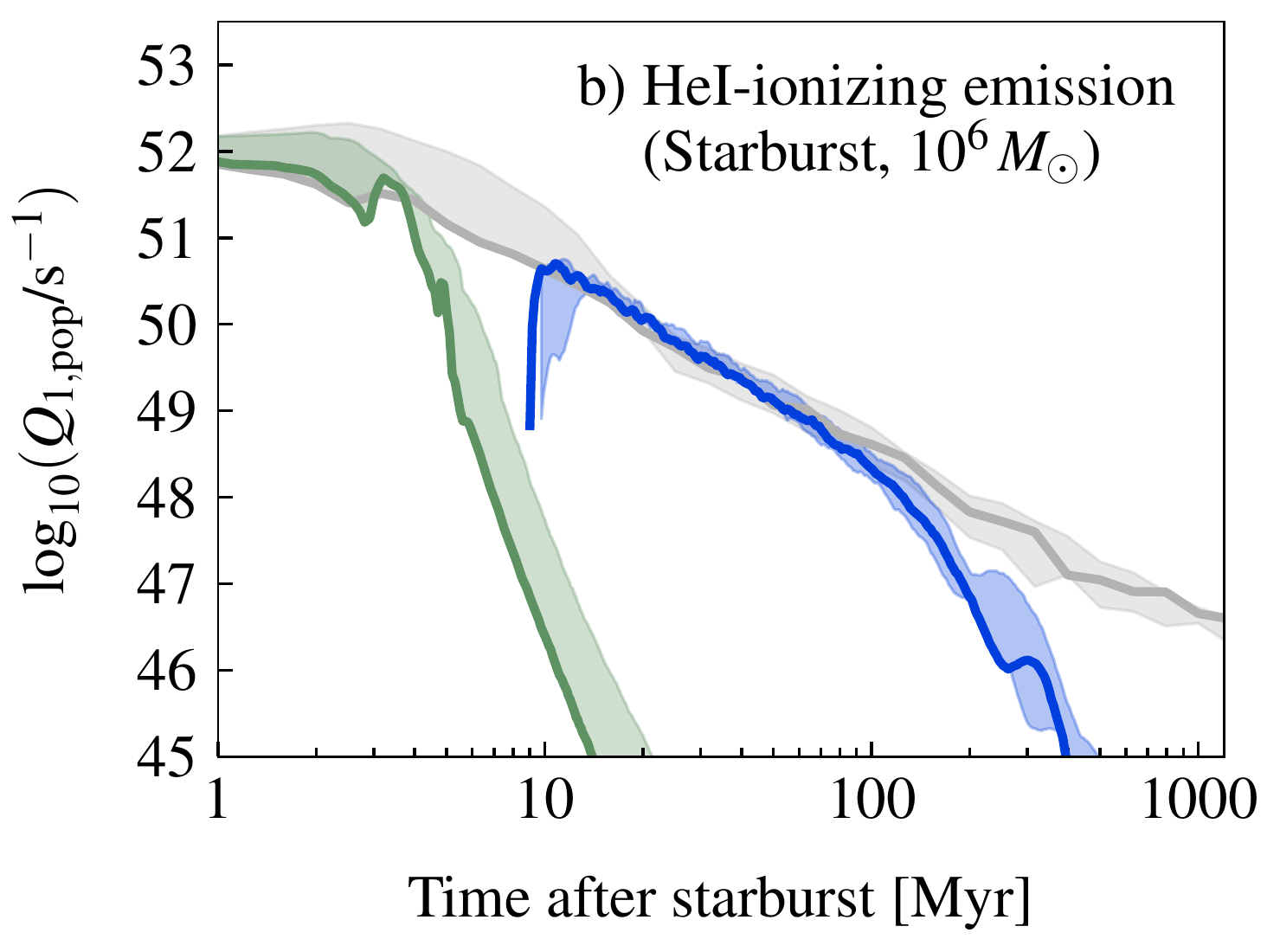}
\hspace{5ex}
\includegraphics[width=.42\hsize]{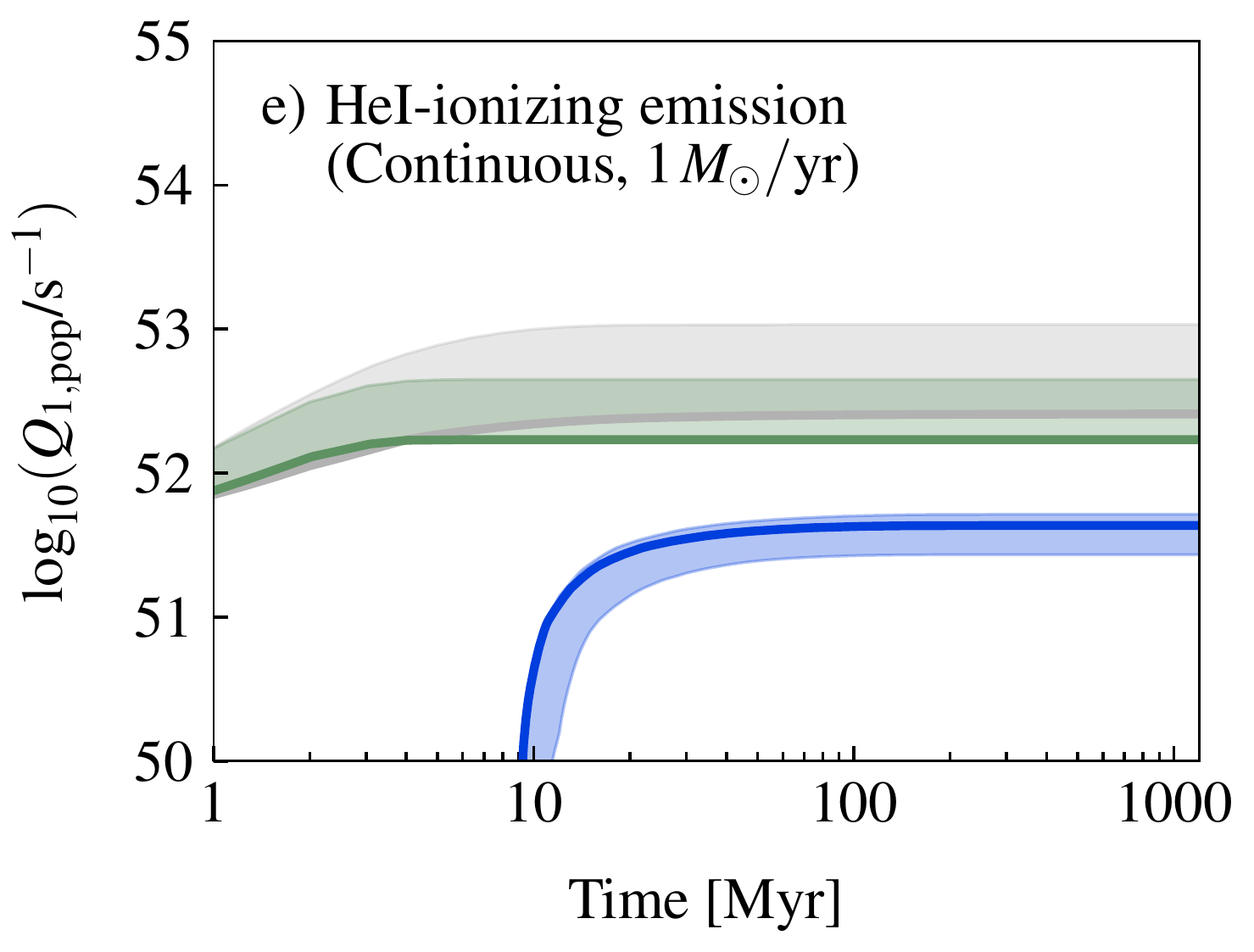}
\includegraphics[width=.42\hsize]{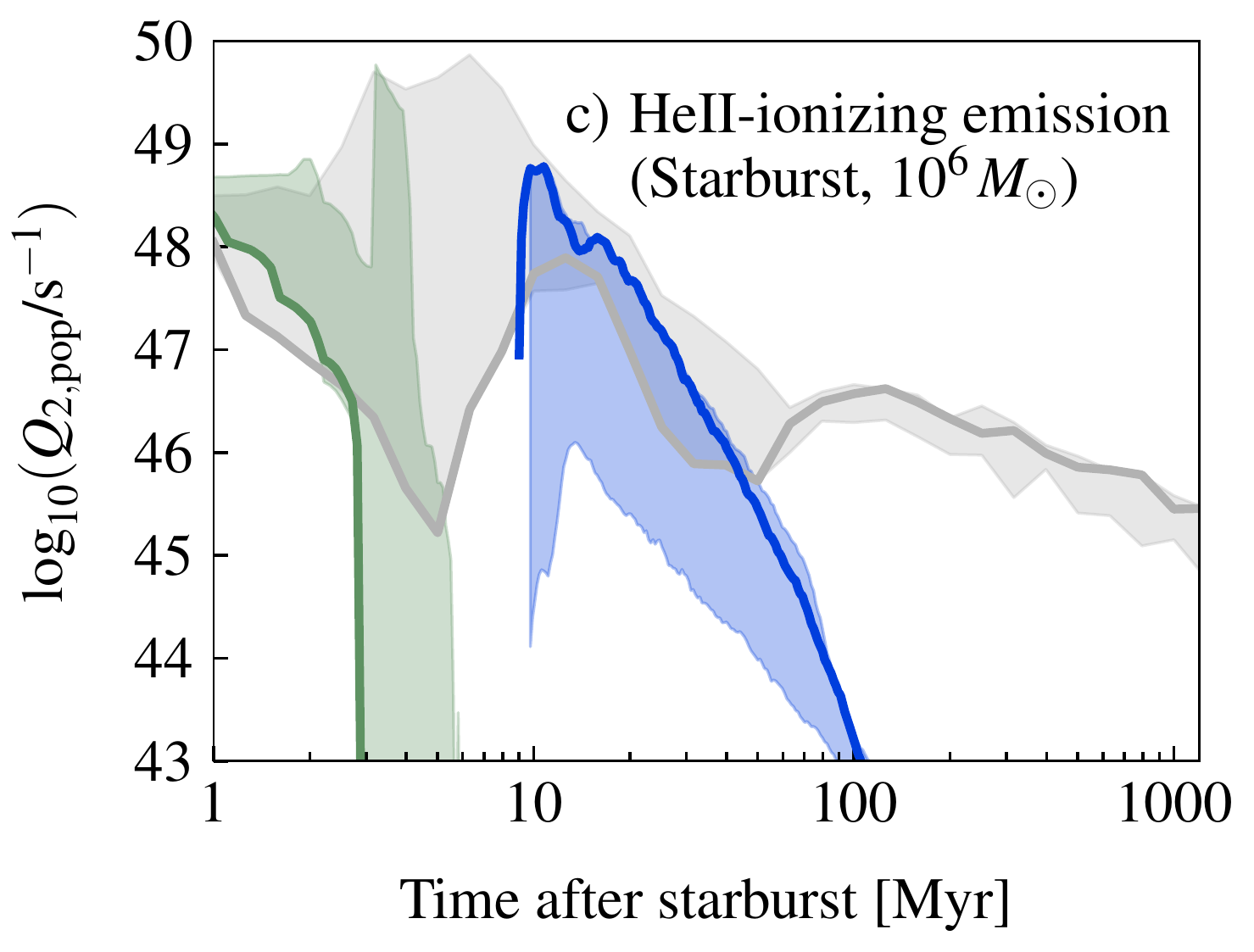}
\hspace{5ex}
\includegraphics[width=.42\hsize]{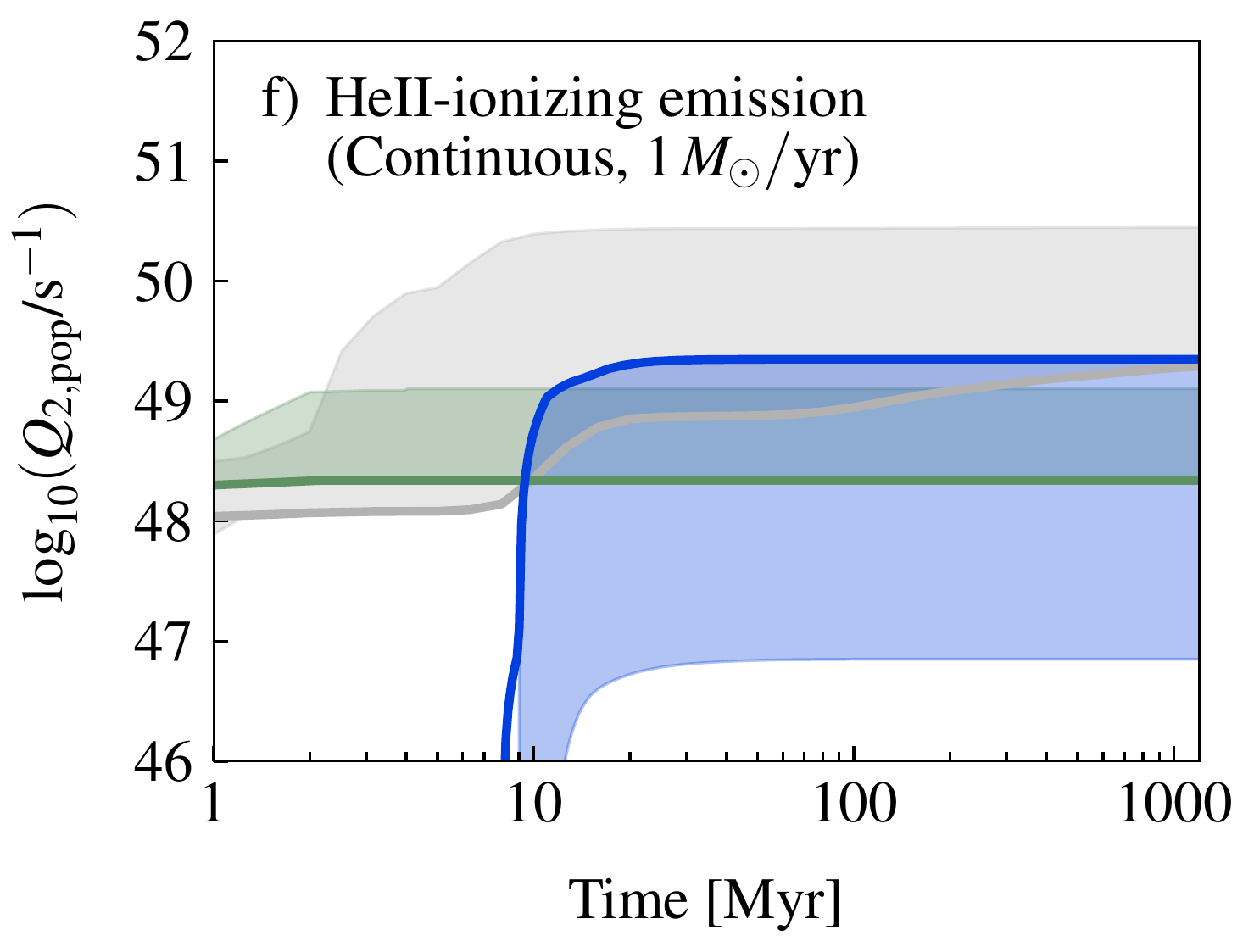}
\caption{Emission rates of ionizing photons from stellar populations as a function of time. The contribution from stripped stars is shown in blue, while green represents the contribution from the massive main sequence and WR stars in the stellar population (models from \code{Starburst99}, see \secref{sec:model_pop_stripped_SB99}). For reference, we show the predictions from \code{BPASS} models in which binary interactions are included using gray color. The solid lines correspond to the predictions from a population with solar metallicity, while the shaded regions of the same color represent the effect of lowering the metallicity, from solar down to $Z = 0.0002$ for the stripped stars (see \appref{app:metallicity} and \tabref{tab:Z_combination}). The left column shows the emission rates from a co-eval stellar population with initially $10^{6}\Msun$ in stars, and the right column shows the emission rates from a stellar population in which stars form at the constant rate of 1\Msunyr. The top, middle, and bottom rows show the emission rates of \HI-, \HeI-, and \HeII-ionizing photons, respectively (\Qz, \Qo, and \Qt). 
}
\label{fig:pop_ionizing_time}
\end{figure*}

We compare the expected contribution from the stripped stars with that from the massive single stars in the same stellar population. For this comparison, we use \code{Starburst99} to represent the emission from the massive stars, as described in \secref{sec:model_pop_stripped_SB99}. We also show the predictions from the \code{BPASS} models.

\subsection{Predictions for co-eval stellar populations}


In \figref{fig:pop_ionizing_time}a, we show the emission rate of \HI-ionizing photons for a co-eval stellar population as a function of time after starburst. The massive stars in the stellar population primarily emit their \HI-ionizing radiation during the first 5~Myr. Stripped stars play a significant role at later times. 

At $\sim 10$~Myr, when the first stripped stars have been created in our simulations, they emit ionizing photons with a rate of $\sim 10^{51}$ photons per second for a burst of $10^6\Msun$ formed stars. This is an emission rate of about a factor of ten higher than what massive single stars produce at that time. 
About 100~Myr after starburst, the emission rate of \HI-ionizing photons from stripped stars has decreased to $\sim 10^{49}$~s$^{-1}$, which corresponds to the typical emission rate of one WR-star or one massive O-star \citep{2002MNRAS.337.1309S, 2005A&A...436.1049M, 2014A&A...564A..30G} and is several orders of magnitudes higher than expected from single stars with an age of 100~Myr. The emission of \HI-ionizing photons keeps decreasing with time, and the trend suddenly steepens around 300~Myr. This is because the stripped stars that are present in the population are subdwarfs that have hydrogen-rich surfaces as a result of diffusion processes, as we discussed briefly in \secref{sec:SED_coeval}, see also Sect.~3 of \citetalias{2018A&A...615A..78G}. This results in a sharper Lyman cut-off in their spectra and thus a drop-off in the contribution of \HI-ionizing photons. After $\sim 1$~Gyr, the total ionizing emission rate is similar to that of one early B-type star \citep[$\Qz \sim 10^{45}$~s$^{-1}$, cf.][]{2002MNRAS.337.1309S}. 

Metallicity does not significantly affect the emission rate of \HI-ionizing photons from stripped stars, as can be seen from the width of the shading in \figref{fig:pop_ionizing_time}a. Because stripped stars have high temperatures ($T_{\rm eff} > 20\, 000$~K) at all metallicities, the radiation peaks in the \HI-ionizing wavelengths and \Qz is, therefore, less dependent on metallicity.
For stripped stars with progenitor masses that are lower than about 4\Msun, the emission rate of \HI-ionizing photons increases with decreasing metallicity. However, because the time of stripping and duration of the stripped phase decreases with decreasing metallicity, the total emission rate of \HI-ionizing photons from stripped stars in a population remains similar. 
The impact of metallicity on the \HI-ionizing emission from main-sequence stars is larger than for stripped stars. The \HI-ionizing emission rate from massive stars increases by up to a factor of five when metallicity is lowered. The boost of ionizing photons at late times due to the presence of stripped stars is independent of metallicity.

The presence of stripped stars appear to cause a sudden increase in \HI-ionizing emission around 10~Myr. This is likely an artifact of our models not reaching higher progenitor masses than 20\Msun and therefore do not account for stripped stars that are formed earlier than 10~Myr.

For comparison, we also show predictions from the \code{BPASS} models in \figref{fig:pop_ionizing_time}a. They closely follow the predictions of \code{Starburst99} during the first 5~Myr but show a boost of ionizing photons at later times, similar to our models of stripped stars \citep[see also \citealt{2016MNRAS.456..485S} and][]{2016MNRAS.457.4296W}. Our predictions for the contribution of stripped stars follow the trend of the \code{BPASS} models after 10~Myr. 
The \code{BPASS} models predict a shallower decrease after $\sim 300$~Myr compared to our models. We expect that this is due to differences in how we treat the atmospheres of stars in binary systems.


Stripped stars emit \HeI-ionizing photons at a rate that is about five times lower than that of \HI-ionizing photons. In \figref{fig:SED_population} the emitted luminosities appear to be of similar order for \HI- and \HeI-ionizing radiation. However, the difference in the emission rates of \HeI- and \HI-ionizing photons is larger since photons with shorter wavelengths are more energetic. The difference in emission rates of \HeI- and \HI-ionizing photons can be seen by comparing the panels b and a in \figref{fig:pop_ionizing_time}. The diagrams show that the decline of \Qo closely follows that of \Qz. This is because the temperature of the stripped stars that are responsible for the ionizing emission does not change sufficiently to significantly affect the relative emission of \HeI- to \HI-ionizing photons. 

Once they are formed, stripped stars dominate the emission of \HeI-ionizing photons. The \HeI-ionizing emission rate predicted for single stars by \code{Starburst99} decreases much more steeply with time than it does for stripped stars. Comparing to the emission of \HeI-ionizing photons from the main-sequence and WR stars, we see that stripped stars boost the emission rate by four orders of magnitudes already 10~Myr after starburst. 

Changing the metallicity does not significantly alter the emission rate of \HeI-ionizing photons from stripped stars. 
At late times, after around 300~Myr the evolution of \Qo shows a feature, a temporary rise of the emission rate. The predictions for different metallicities deviate in this region as can be seen from the spread in the shading. This feature originates in the treatment of diffusion processes in surface layers of subdwarfs as discussed earlier. 

The prediction of \Qo from \code{BPASS} closely follows the predictions from stripped stars between 10~Myr and 100~Myr after starburst. At early times, $\lesssim 5$~Myr, we see that the predictions from \code{BPASS} closely follow those from \code{Starburst99} at solar metallicity. However, for low metallicities, $Z \leq 0.004$, the \code{BPASS} models show an increase at these early times. The increase is particularly prominent between 3~Myr and 20~Myr after starburst. This is a result of the treatment of stars that have gained mass or merged through binary interaction, causing them to rotate rapidly. 
Rapid rotation is thought to give rise to mixing processes in the stellar interior, causing the stars to evolve chemically homogeneously \citep[][]{1987A&A...178..159M, 2005A&A...443..643Y, 2007A&A...465L..29C}. 
Chemically homogeneous stars are thought to be very hot and bright and, if they exist, they can contribute with a significant amount of ionizing photons \citep{2011A&A...530A.115B, 2015A&A...573A..71K, 2015A&A...581A..15S, 2019A&A...623A...8K}. 
Other sources that could affect the emission of \HI-ionizing photons from stellar populations are white dwarfs \mbox{\citep{1984ApJ...287..315P}}, rejuvenated mass-gainers \citep{2009MNRAS.395.1822C, 2014ApJ...782....7D}, and central stars in planetary nebulae or post-AGB stars \citep[e.g.,][]{2016A&A...588A..25M}.


Our models predict that stripped stars are important contributors of the \HeII-ionizing photons emitted by stellar populations. \figref{fig:pop_ionizing_time}c shows that stripped stars reach emission rates of $\sim 10^{48.5}$~s$^{-1}$, which is similar to the emission rates from massive stars. Stripped stars maintain their high emission rate for longer because of their longer lifetimes and because they form from progenitors with a range of ages. The \HeII-ionizing emission is strongly temperature dependent. The decline of \HeII-ionizing emission with time is, therefore, steeper for \Qt than for \Qz or \Qo. After about 50~Myr, the emission rate of \HeII-ionizing photons from stripped stars falls below to $10^{45}$~s$^{-1}$. 

Metallicity affects the emission rate of \HeII-ionizing photons from stripped stars, causing variations that span two orders of magnitude. The largest deviation occurs at the lowest metallicities, where we find that the stripping process fails to remove all the hydrogen, resulting in cooler stars \citepalias[][\citealt{2017ApJ...840...10Y}]{2017A&A...608A..11G}. The emission rate of \HeII-ionizing photons is determined in the steep Wien-part of the stellar spectrum, meaning that a small shift in temperature results in a large shift in \Qt. The emission rate of \HeII-ionizing photons is also dependent on parameters of the stellar wind and can increase by several orders of magnitudes if, for example, the mass-loss rate is just a few factors lower \citepalias{2017A&A...608A..11G}.


A feature in the otherwise smooth decline of \Qt for stripped stars is seen in \figref{fig:pop_ionizing_time}c about 15~Myr after starburst. This feature occurs because of the change of the treatment of wind clumping in the atmospheres of the stripped stars (see \citetalias{2018A&A...615A..78G} for details). Future detailed spectral analysis of observed stripped stars is necessary to constrain the properties of the stellar winds.

The \code{BPASS} predictions for \Qt roughly follow the predictions of \code{Starburst99} for the first $\sim 3$~Myr. Between 10~Myr and 50~Myr, they are consistent with our predictions for stripped stars. After that, \code{BPASS} predicts an emission rate that is several orders of magnitude higher and that stays high for the remaining considered time, probably because of the adopted atmosphere models for central stars in planetary nebulae and the inclusion of the emission from X-ray binaries (priv.\ communication with JJ Eldridge). 
The models of \code{BPASS} predict that metallicity variations give rise to a large range of values for \Qt before about 10~Myr has passed. Similar to the case of \Qo, we believe that these variations are due to rapidly rotating stars. 
Stars that likely play important roles as emitters of \HeII-ionizing photons are accreting white dwarfs and X-ray binaries \citep{2015MNRAS.453.3024C, 2013ApJ...764...41F, 2017ApJ...840...39M, 2019A&A...622L..10S}. These binaries reside in late evolutionary stages of interacting binaries, where one star has already died, possibly after interaction already occurred, and the second star now fills its Roche-lobe and transfers material to the compact object. 

\subsection{Predictions for continuous star formation}

We show the predicted emission rates of ionizing photons for continuous star formation in \figref{fig:pop_ionizing_time}d, e, and f. Once equilibrium is reached, stripped stars emit \HI-ionizing photons at a rate of $10^{51.8}$~s$^{-1}$. This is about 5~\% of the emission rate from massive stars ($10^{53.1}$~s$^{-1}$). As \figref{fig:pop_ionizing_time}d shows, the massive stars dominate the emission of \ion{H}{I}-ionizing photons in the case of continuous star formation.

Stripped stars emit \HeI-ionizing photons at a rate that is about five times lower than the emission rate from massive stars.
The contribution from stripped stars corresponds to about 15~\% of the total stellar emission of \HeI-ionizing photons. Depending on the assumed metallicity, the contribution varies somewhat (between 10-20~\%). At lower metallicity, the massive main sequence stars are hotter and therefore contribute with a larger fraction compared to the stripped stars.

The emission of \HeII-ionizing photons shown in \figref{fig:pop_ionizing_time}f is dominated by the contribution from stripped stars as they emit about five times more \HeII-ionizing photons than the massive stars. Stripped stars reach emission rates of \HeII-ionizing photons of $10^{49}$~s$^{-1}$, while the massive stars only reach an emission rate of $10^{48.3}$~s$^{-1}$. 
We note that the emission rate of \HeII-ionizing photons is sensitive to metallicity variations. At very low metallicity ($Z \sim 0.0002$), the emission rate from stripped stars decreases by two orders of magnitude. Conversely, the emission rate from massive stars increases by a factor of five at very low metallicity.

\section{Implications for observable quantities}\label{sec:observables}

In this section, we discuss the implications of accounting for stripped stars for various observable quantities commonly used to describe unresolved stellar populations. We summarize the values for the considered quantities in \tabref{tab:observable_parameters} at several snapshots after a starburst of $10^6 \Msun$ and also for continuous star formation, taken after 500~Myr.

\begin{sidewaystable*}
\begin{center}
\caption{Values of observable quantities for models of stellar populations including stripped stars. We use curly brackets to show values for only the stripped stars in the population and regular parentheses to show the values for single star populations \citep[predicted by \code{Starburst99},][]{1999ApJS..123....3L, 2010ApJS..189..309L}. }
\label{tab:observable_parameters}
{\small
\input{Table_parameters_Z0.014.tex}
}
\end{center}

{\textbf{Notes.} The presented quantities are the following. First, the emission rates of \HI-, \HeI-, and \HeII-ionizing photons, which we refer to as \Qz, \Qo, and \Qt, respectively (\secref{sec:ionizing_time}). Then, the production efficiencies of \HI-, \HeI-, and \HeII-ionizing photons, labeled \xii, \xiio, and \xiit, respectively (\secref{sec:xi_ion}). For calculating the production efficiencies, the UV luminosity of the following column was used (see \eqref{eq:xi_ion} and also \secref{sec:beta}). The next column displays the ionization parameter, $U$ (\secref{sec:U}), and the last column the slope of the UV continuum, $\beta$ (\secref{sec:beta}). 
We assume the gas density to be $n_{\text{H}} = 100$~cm$^{-3}$ when calculating the ionization parameter. Snapshots for which no ionizing radiation was published are marked with '--'.}
\end{sidewaystable*}

\subsection{Diagnostics of the budget of ionizing photons}\label{sec:diagnostic_ionizing}

\subsubsection{Production efficiency of ionizing photons, $\xi_{\text{ion}}$}\label{sec:xi_ion}

The production efficiency of ionizing photons is a quantity that can be measured observationally. It relates the emission rates of ionizing photons to the UV luminosity and is, therefore, a parameter that describes the strength of the ionizing emission independent on the stellar mass or star formation rate of the population. The production efficiency of hydrogen-ionizing photons has already been measured for a large number of unresolved stellar populations \citep{2013ApJ...768...71R, 2015MNRAS.454.1393S, 2016ApJ...831..176B, 2017MNRAS.465.3637M, 2018ApJ...855...42S}.

The production efficiency of ionizing photons is defined as follows:
\begin{equation}\label{eq:xi_ion}
\xi_{\text{ion}} = \dfrac{Q_{\text{pop}}}{L_{\nu} (1500\, \AA)},
\end{equation}
where $Q_{\text{pop}}$ is the emission rate of ionizing photons and $L_{\nu} (1500\, \AA)$ is the luminosity at the wavelength 1500\,\AA\ in units of erg~s$^{-1}$~Hz$^{-1}$. We use \Qz, \Qo, and \Qt in \eqref{eq:xi_ion} to calculate the production efficiencies of \HI-, \HeI-, and \HeII-ionizing photons, which we refer to as \xii, \xiio, and \xiit, respectively. We average the UV luminosity between 1450\,\AA\ and 1550\,\AA\ to estimate the continuum luminosity and avoid fluctuations caused by spectral features. 

\begin{figure*}
\centering
\includegraphics[width=.43\hsize]{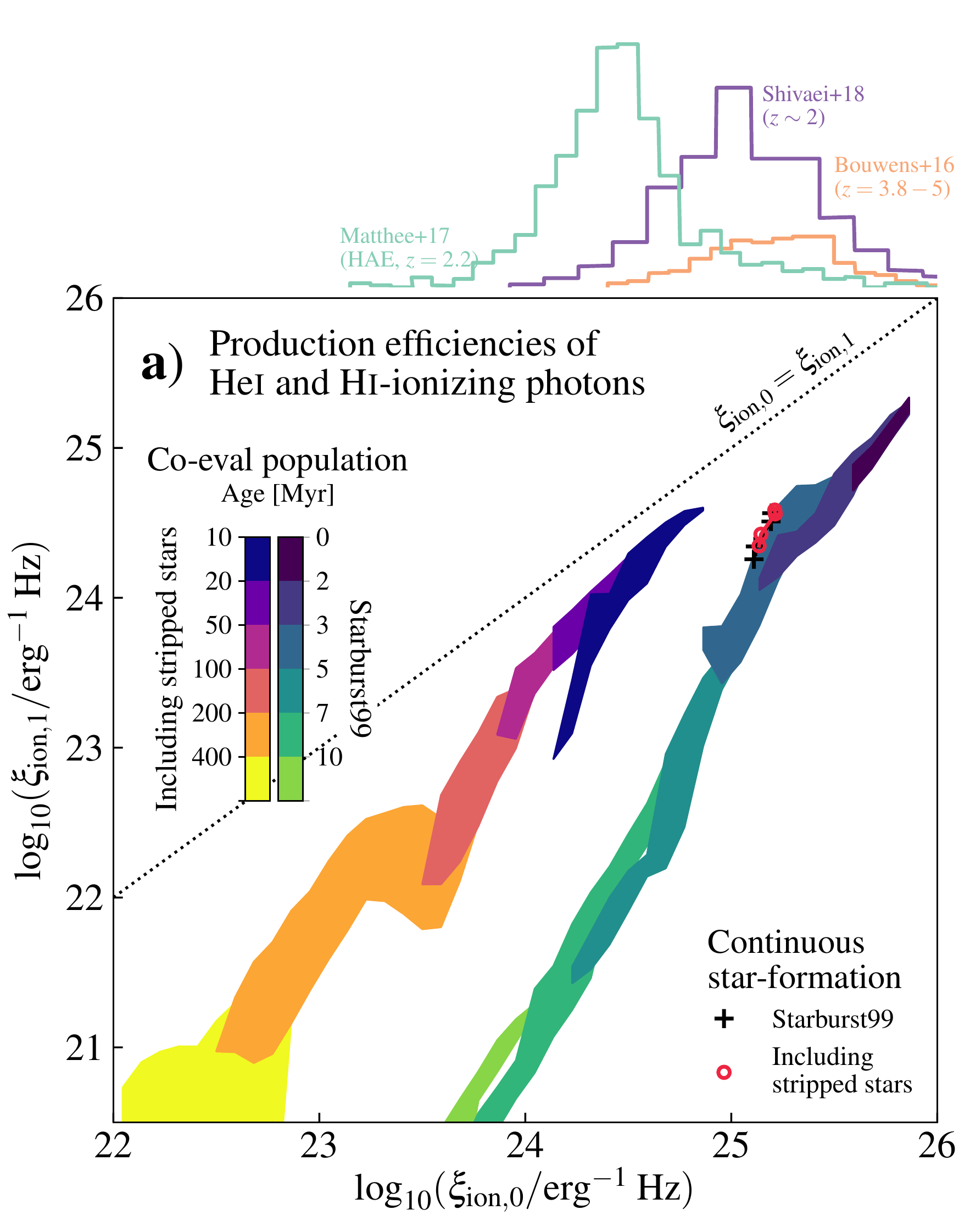}
\hspace{5ex}
\includegraphics[width=.43\hsize]{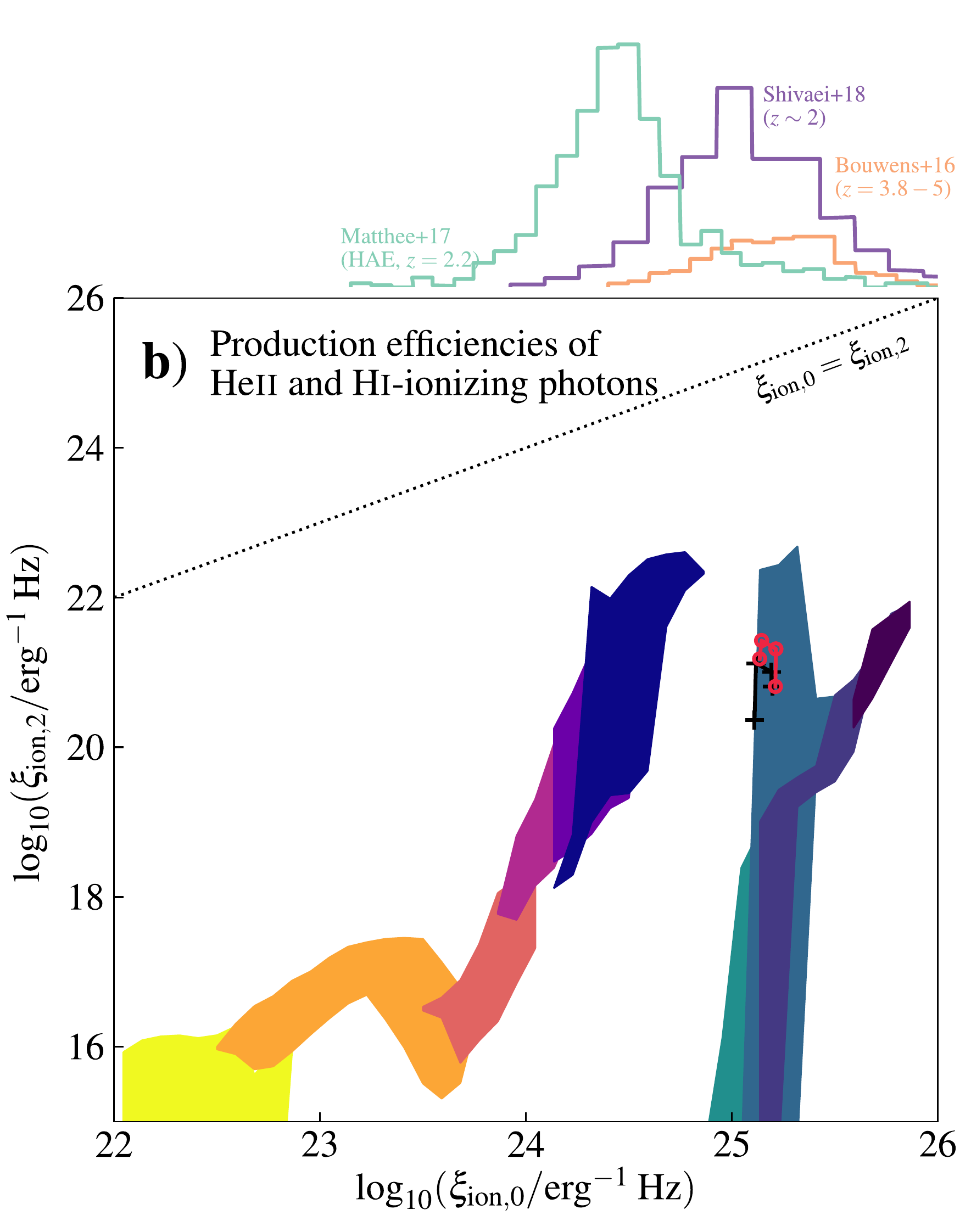}
\vspace{5mm}
\caption{Diagrams showing the hardness of the ionizing part of the spectrum of stellar populations, constructed by the production efficiencies of \HI- and \HeI- (Panel a) or \HeII-ionizing photons (Panel b) (\xii, \xiio, \xiit). The colored regions represent the time spans indicated by the color bars for co-eval stellar populations and cover the predictions from all metallicities (from solar metallicity down to $Z = 0.0002$ for stripped stars, see \appref{app:metallicity} and \tabref{tab:Z_combination}). Green shades show predictions for young, single star populations, while the purple and red shades represent the hardness of stellar populations in which stripped stars are included. The case of constant star formation is shown with solid lines and markers (pluses in black for single stars and circles in red for when stripped stars are included), taken after 500~Myr. Above the diagrams, we show the distribution of measured \xii for three samples of observed unresolved stellar populations from intermediate to high-redshift by \citet{2016ApJ...831..176B}, \citet{2017MNRAS.465.3637M}, and \citet{2018ApJ...855...42S}.}
\label{fig:xi_ion}
\end{figure*}

For co-eval stellar populations, the ionizing radiation that stripped stars produce causes the production efficiencies of ionizing photons to remain at high values for much longer than what is predicted from single star populations. The integrated spectrum is harder if stripped stars are present, which can be recognized by comparing the production efficiency of \HI-ionizing photons with either that of \HeI- or \HeII-ionizing photons. This is visualized in the hardness diagrams shown in \figref{fig:xi_ion}.

\figref{fig:xi_ion}a shows that the same range of values for the production efficiency of \HI-ionizing photons can be produced by either massive stars in a stellar population younger than 10~Myr or stripped stars in a stellar population that is up to ten times older \citep[cf.][]{2016MNRAS.458L...6W}. In the figure, a clear separation is visible between a population that contains stripped stars and one that does not. The difference is due to the harder ionizing spectra that stripped stars introduce, which shift the production efficiencies of helium-ionizing photons to higher values relative to what is expected from a single star population. In the case of \xiit, the separation is several orders of magnitude. For continuous star formation, the role of stripped stars is small for the production efficiencies of ionizing photons but could be relevant for \xiit, as can be seen from \figref{fig:xi_ion}b and in \tabref{tab:observable_parameters}. 

For reference, we also show three distributions of measured \xii in observational samples of galaxies on top of the diagrams in \figref{fig:xi_ion}. These samples are of distant galaxies with various classifications and span a range of redshifts \citep{2016ApJ...831..176B, 2017MNRAS.465.3637M, 2018ApJ...855...42S}. The observed galaxies have a broad range of \xii. For consistency, we chose to show the samples with the dust-correction made in the same way when measuring the UV luminosity (assuming a \citealt{1994ApJ...429..582C} slope for the dust extinction). We note that the method to account for the dust correction has an impact on the distribution of the estimated \xii \citep[][see also \citealt{2011ApJ...741..124H, 2011ApJ...737...67M}]{2017MNRAS.465.3637M}.

An interesting test of the underlying stellar population would be to measure the production efficiencies of helium-ionizing photons for observed galaxies, which would enable to place them individually in the hardness diagrams. We believe that this would provide a very valuable test for the models of stellar populations, and in particular for the impact of binary stellar evolution.

\subsubsection{Ionization parameter, $U$}\label{sec:U}

The ionization parameter is traditionally used to quantify the degree of ionization a stellar population causes on the surrounding nebula as it compares the flux of ionizing photons to the density of the surrounding gas \citep[e.g.,][]{1989agna.book.....O}. 
As a result, populations with the same ionization parameter typically show very similar nebular spectra even though they may have different stellar masses or star formation rates \citep[e.g.,][]{2000ApJ...542..224D, 2014MNRAS.442..900N}.

We follow the standard definition of the ionization parameter, $U$, \citep[e.g.,][see also \citealt{2013ApJ...774..100K}]{1990ARA&A..28..525S}, where the isotropically emitted ionizing radiation from a central source is compared to the density of the gas surrounding the source:
\begin{equation}\label{eq:U}
U \equiv \dfrac{\Qz}{4\pi R_S^2 n_H c} = \dfrac{\alpha_B ^{2/3}}{3^{2/3}c} \left( \dfrac{\Qz \epsilon^2 n_H}{4\pi} \right)^{1/3}.
\end{equation}
In \eqref{eq:U}, $n_H$ is the number density of hydrogen in the gas, $\epsilon$ is the volume filling factor of the gas, $\alpha_B$ is the recombination coefficient for hydrogen, and $c$ is the speed of light. In the last equality of \eqref{eq:U}, we expanded the Str\"{o}mgren radius, $R_S = [3\Qz/(4\pi n_H^2 \alpha_B \epsilon)]^{1/3}$ \citep{1939ApJ....89..526S}. When calculating the ionization parameter, we account for clumping in the nebula by assuming $\epsilon = 0.1$, following \citet{2013ApJ...769...94Z}. We assume a typical gas temperature of $10\,000$~K, which leads $\alpha_B = 2.6 \times 10^{-13}$~cm$^3$~s$^{-1}$ \citep[Case~B type recombination,][]{2006agna.book.....O}. We adopt the emission rates of ionizing photons presented in \secref{sec:ionizing_time} and assume a range of gas densities from $n_{\text{H}} = 10$ to $10^4$~cm$^{-3}$.

\begin{figure*}
\begin{minipage}{\textwidth}
\centering
\vspace{3mm}
\includegraphics[width=.9\textwidth]{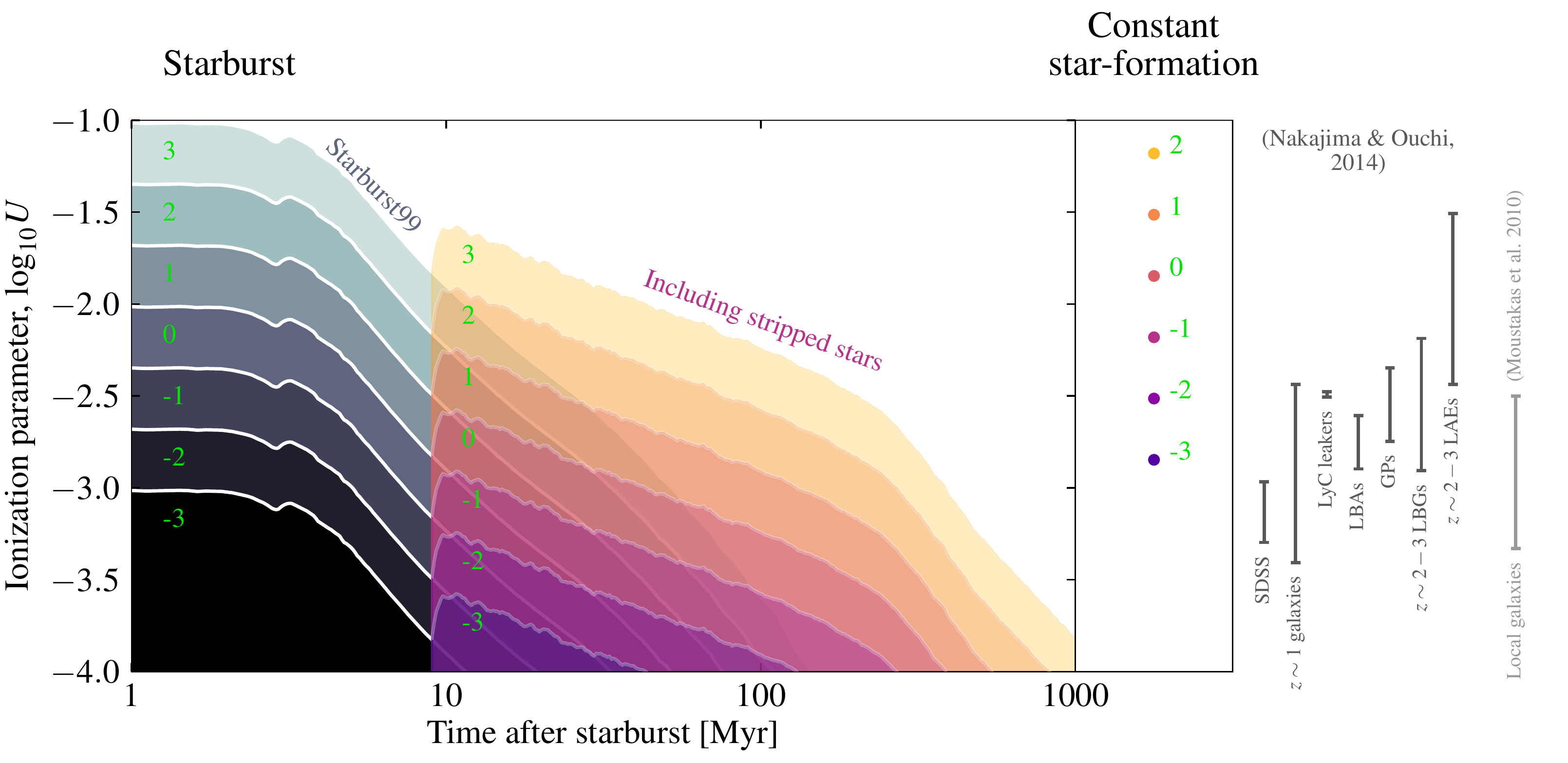}
\vspace{4ex}

\begin{minipage}[b]{\textwidth}
\centering
{\footnotesize
\begin{tabular}{|c|c|c|c|c|c|c|c|}
\multicolumn{6}{l}{{\small Explanation of the green labels.}} \\ 
\multicolumn{6}{l}{}\\
\toprule
Starburst & $10^3 \Msun$ & $10^4 \Msun$ & $10^5 \Msun$ & $10^6 \Msun$ & $10^7 \Msun$ & $10^8 \Msun$ & $10^9 \Msun$ \\
\midrule
Constant & $0.001 \Msunyr$ & $0.01 \Msunyr$ & $0.1 \Msunyr$ & $1 \Msunyr$ & $10 \Msunyr$ & & \\
\midrule
10~cm$^{-3}$ & -- & -3 & -2 & -1 & 0 & 1 & 2 \\
\midrule
100~cm$^{-3}$ & -3 & -2 & -1 & 0 & 1 & 2 & 3 \\
\midrule
1\,000~cm$^{-3}$ & -2 & -1 & 0 & 1 & 2 & 3 & --\\
\midrule
10\,000~cm$^{-3}$ & -1 & 0 & 1 & 2 & 3 & -- & -- \\
\bottomrule
\end{tabular}
}
\end{minipage}
\end{minipage}
\vspace{3mm}
\caption{\textit{Top:} The ionization parameter computed for co-eval stellar populations as a function of time and for constant star formation, taken after 500~Myr. We show the predictions for stellar populations containing only single stars in gray shades and the model when stripped stars are included in purple shades. For constant star formation, the contribution from stripped stars is about 2\% and therefore we do not show the markers for single star populations as they overlap. We show measurements of the ionization parameter for groups of observed galaxies to the right of the diagram \citep{2010ApJS..190..233M, 2014MNRAS.442..900N}.
\textit{Bottom:} The table explains which gas density and stellar mass for co-eval stellar populations that correspond to which contour in the diagram using numbers as labels. In the case of constant star formation, the numbers are correlated with combinations of the gas density and star formation rate instead.}
\label{fig:logU}
\end{figure*}

Our models show that stripped stars increase the ionization parameter for stellar populations in which star formation has stopped at least 10~Myr ago, as shown in \figref{fig:logU}. The stripped stars cause the ionization parameter to remain at high values for an extended time period. We note, however, that it is likely that the density of the surrounding gas changes with time, which will consequentially alter tthe effect of stripped stars on the ionization parameter.
In the case of constant star formation, stripped stars increase $U$ by about 2\%. 

The ionization parameter depends on the gas density and the emission rate of ionizing photons. The latter is closely related to the stellar mass in case of co-eval stellar populations and the star formation rate in case of constant star formation. We, therefore, compute the evolution of the ionization parameter for a range of gas density, stellar mass, and star formation rate combinations, as seen in \figref{fig:logU}. For reference, we show the measured ranges of several samples of observed galaxies as vertical bars on the right side of the figure \citep{2010ApJS..190..233M, 2014MNRAS.442..900N}. These galaxies are grouped roughly according to their properties and redshift. The local galaxy samples are the SDSS galaxies, the Green Pea galaxies (GPs), the Lyman Continuum (LyC) leakers, and the Lyman-Break Analogs (LBAs) summarized by \citet{2014MNRAS.442..900N} together with the local galaxy sample presented by \citet{2010ApJS..190..233M}. The samples of distant galaxies are the $z\sim 1$ galaxies, the Lyman Break Galaxies (LBGs) and the Lyman Alpha Emitters (LAEs) summarized in \citet{2014MNRAS.442..900N}. The individual galaxies are likely to experience a bursty star formation rate that can resemble the behavior of single starbursts at times (see \appref{app:SFH}, \citealt{2014MNRAS.444.3466S} and \citealt{2019arXiv190504314C}). It is, however, unlikely that they experience constant star formation as a very low star formation rate is inferred in most cases. As stripped stars prolong the time during which a stellar population can create ionizing photons, they could play an important role in shaping the distribution of ionization parameters found in the observed samples. Even though we likely have over-estimated the contribution from stripped stars to the ionization parameter because of the decreasing gas density, they remain interesting to consider since the observed galaxies show a large spread in measured ionization parameters. Assuming that the gas density has decreased by a factor of ten by the time stripped stars are formed, means that we have over-estimated the ionization parameter merely by about a factor of three, while the observed ionization parameters range over about two orders of magnitude. 

Co-eval stellar populations, such as individual star clusters, have been observed to have ionization parameters that coincide with the distribution for the ionization parameters for the observed galaxies mentioned above. \cite{2019MNRAS.482..384X} show that ionization parameters for clusters are typically between $-2.5$ and $-4$. This suggests that the comparison with a co-eval stellar population for the observed galaxies is justified.

We do not expect that stripped stars contribute significantly to the ionizing emission from populations with very high measured values for the ionization parameter of $\log_{10} U \gtrsim -2$ as observed by, for example, \citet{2010ApJ...719.1168E} and \citet{2018ApJ...865...55L}. When stripped stars dominate the ionizing emission, such high ionization parameters require that the galaxy is of high mass ($\gtrsim 10^{8} \Msun$) and that star formation has halted about 10~Myr ago (see \figref{fig:logU}). Galaxies with such high ionization parameters are observed to have ongoing star formation.

\subsection{Impact on the UV luminosity and the UV continuum slope, $\beta$}\label{sec:beta}



The luminosity, $L_{\nu}$, in the ultraviolet wavelengths has traditionally been used as a diagnostic for the star formation rate of stellar populations \citep{1998ARA&A..36..189K} as the wavelength range is dominated by the emission from young and massive stars. Despite stripped stars are very hot, they do not significantly impact the UV luminosity in stellar populations that form stars at a constant rate or in which star formation has halted less than 500~Myr ago. We display the results for $L_{\nu}$ at 1500~\AA\ in \tabref{tab:observable_parameters}. 

The slope of the UV continuum, $\beta$, can be used to infer dust attenuation of stellar populations \citep[e.g.,][]{1999ApJ...521...64M}. Similar to the UV luminosity, the slope of the UV continuum is not affected by the presence of stripped stars unless star formation has halted more than about 100~Myr ago. To quantify the effect, we take the common approach and define the UV continuum slope as the exponent in a power-law: $L_{\lambda} \propto \lambda^{\beta}$, between 1250\,\AA\ and 2600\,\AA\ \citep{1994ApJ...429..582C}. 

We find that stripped stars do not significantly affect the slope of the UV continuum if the slope is steep, $\beta \lesssim -0.5$. 
For such cases, the UV is dominated by hot main-sequence stars. For shallower slopes of $\beta \sim -0.5$, stripped stars change $\beta$ by at least 0.1, and for even shallower slopes stripped stars can dominate the UV radiation. This resembles the observed phenomenon called the UV-upturn \citep{1988ApJ...328..440B}, which has been considered to originate from subdwarfs that are formed late after star formation has ended, for example, through binary interaction in low-mass stars \citep{2007MNRAS.380.1098H}. In our models, the UV slope becomes shallower with time for stellar populations in which stars are no longer forming.
In the case of constant star formation, the UV is dominated by radiation from massive stars and the effect of stripped stars on the UV slope is negligible. 
The effect of metallicity is small on both the UV luminosity and the slope of the UV continuum. We show the results for lower metallicity in \tabrefthree{tab:ob_006}{tab:ob_002}{tab:ob_0002}.

We conclude that both the UV luminosity and the slope of the UV continuum are un-affected by the presence of stripped stars in stellar populations in which star formation is ongoing or that are younger than about 100~Myr. Therefore, in such stellar populations, the method of inferring the star formation rate using the UV luminosity remains the same as well as inferring the dust attenuation using the UV continuum slope \citep[cf.][]{2018ApJ...853...56R}.

\subsection{Impact on spectral features}\label{sec:nebular_ionization}

We find that the stellar emission-line contribution from stripped stars is not distinguishable in the integrated spectrum of stellar populations. An emission feature that is observed to be very strong among stellar populations is $\HeII \, \lambda 1640$ \citep[e.g.,][]{2015ApJ...801L..28K} and the equivalent width of this line increases at most by 1~\AA\ when stripped stars are included. This maximum increase occurs when the most massive stripped stars appear in co-eval and high-metallicity stellar populations. This also contrasts with the observed increase of the \HeII~$\lambda 1640$ feature with decreasing metallicity \citep[e.g.,][]{2012MNRAS.421.1043S}. We note that higher wind mass-loss rates or slower stellar winds would increase the equivalent widths of the emission lines from stripped stars, as they are mainly formed by recombination in the stellar wind (see \secref{sec:modeling} and \citetalias{2018A&A...615A..78G} for a discussion). \citet{2012MNRAS.419..479E} found that \code{BPASS} suggests that the \HeII~$\lambda 1640$ feature could be significantly enhanced if rotation of massive stars leading to chemically homogeneous evolution is invoked. However,  the observed \HeII~$\lambda 1640$ emission is sometimes even stronger than what can be explained with rotating stars \citep{2019A&A...624A..89N}.

The impact from stripped stars on the nebular spectrum is likely more interesting because their ionizing radiation affects the ionization state of the gas and thus also the spectral features emitted by the nebula. Nebular features are important diagnostics for the nature of observed galaxies and can, for example, be used to determine whether the ionizing source is stellar or quasar \citep[][see also \citealt{2015MNRAS.449..559S}]{2016MNRAS.456.3354F, 2016MNRAS.462.1757G}. Here, we discuss likely effects that stripped stars have on the nebula based on simple considerations of the hardness of the ionizing spectrum. We assume that the gas density is sufficiently high for the ionizing emission from stripped stars to produce nebular emission lines. However, we note that the gas density likely has decreased when stripped stars are formed and the nebular features may therefore be relatively faint. Detailed modeling of the nebula is required to understand the strength of the nebular features caused by stripped stars. This is out of the scope of this study, but the topic of a forthcoming paper.


\begin{figure*}
\centering
\includegraphics[width=.95\hsize]{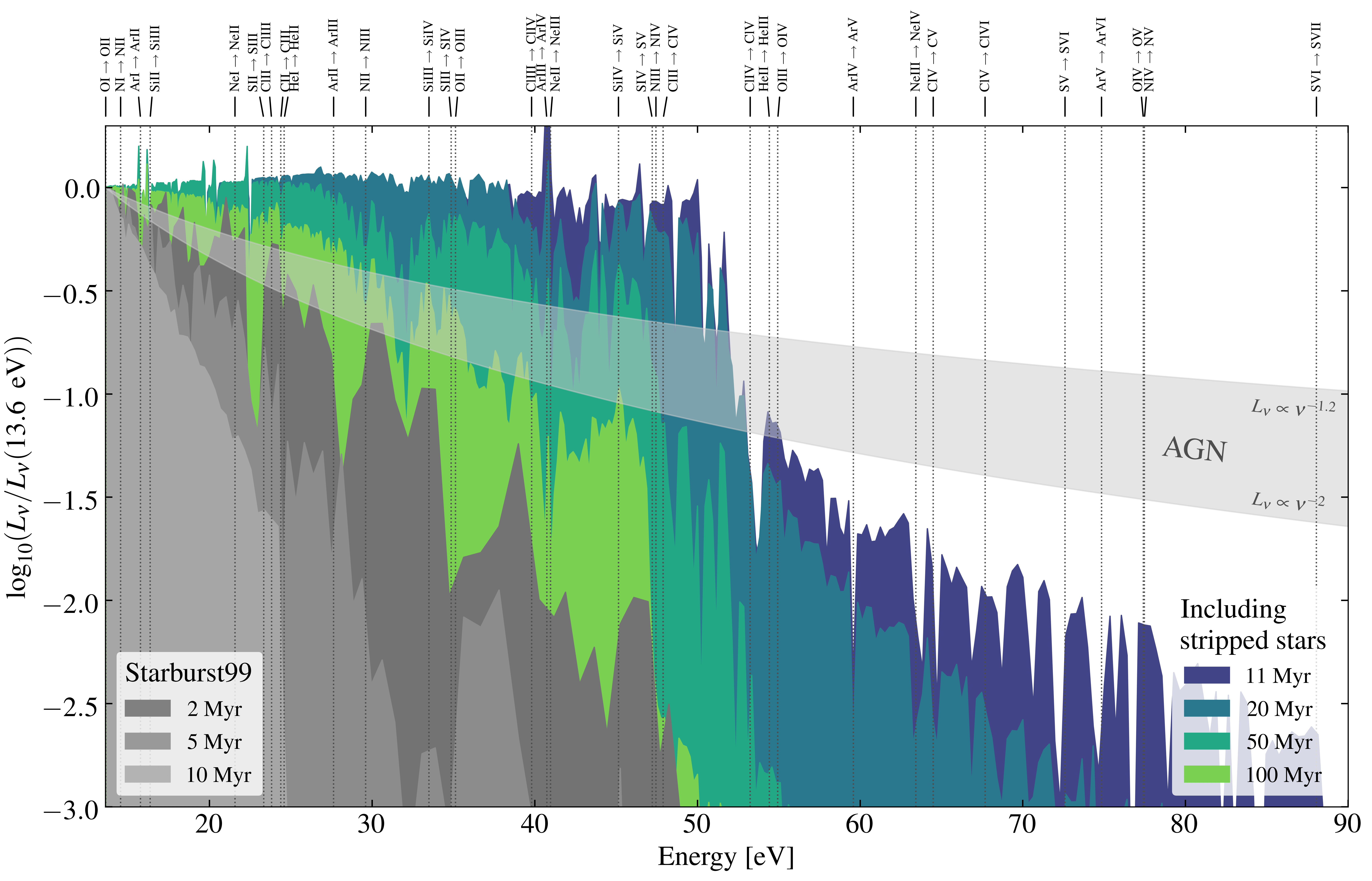}
\caption{Shape of the ionizing part of the spectra of co-eval stellar populations compared to that of AGN, shown by normalizing the spectra at the \HI-ionizing threshold at the photon energy 13.6~eV. We use gray for the spectra of stellar populations that only contain single stars and green for spectra of stellar populations in which stripped stars are included. Expected spectral slopes for AGN are indicated in transparent gray. We mark the ionization thresholds of different ionization stages for a variety of elements above the figure. }
\label{fig:ionization_species}
\end{figure*}

\figref{fig:ionization_species} shows the shape of the ionizing part of the spectra of co-eval stellar populations compared to AGN (cf.~\citealt{2014ApJ...795..165S, 2015MNRAS.454.1393S, 2016MNRAS.456.3354F}). We assume that the spectra of AGN are characterized by a power-law, $L_{\nu} \propto \nu^{\alpha}$, with a slope of $-2 < \alpha < -1.2$ \citep[e.g.,][]{2016MNRAS.456.3354F}. The figure shows that the spectra from single star populations always are softer than that of AGN as they are characterized by steep spectral slopes of $\alpha \lesssim -2.5$ \citep[see also][]{2019ApJ...874..154D}. When stripped stars are present, the ionizing spectrum for photon energies lower than about 50~eV is harder. The slope is even close to flat for stellar populations younger than 50~Myr. The spectrum becomes softer than what young and massive stars can produce first after more than 100~Myr after a starburst.

The ionizing part of the integrated spectrum is sufficiently hard when stripped stars are included for nebular oxygen to be ionized to \Otwop and carbon to be ionized to \Cthreep, while the ionizing radiation from massive stars favors lower ionization stages such as \Op and \Ctwop. The high ionization of carbon has been detected at high-redshift \citep[e.g., the \CIV~$\lambda 1548$~\AA\ feature,][]{2015MNRAS.454.1393S}, making stripped stars interesting to consider as sources of such energetic photons. We also expect that stripped stars can give rise to high ratios of the nebular emission lines of  \OIII to \OII, which is a ratio used as an ionization measure and commonly labeled O32 \citep[see e.g.,][for observed distributions of O32]{2015ApJ...801...88S, 2017ApJ...836..164S}. Full ionization of nebular helium is favored for the hard ionizing radiation of AGN compared to that of stellar populations. However, when stripped stars are included more \HeII-ionizing photons are produced than when only single stars are considered (see \secref{sec:ionizing_time}), which potentially could give rise to nebular \HeII features, however likely not as strong as observed \citep[cf.][]{2018MNRAS.tmp.1827K}. Promising sources responsible for the strong nebular emission of \HeII-ionizing photons are, for example, X-ray binaries \citep{2019A&A...622L..10S}.

We expect that stripped stars affect the location of stellar populations in the BPT diagram \citep{1981PASP...93....5B}, which is a commonly used diagnostic diagram used to distinguish between star-forming galaxies and galaxies that host an AGN \citep{2001ApJ...556..121K, 2003MNRAS.346.1055K}. For example, given the likely impact of stripped stars on the strength of \OIII lines, we expect that the ratio of $[\OIII]/$H$\beta$ (the vertical axis of the BPT diagram) is affected. This is consistent with a recent study by \citet[][see also \citealt{2014MNRAS.444.3466S}]{2018MNRAS.477..904X} who used \code{BPASS} models and found that populations that contain stripped stars may be located in the same region as small dwarf galaxies with strong ionizing emission are found \citep[e.g., the Green Pea galaxies,][see also the low-mass galaxies of \citealt{2015ApJ...801...88S}]{2009MNRAS.399.1191C}. This questions the age estimate and the stellar content of such stellar populations and similar to \HII regions.

For lower metallicity, the spectra of massive stars become harder and the spectra of stripped stars become softer. However, a distinct difference in the hardness between the spectra of massive stars and those that stripped stars introduce is visible for metallicities larger than $Z = 0.0002$. This is visible in \figref{fig:neb_Z}, which shows figures analogous to \figref{fig:ionization_species}, but for lower metallicities. In the case of $Z = 0.0002$, stripped stars produce as hard ionizing emission as the very youngest massive stars, which is interesting as the impact from stripped stars is present for up to 100~Myr, while the massive stars die after about 5~Myr.

In the case of constant star formation, massive stars dominate and the result is a softer ionizing spectrum. Our models suggest that stripped stars primarily contribute with photons more energetic than 50~eV, but the increase is small and possibly difficult to distinguish.


Stripped stars produce photons capable of ionizing hydrogen and give rise to nebular H$\alpha$ emission \citep{2018MNRAS.477..904X}. 
The contribution from stripped stars to nebular H$\alpha$ is expected to be smaller than from massive stars for stellar populations in which stars form at a constant rate, since the line luminosity is related to the emission rate of \HI-ionizing photons \citep[see e.g.,][]{1995ApJS...96....9L, 2003AA...397..527S}. However, in stellar populations in which star formation has halted, stripped stars can be responsible the formation of nebular H$\alpha$. 

Nebular H$\alpha$ is often used to infer the star formation rate of stellar populations as most \HI-ionizing photons are produced by young and massive stars \citep{1998ARA&A..36..189K}. Our models predict that this method is reliable unless star formation has stopped more than about 10~Myr ago.

\section{Summary \& conclusions}\label{sec:summary_conclusions}

We analyzed the radiative contribution from stars stripped in binaries to the spectral energy distribution and ionizing emission of stellar populations. To do this, we simulated a stellar population including a realistic fraction of binaries to estimate at what time stripped stars are expected to be present. For the spectra, we used our previously computed grid of custom-made atmosphere models \citepalias{2018A&A...615A..78G}.

Our main focus is the ionizing emission since stripped stars emit the majority of their radiation in the ionizing wavelengths. They thus constitute a stellar source of ionizing photons that is neglected in models that only account for single stars. We compare the emission rates of hydrogen and helium ionizing photons from stripped stars to the emission rates from massive single main-sequence stars and WR stars as predicted by \code{Starburst99}. We also compare our results with predictions from the binary spectral synthesis code \code{BPASS}.

We quantify the effects of stripped stars on observable properties for co-eval stellar populations and continuous star formation. These include the production efficiencies of hydrogen and helium ionizing photons (\xii, \xiio, and \xiit), the ionization parameter ($U$), the UV luminosity, and the UV-continuum slope ($\beta$). We find that the presence of stripped stars increases the hardness of the ionizing part of the integrated spectrum. The harder spectrum may be distinguishable when both the production efficiency of hydrogen and helium ionizing photons are inferred. It is also likely that the hard ionizing spectrum gives rise to a different ionization structure of the surrounding nebula and thus a characteristic combination of nebular emission lines.
\newline

\noindent Our main findings are the following: 
\begin{enumerate}
\item Stripped stars are important as an additional stellar source of ionizing radiation. For co-eval stellar populations, stripped stars dominate the ionizing output by several orders of magnitude once they are created, which in our simulations is at an age of $\sim 10$~Myr. In the case of continuous star formation their emission rates of \HI-, \HeI-, and \HeII-ionizing photons reach levels as high as 5, 15, and 500\% compared to what is expected from massive stars alone. \\

\item The ionizing emission from stripped stars is primarily important in stellar populations in which star formation has recently halted. The reason is that stripped stars are created with a delay after the start of star formation. The emission of ionizing photons from stripped stars is significant after about 10~Myr and up to 100~Myr after star formation stopped.\\

\item Our models indicate that stripped stars only impact the integrated spectrum of a stellar population in the ionizing wavelengths.  \\


\item Stripped stars introduce a characteristic hardness in the ionizing part of the spectral energy distribution. The effect on observable properties is not only an increase in the production efficiency of \HI-ionizing photons (\xii), but a relatively larger increase in the production efficiency of \HeI- and \HeII-ionizing photons (\xiio and \xiit). Current measurements of \xii agree well with our predictions. We argue that future measurements of \xiio and \xiit will provide stringent tests for the theoretical models.\\

\item The presence of stripped stars in co-eval populations also affects the commonly used ionization parameter ($U$), causing it to remain between $-3.5 \lesssim \log_{10} U \lesssim -2.5$ for a few 100~Myr. This is a range that often is observed for stellar populations. In the case of continuous star formation, stripped stars only affect the ionization parameter by up to 2\%. \\

\item The ionizing radiation introduced by stripped stars is sufficiently hard to ionize the nebula to high ionization states, which potentially are visible via nebular emission lines if gas is still present at these times. Stripped stars likely have the largest relative impact on the nebular spectrum for stellar populations in which star formation recently halted. In these cases, it is likely that the nebula reaches high ionization states, such as \Otwop, \Cthreep, and possibly also fully ionized helium. Because of the hard, integrated spectrum, we expect high ratios of nebular emission of \OIII to \OII (O32). This means that the location of the stellar population in the BPT diagram is affected as well \citep[cf.][]{2018MNRAS.477..904X}. \\

\item Stripped stars do not significantly affect the UV luminosity or the slope of the UV continuum ($\beta$) in populations in which stars form at a constant rate or in populations that are younger than $\sim$100~Myr. This suggests that in such populations, the UV luminosity can still be used as a reliable diagnostic for the star formation rate and the UV continuum slope can be used to make inferences about dust attenuation. In co-eval populations with ages between about 200~Myr and 1~Gyr, stripped stars significantly affect or even dominate the output of UV radiation, which significantly affects the UV continuum slope. However, at this point the slope is already close to flat, $\beta \gtrsim -0.5$. We note that we only consider the effect from stripped stars, while in reality other products of binary interaction may affect the ultraviolet part of the integrated spectral energy distribution.\\


\item For co-eval stellar populations, metallicity only modestly affects the emission of ionizing photons from stripped stars. However, in the case of continuous star formation, metallicity become more important. 
With lower metallicity, the stripped stars are cooler and their effect on the spectral hardness is therefore smaller. However, other products of binary interaction, such as rapidly spinning accretor stars, may still have large effects \citep[cf.][]{2017PASA...34...58E}.

\end{enumerate}

\noindent Our results are mostly consistent with the results from the more complex binary population and spectral synthesis code \code{BPASS} for the parts that we expect are due to the presence of stripped stars (differences discussed in \appref{app:BPASS_comparison}). Our simulations have enabled an insightful analysis for the impact of stars stripped in binaries in particular. With our models, we show which differences that likely are due to stripped stars. We have discussed in detail (\secref{sec:modeling}) and shown (\appref{app:stripped_phase}) that a simple model is sufficient for representing the emission from stripped stars. It has also allowed us to focus on refining detailed evolutionary models and spectral models especially made for stripped stars, to make sure that these objects are modeled according to our expectations.

Our models are publicly available on the CDS database, where we provide electronic tables with the contribution from stripped stars to the spectral energy distribution and to the emission rates of \HI-, \HeI-, and \HeII-ionizing photons. Our models can also be obtained via the \code{Starburst99} online interface.


\begin{acknowledgements}
We thank the anonymous referee for a constructive report that helped improve the manuscript.
The authors also acknowledge various people for helpful and inspiring discussion at various stages during the preparation of this manuscript, including Evan Bauer, Jared Brooks, Maria Drout, JJ Eldridge, Chris Evans, Rob Farmer, Miriam Garcia, Stephan Geier, Zhanwen Han, Edward van den Heuvel, Stephen Justam, Lex Kaper, Alex de Koter, S{\o}ren Larsen, Danny Lennon, Pablo Marchant, Colin Norman, Philipp Podsiadlowski, Onno Pols, Mathieu Renzo, Hugues Sana, Tomer Shenar, Nathan Smith, Elizabeth Stanway, Silvia Toonen, Jorick Vink, Manos Zapartas and the VFTS collaboration. 
This work was carried out on the Dutch national e-infrastructure with the support of SURF Cooperative. The authors acknowledges John Hillier for making his code, CMFGEN, publicly available. YG thank Martin Heemskerk for providing computing expertise and support throughout the project and Alessandro Patruno for allowing us to use the Taurus computer. 
SdM has received funding under the European Union's Horizon 2020 research and innovation program from the European Research Council (ERC) (Grant agreement No.\ 715063). JHG acknowledges support from the Irish Research Council New Foundations Award 206086.14414 "Physics of Supernovae and Stars".
This work made use of v2.2.1 of the Binary Population and Spectral Synthesis (BPASS) models as last described in \citet{2017PASA...34...58E} and \citet{2018MNRAS.479...75S}.
\end{acknowledgements}


\bibliographystyle{aa.bst}
\bibliography{references_bin.bib}


\appendix


\section{Effect of metallicity}\label{app:metallicity}

\subsection{Number of stripped stars}

\begin{table}
\centering
\caption{Combinations of metallicities that we use to represent Solar metallicity and environments similar to the Large and Small Magellanic Clouds. We also show the combination for an extremely low-metallicity environment. The stripped stars, \code{Starburst99} and \code{BPASS} have models at different metallicities and the same number can therefore not be chosen to represent an environment.}
\label{tab:Z_combination}
\begin{tabular}{lccc}
\toprule\midrule
Environment & Stripped stars & \code{Starburst99} & \code{BPASS} \\
\midrule
Solar & 0.014 & 0.014 & 0.014 \\
LMC-like & 0.006 & 0.008 & 0.006 \\
SMC-like & 0.002 & 0.002 & 0.002 \\
Low-Z & 0.0002 & 0.001 & 0.0001 \\
\bottomrule
\end{tabular}
\end{table}

\begin{figure*}
\centering
\includegraphics[width=0.45\textwidth]{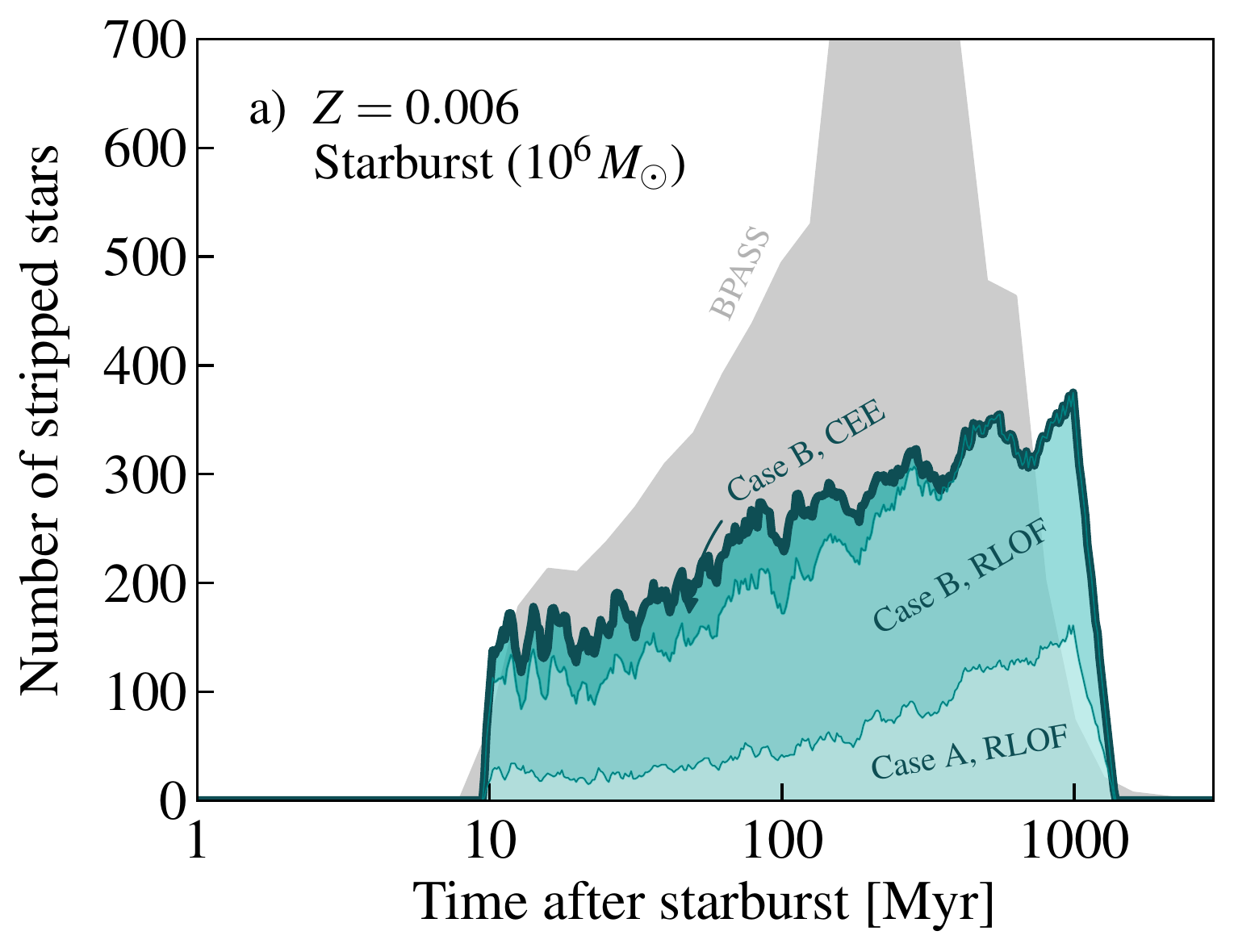}
\hspace{5mm}
\includegraphics[width=0.45\textwidth]{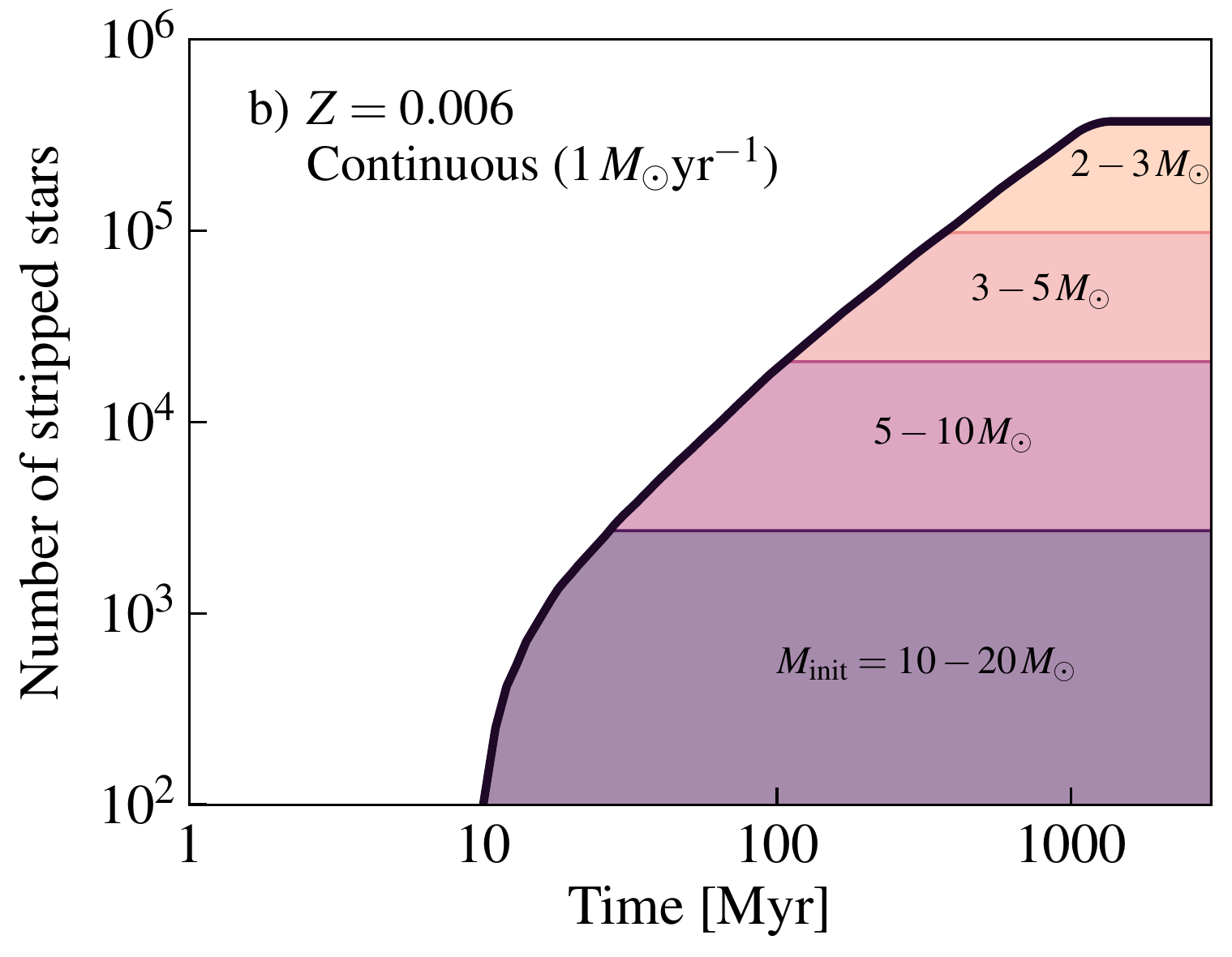}
\includegraphics[width=0.45\textwidth]{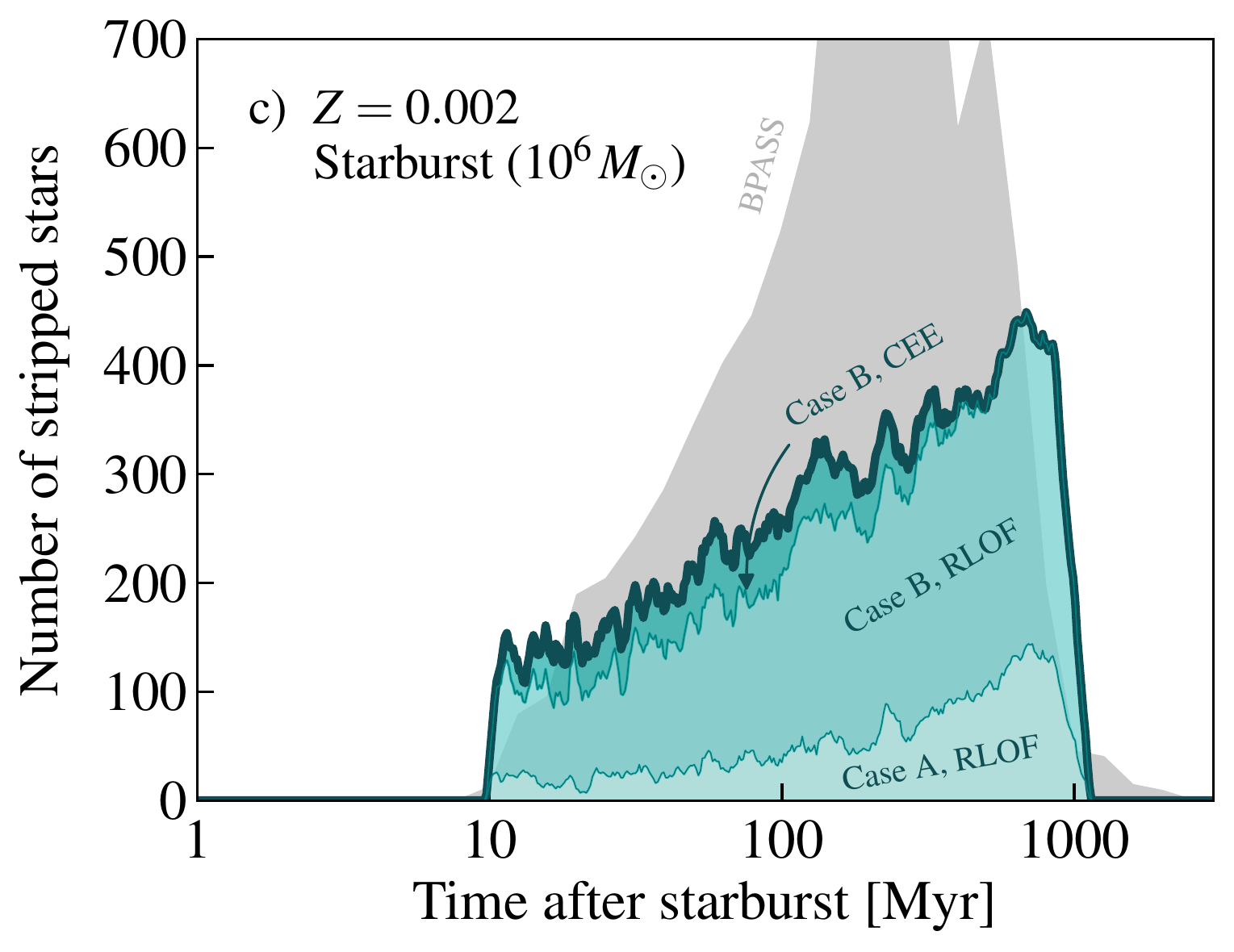}
\hspace{5mm}
\includegraphics[width=0.45\textwidth]{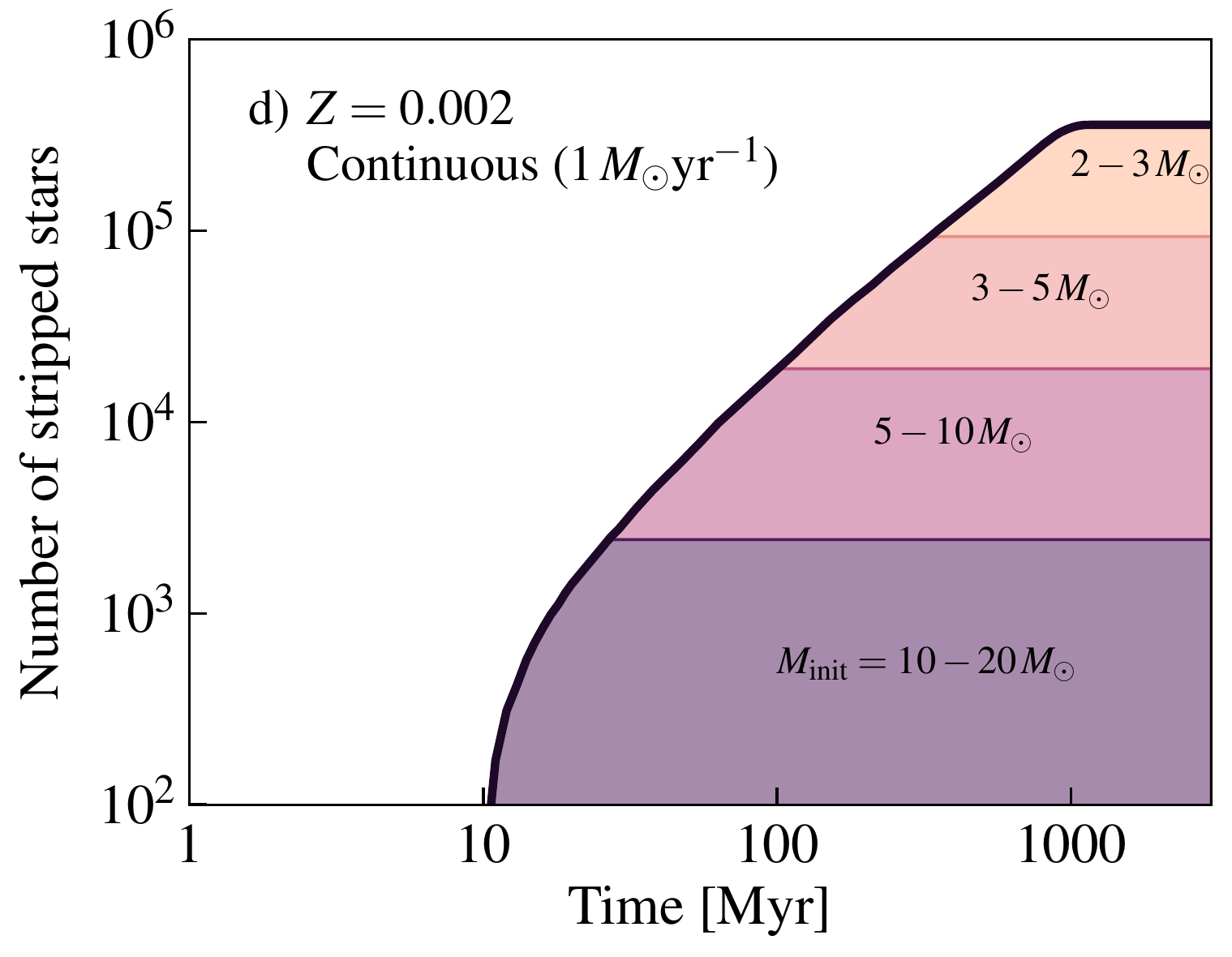}
\includegraphics[width=0.45\textwidth]{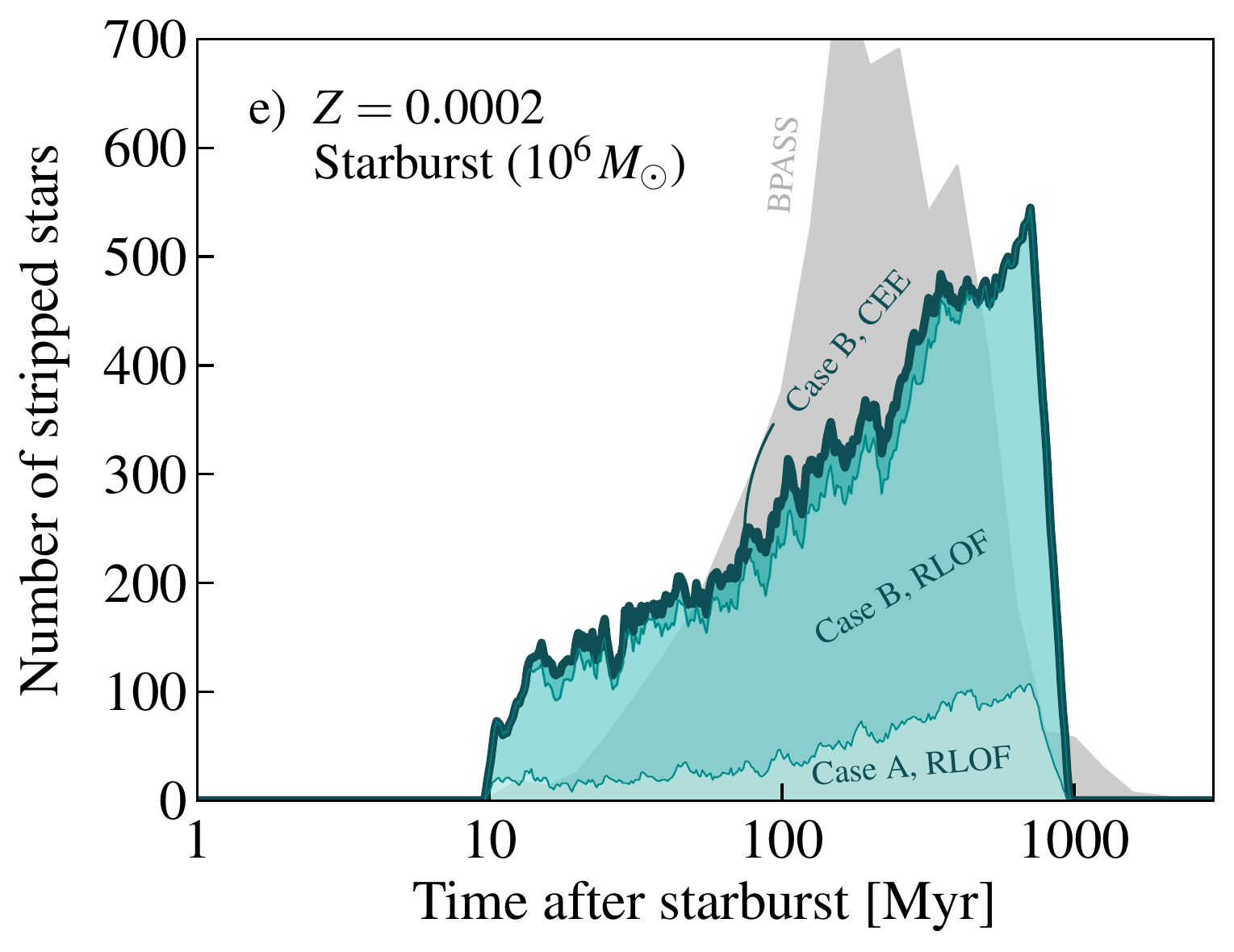}
\hspace{5mm}
\includegraphics[width=0.45\textwidth]{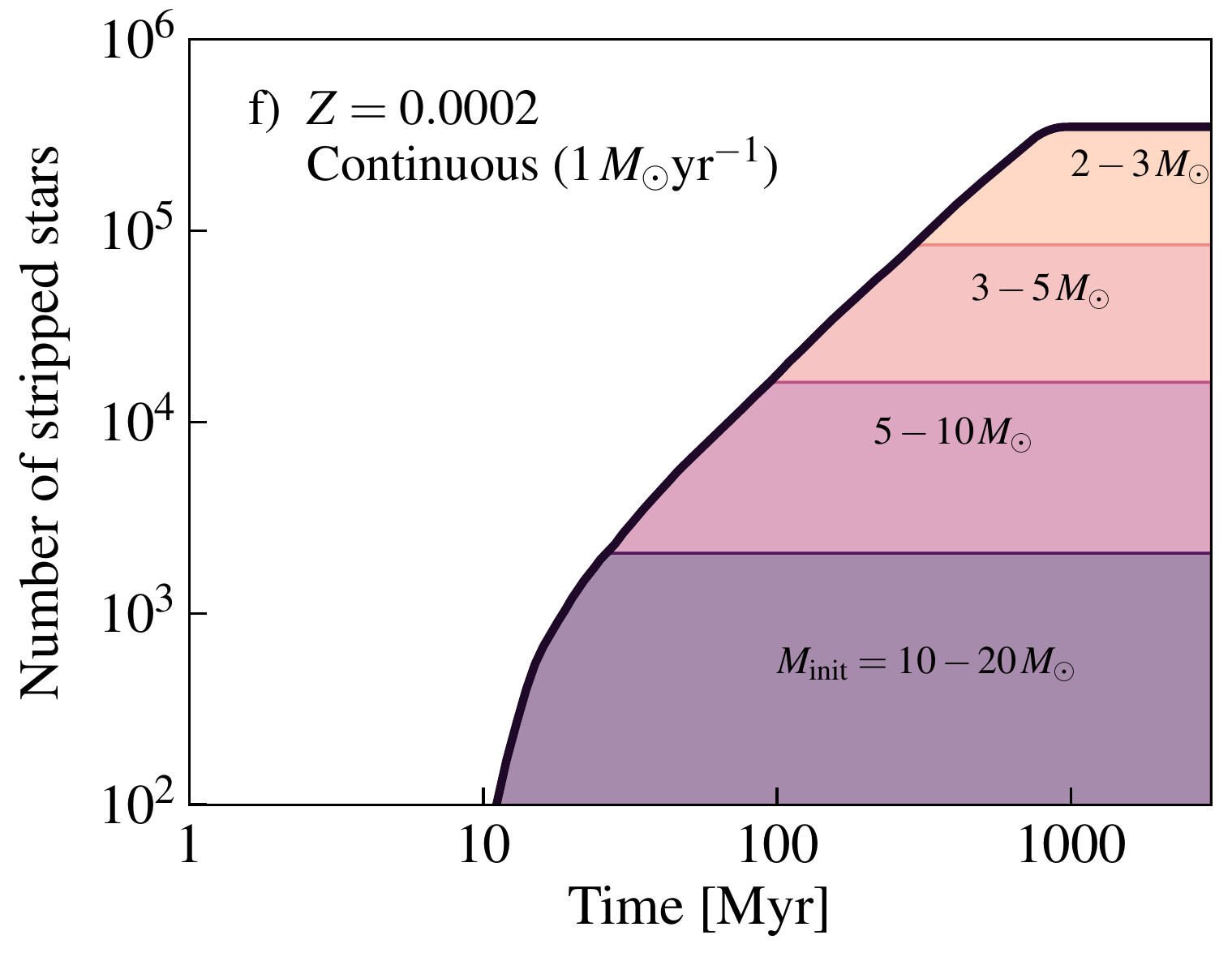}
\caption{Number of stripped stars present in a co-eval stellar population as function of time (left column) and for continuous star formation (right column). These panels are analogous to those in \figref{fig:nbr_time} but for populations with metallicity $Z = 0.006$ (top row), 0.002 (middle row), and 0.0002 (bottom row).}
\label{fig:nbr_time_Z}
\end{figure*}


\noindent Envelope-stripping through binary interaction is not strongly dependent on metallicity \citepalias{2017A&A...608A..11G}, meaning that stripped stars can be formed both at low and at high metallicity. This is not the case for wind-stripped WR stars and envelope-stripping has therefore been considered a formation channel for WR stars in low-metallicity environments \citep[e.g.,][]{2016A&A...591A..22S}. 

\figref{fig:nbr_time_Z} shows the number of stripped stars present in stellar populations at metallicities of $Z = 0.006$, 0.002, and 0.0002. The independence on metallicity for the formation of stars stripped in binaries is visible since both in the co-eval case and for continuous star formation, metallicity does not significantly change the number of stripped stars that are formed. 
In the case of solar metallicity (see \figref{fig:nbr_time}) between around 150 and 400 stripped stars are present in the co-eval stellar population at the times when stripped stars are produced. In the case for very low metallicity ($Z = 0.0002$), the numbers are between 150 and 500. With lower metallicity, the contribution from Case~B type mass transfer somewhat increases while the contribution from Case~A type mass transfer somewhat decreases. The reason is that the expansion during the Hertzsprung gap is relatively larger for lower mass stars at lower metallicity. 
In agreement with the co-eval stellar population, the number of stripped stars in a population is very similar independent on metallicity for the case of continuous star formation. 

The number of low-luminosity WR stars computed in \code{BPASS} is also shown for the co-eval stellar populations in \figref{fig:nbr_time_Z}. The number of these objects is larger for metallicities lower than solar compared to at solar metallicity. The reason could be that other evolutionary processes than envelope-stripping contribute to the formation of what is referred to as low-luminosity WR stars in \code{BPASS}.

\subsection{Integrated spectrum and ionizing emission}

\begin{figure*}
\centering
\includegraphics[width=0.33\textwidth]{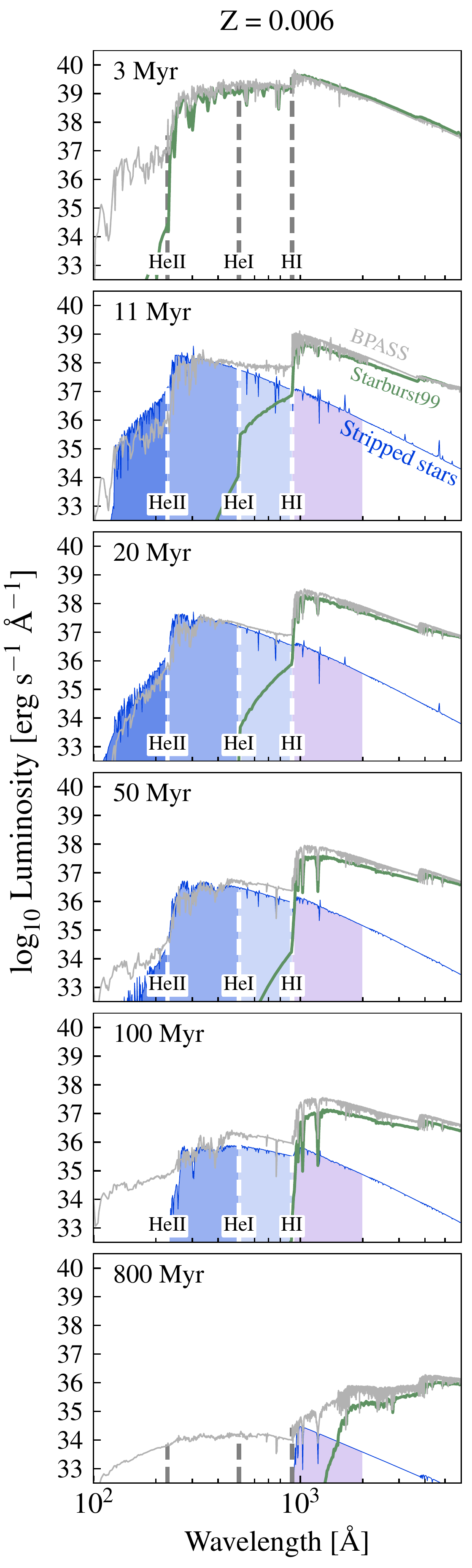}
\includegraphics[width=0.33\textwidth]{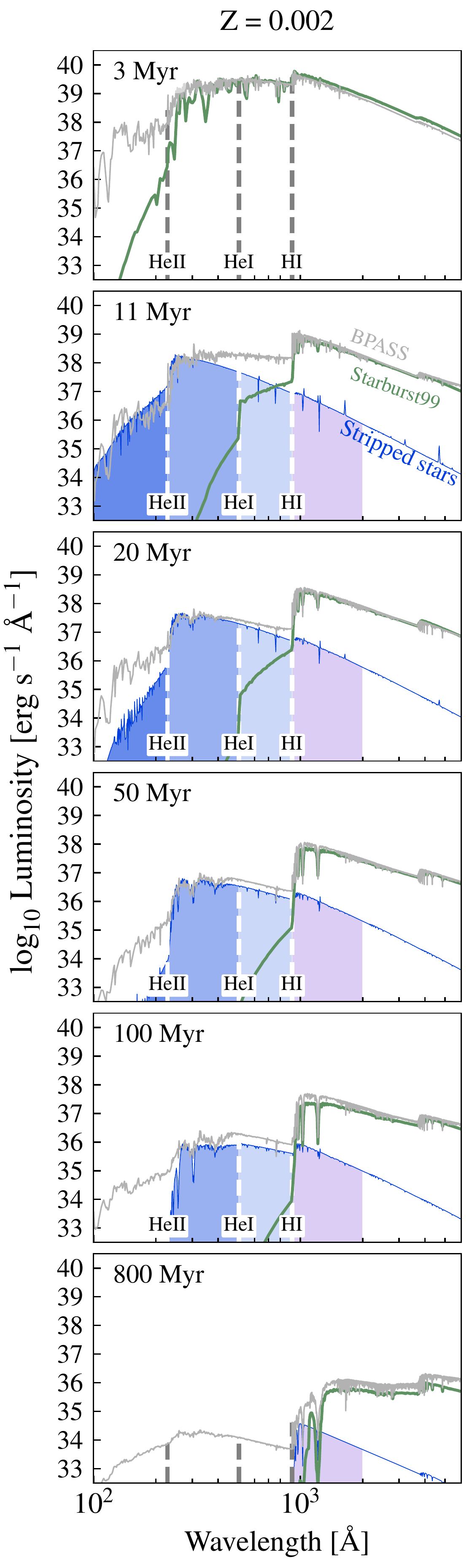}
\includegraphics[width=0.33\textwidth]{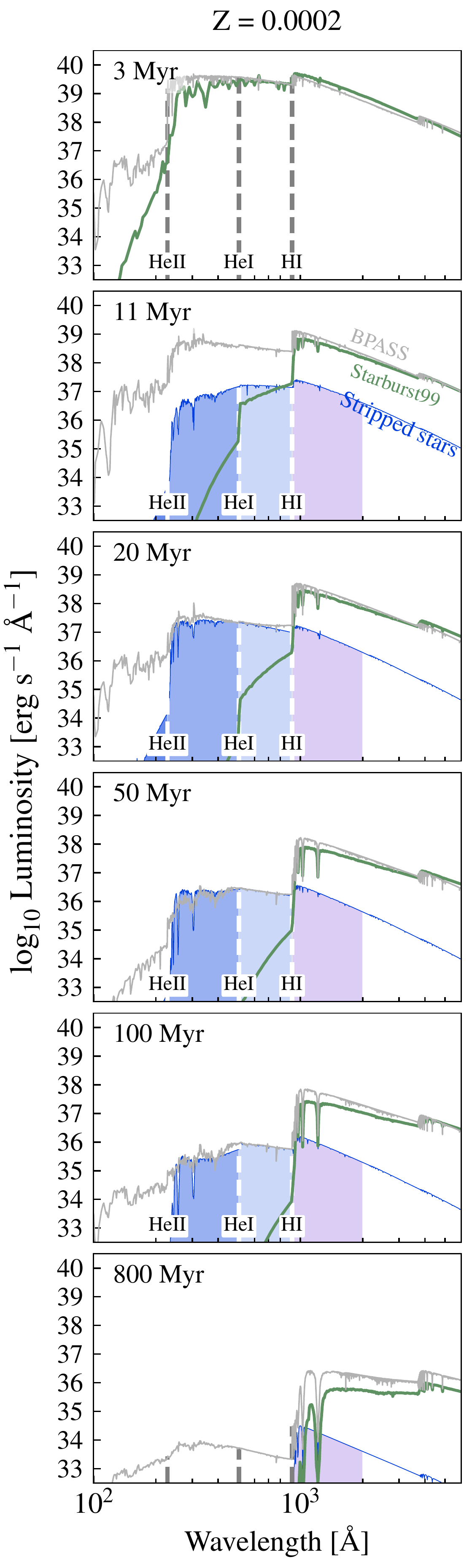}
\caption{Spectral energy distribution of a co-eval stellar population with initially $10^6 \Msun$ stars, here shown for metallicities $Z = 0.006$, 0.002, and 0.0002 (horizontally) and for increasing time after starburst (vertically). 
The figure is analogous to \figref{fig:SED_population}. See \tabref{tab:Z_combination} for the metallicities of the \code{Starburst99} models.}
\label{fig:SED_Z}
\end{figure*}

\begin{figure*}
\centering
\includegraphics[width=.33\textwidth]{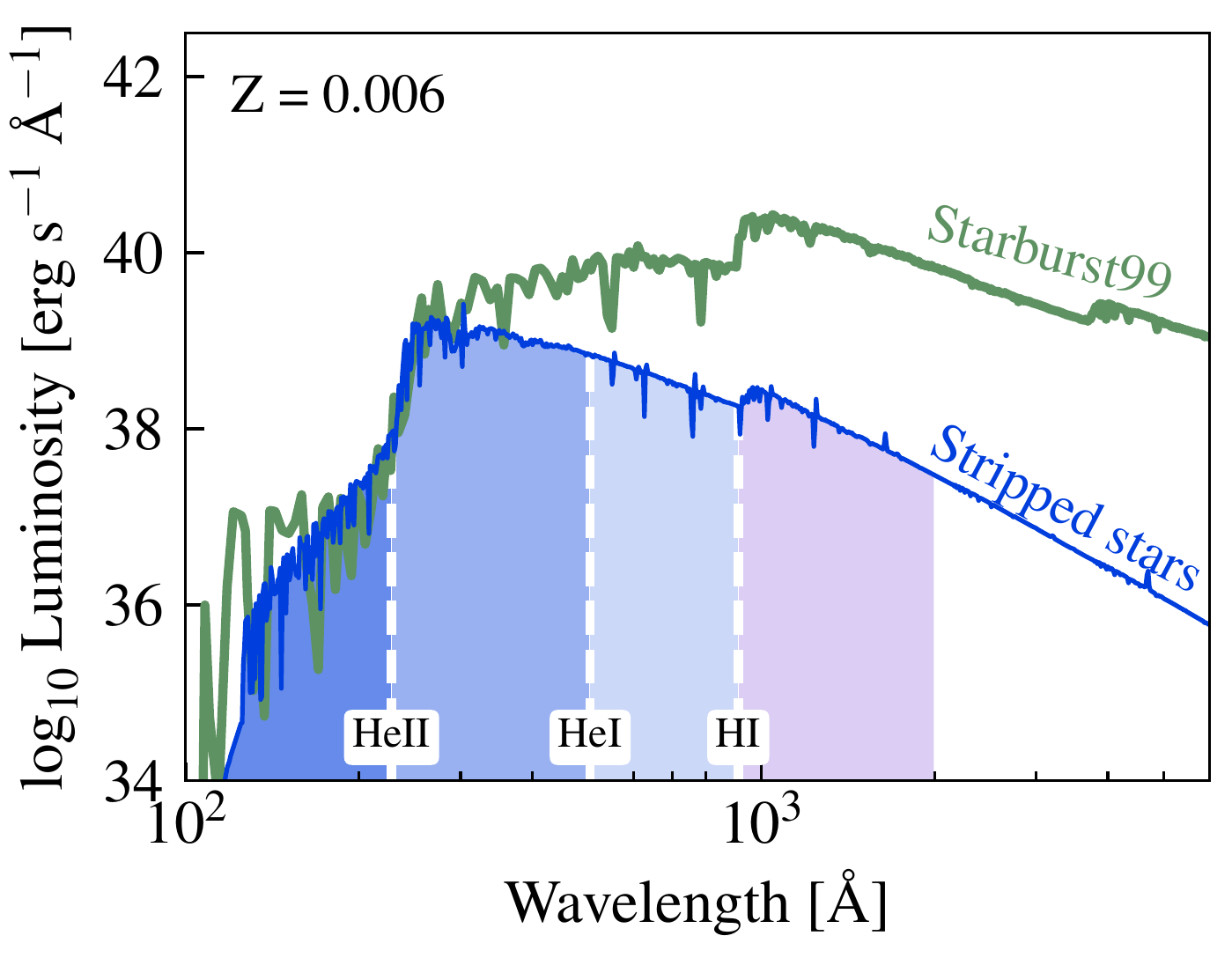}
\includegraphics[width=.33\textwidth]{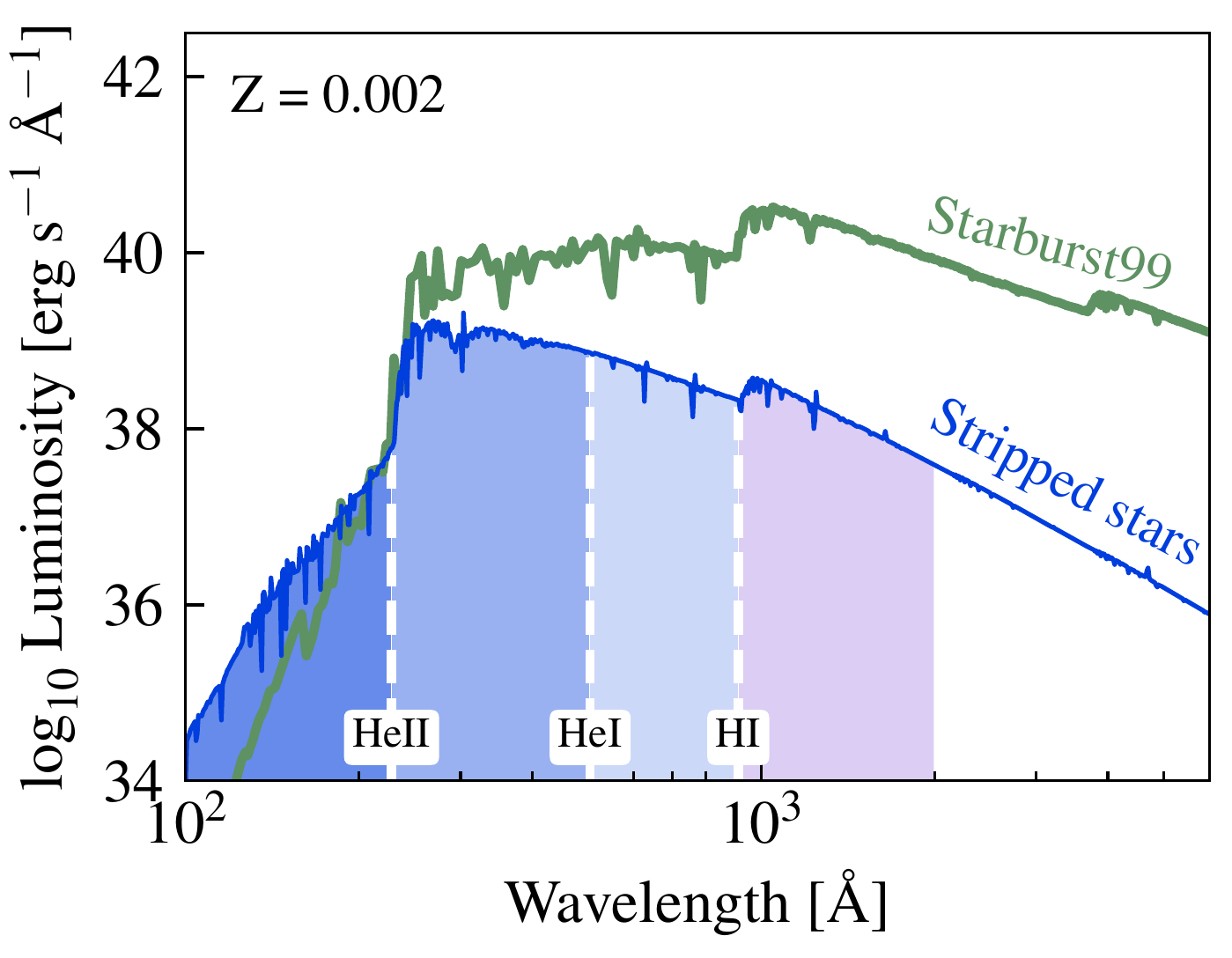}
\includegraphics[width=.33\textwidth]{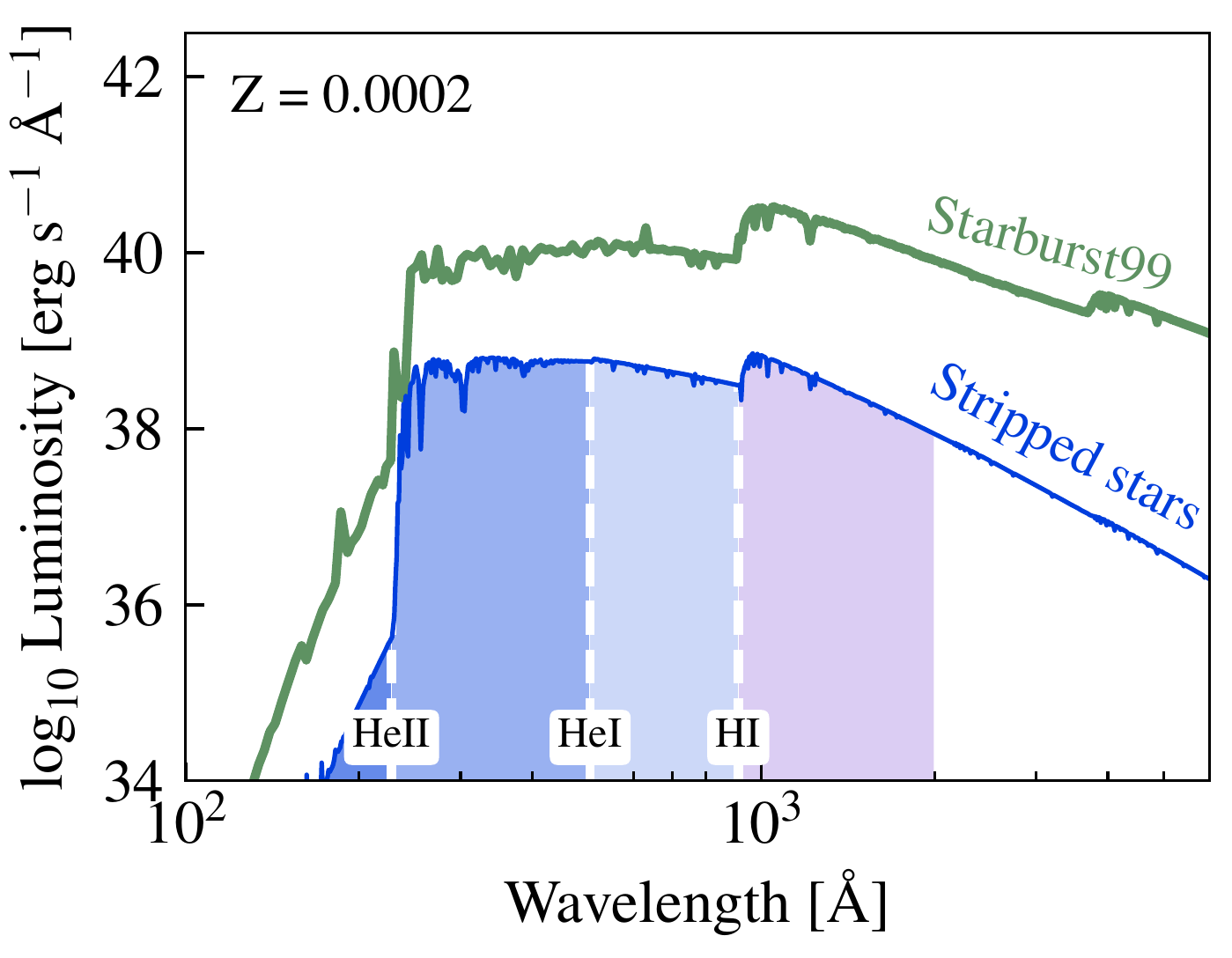}
\caption{Spectral energy distribution in the case of constant star formation for metallicities $Z = 0.006$, 0.002, and 0.0002 (for the metallicities of the \code{Starburst99} models, see \tabref{tab:Z_combination}). The models are for 1\Msunyr and are taken 500~Myr after star formation started. The figures are analogous to \figref{fig:SED_cont_500}.}
\label{fig:SED_Z_cont}
\end{figure*}

\noindent Main-sequence stars are more luminous and hotter at low metallicity than at high metallicity. Their emission rates of ionizing photons therefore increase with decreasing metallicity. 
The emission rates of ionizing photons from stripped stars are also affected by their luminosity and effective temperature. The luminosity of stripped stars increases with lower metallicity, but the effective temperature decreases \citepalias{2017A&A...608A..11G, 2018A&A...615A..78G}. The reason is primarily that envelope-stripping is less efficient at lower metallicity, which causes a larger amount of hydrogen to be left after mass transfer. The size of the stripped star is then dependent on how much leftover hydrogen there is, which leads them to be cooler at lower metallicity. Here, we combine the models for stripped stars at metallicities $Z = 0.006$, 0.002, and 0.0002, with the models from \code{Starburst99} for $Z = 0.008$, 0.002, and 0.001, respectively. We compare with the \code{BPASS} models with $Z = 0.006$, 0.002, and 0.0001. We summarize the metallicity combinations that we use for the different models in \tabref{tab:Z_combination}.

 We show the evolution of the contribution from stripped stars to the integrated spectrum of a co-eval stellar population in \figref{fig:SED_Z}. The figure shows that the emission from stripped stars becomes softer when metallicity decreases, while the emission from main-sequence stars becomes harder. As shown in \figref{fig:pop_ionizing_time}, the total effect of metallicity on the emission from stripped stars is small for \Qz and \Qo, but large for \Qt. The reason why the effect is large for \Qt is that \HeII-ionizing photons are created in the steep Wien-part of the spectrum and the emission rate of \HeII-ionizing photons is therefore very sensitive to temperature variations. Since the stripped stars are cooler at lower metallicity, the emission rate of \HeII-ionizing photons drops.

The largest differences between our predictions and those from \code{BPASS} occur at low metallicity. Around 10~Myr after starburst and for $Z \leq 0.004$, \code{BPASS} accounts for chemically homogeneous evolution for the accreting stars that were spun up during mass transfer \citep{2017PASA...34...58E}. The result is that \code{BPASS} predicts higher emission rates of \HI- and \HeI-ionizing photons than our models that only account for the ionizing emission from stripped stars at low metallicity.

For continuous star formation, we show the contribution from stripped stars to the spectral energy distribution in \figref{fig:SED_Z_cont} for the cases of lower metallicity. Comparing with \figref{fig:SED_cont_500}, we find that the contribution from stripped stars is similar for $Z \gtrsim 0.002$, while the softer spectra of stripped stars are clearly visible at metallicity $Z = 0.0002$.

\begin{figure*}
\centering
\includegraphics[width=.65\textwidth]{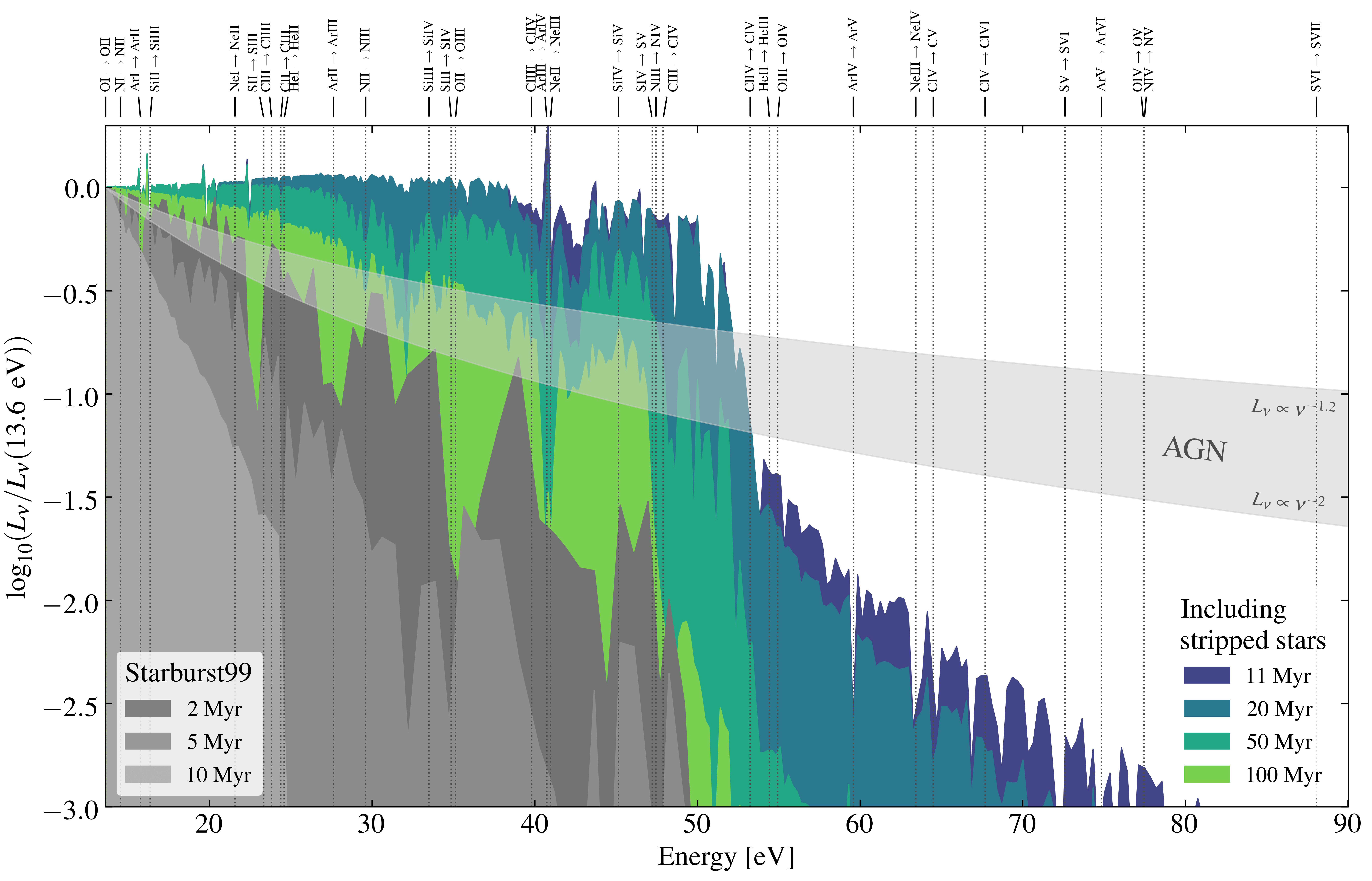}
\includegraphics[width=.65\textwidth]{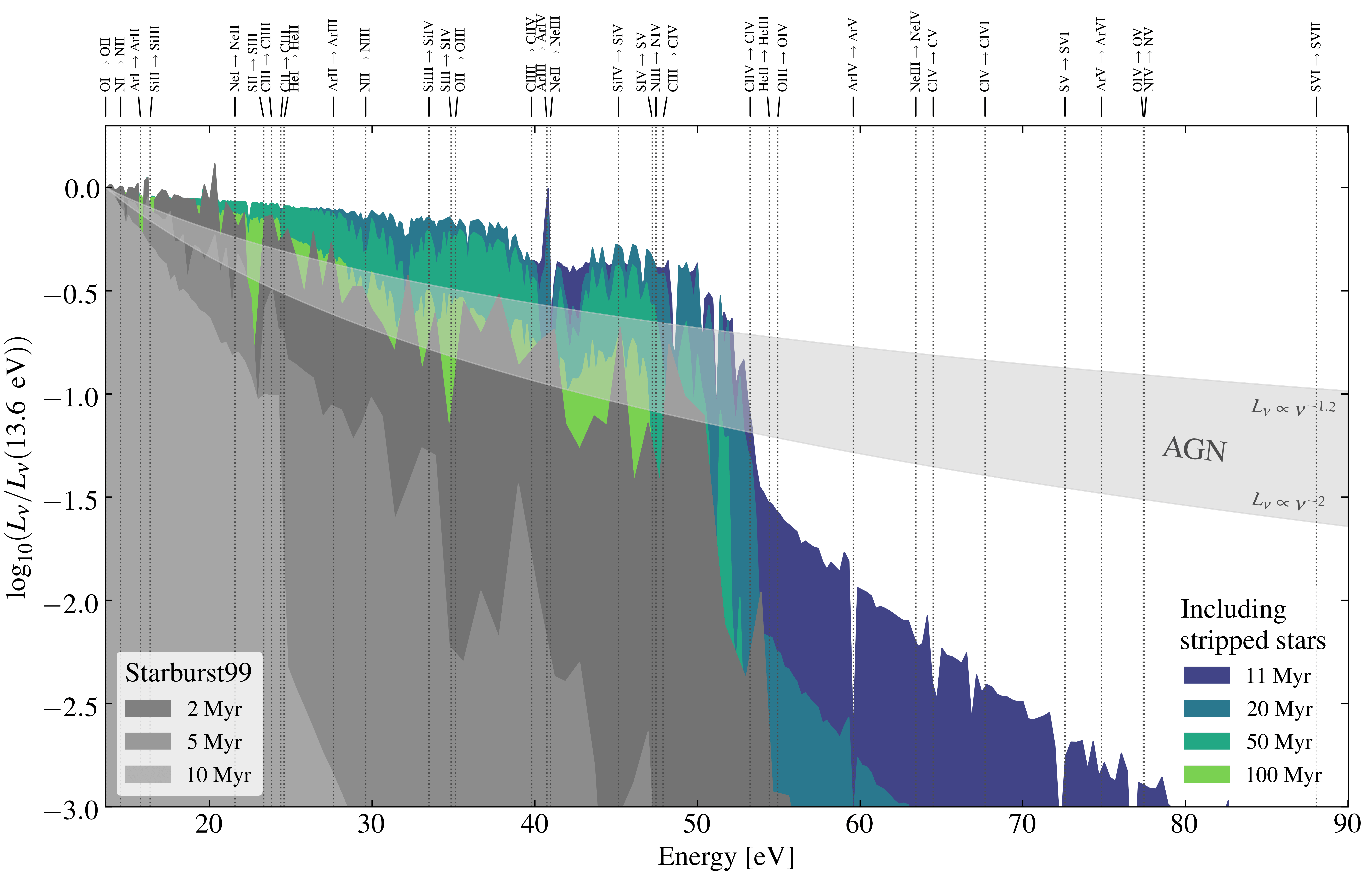}
\includegraphics[width=.65\textwidth]{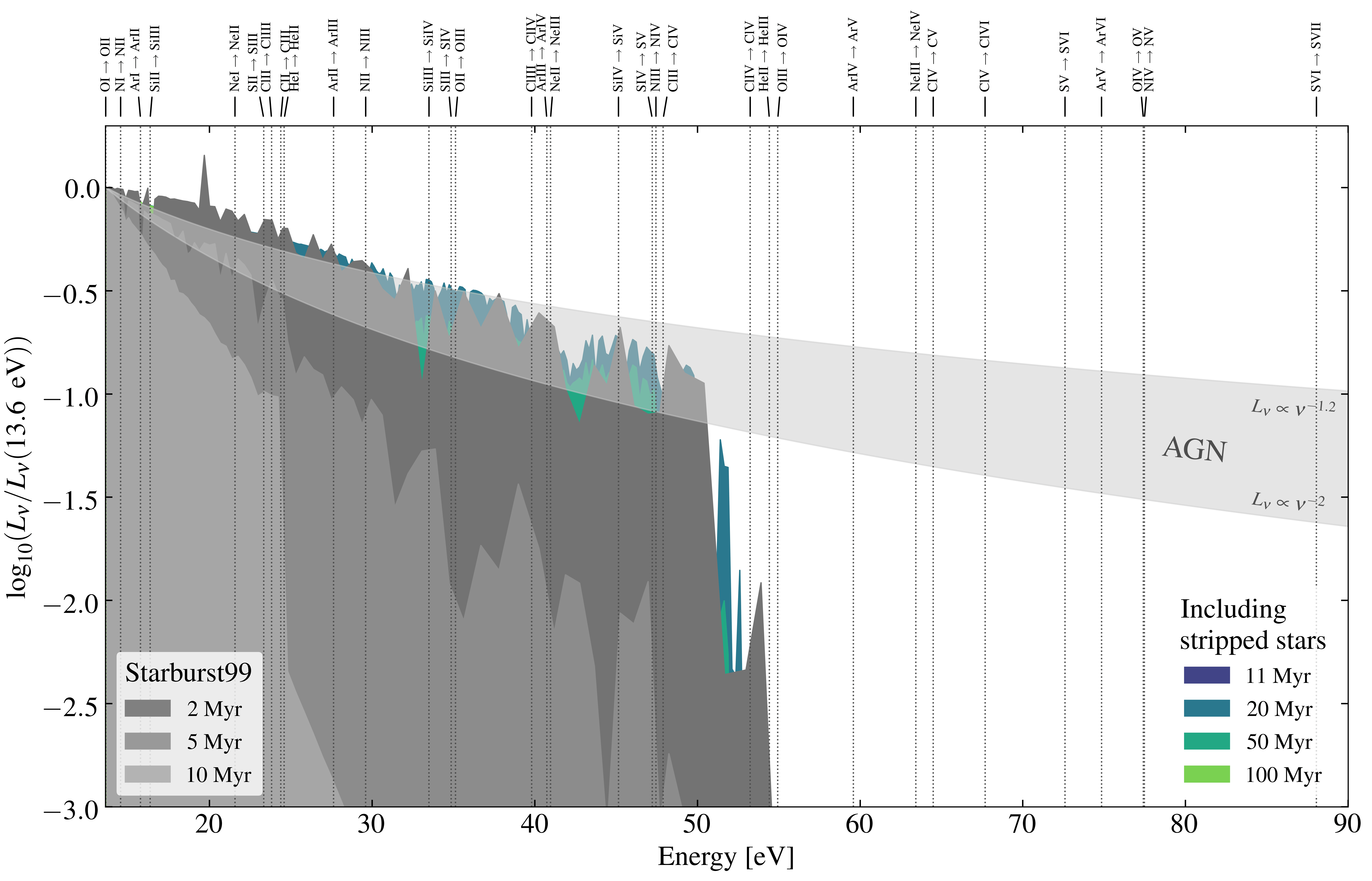}
\caption{Ionizing part of the spectral energy distribution for co-eval stellar populations with metallicities $Z = 0.006$, 0.002, and 0.0002 from top to bottom. The spectra are normalized at the ionization threshold for hydrogen, $13.6$~eV. The models from \code{Starburst99} have here $Z = 0.008$, 0.002, and 0.001 (see \tabref{tab:Z_combination}). See also \figref{fig:ionization_species}. }
\label{fig:neb_Z}
\end{figure*}

The hardness of the ionizing part of the integrated spectrum affects the nebular ionization, as discussed in \secref{sec:nebular_ionization}. We show the ionizing part of the spectra of co-eval stellar populations at low metallicity in \figref{fig:neb_Z}. The spectra of main-sequence stars are seen to become harder and those of stripped stars to become softer with lower metallicity. At $Z = 0.0002$, the hardness is similar for a population containing stripped stars and at an age of $20$~Myr as for a population of only 2~Myr that contains massive main-sequence stars. However, we note that the duration for which massive stars give rise to such hard ionizing spectra is significantly shorter than the duration stripped stars emit ionizing radiation.

\begin{figure*}
\centering
\includegraphics[width=\textwidth]{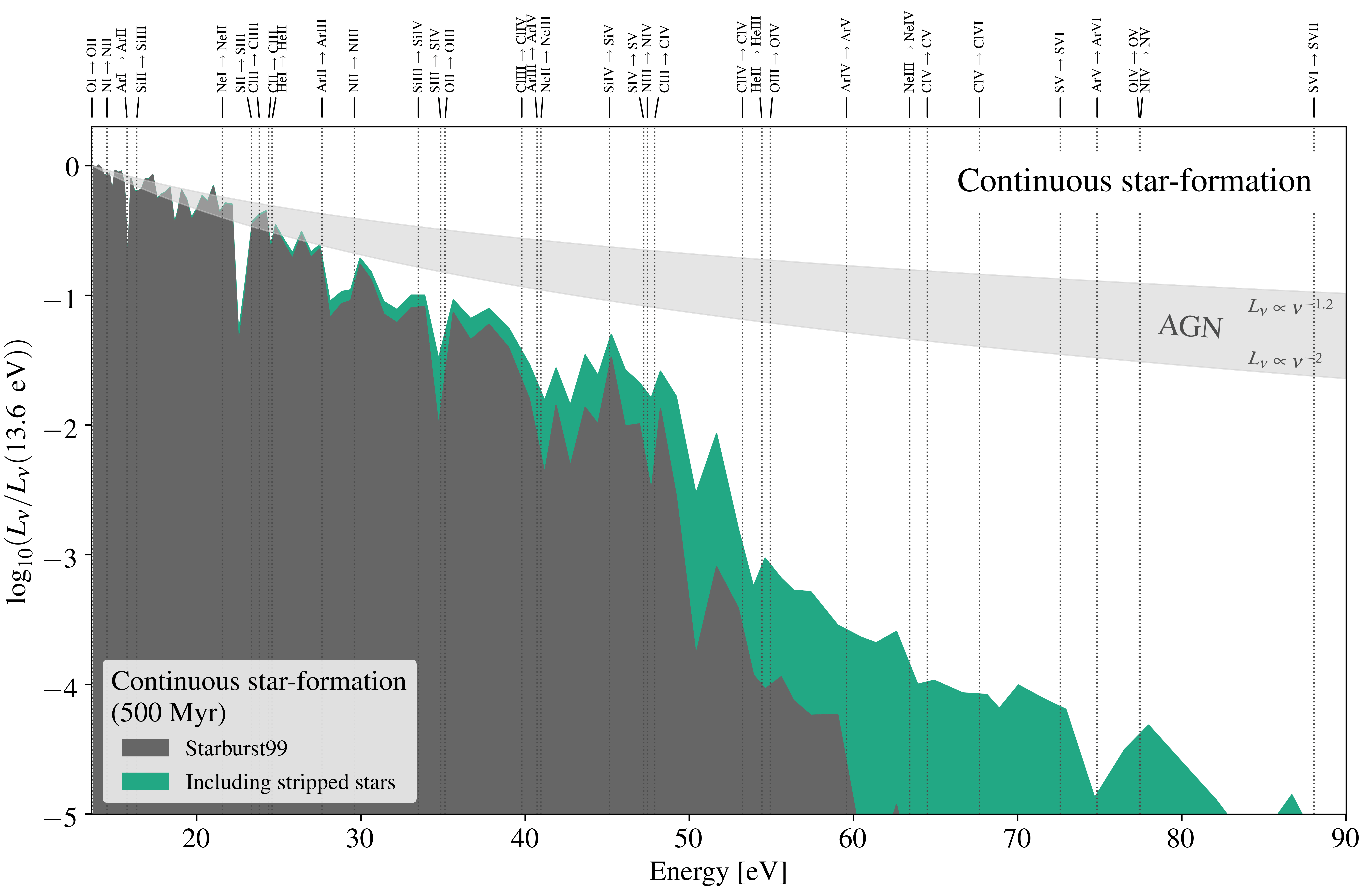}
\caption{Ionizing part of the spectral energy distribution in the case of continuous star formation, shown after 500~Myr. We compare a population containing only single stars (gray) with one containing also stripped stars (green). These models have solar metallicity. See also \figref{fig:ionization_species}. }
\label{fig:neb_Z_cont}
\end{figure*}

As a complement to \figref{fig:ionization_species}, we show the hardness of the ionizing part of the spectrum for a stellar population in which stars have formed at a constant rate for 500~Myr and with solar metallicity. The spectrum is only mildly affected by the presence of stripped stars. The largest differences from a population containing only single stars appear at high photon energies ($\gtrsim 40$~eV). This could lead to stronger nebular emission of \OIII, \CIV, and \HeII than what is expected from single star models.

We present our predictions for properties of stellar populations with metallicities of $Z = 0.006$, 0.002, and 0.0002 in \tabrefthree{tab:ob_006}{tab:ob_002}{tab:ob_0002}. The general trends of metallicity are discussed in the sections that are mentioned in the tables.

\begin{sidewaystable*}
\centering
\caption{Values of observable quantities for models of stellar populations including stripped stars. This table is analogous to \tabref{tab:observable_parameters}, but for a population of stripped stars with $Z = 0.006$. We compare to the \code{Starburst99} models with $Z = 0.008$.}
\label{tab:ob_006}
{\small
\input{Table_parameters_Z0.006.tex}
}
\end{sidewaystable*}

\begin{sidewaystable*}
\centering
\caption{Values of observable quantities for models of stellar populations including stripped stars and at $Z = 0.002$. This table is analogous to \tabref{tab:observable_parameters}, but for a population of stripped stars with $Z = 0.002$. We compare to the \code{Starburst99} models with $Z = 0.002$.}
\label{tab:ob_002}
{\small
\input{Table_parameters_Z0.002.tex}
}
\end{sidewaystable*}

\begin{sidewaystable*}
\centering
\caption{Values of observable quantities for models of stellar populations including stripped stars and at $Z = 0.0002$. This table is analogous to \tabref{tab:observable_parameters}, but for a population of stripped stars with $Z = 0.0002$. We compare to the \code{Starburst99} models with $Z = 0.001$.}
\label{tab:ob_0002}
{\small
\input{Table_parameters_Z0.0002.tex}
}
\end{sidewaystable*}


\section{Effect of star formation history}\label{app:SFH}

Stellar populations are in reality not co-eval or form stars at exactly a constant rate. More realistic stellar populations are either extended starbursts, composed of multiple stellar clusters that make up a bursty star formation rate or quenched of star formation. For simplicity, we considered the simplified cases of a starburst and a stellar population in which stars form at a constant rate in the main part of this study (see \secref{sec:ionizing_time}). If we want to better understand the effect of stripped stars in stellar populations, more realistic star formation histories are needed to compare observations with.

\begin{figure*}
\centering
\includegraphics[width=.325\textwidth]{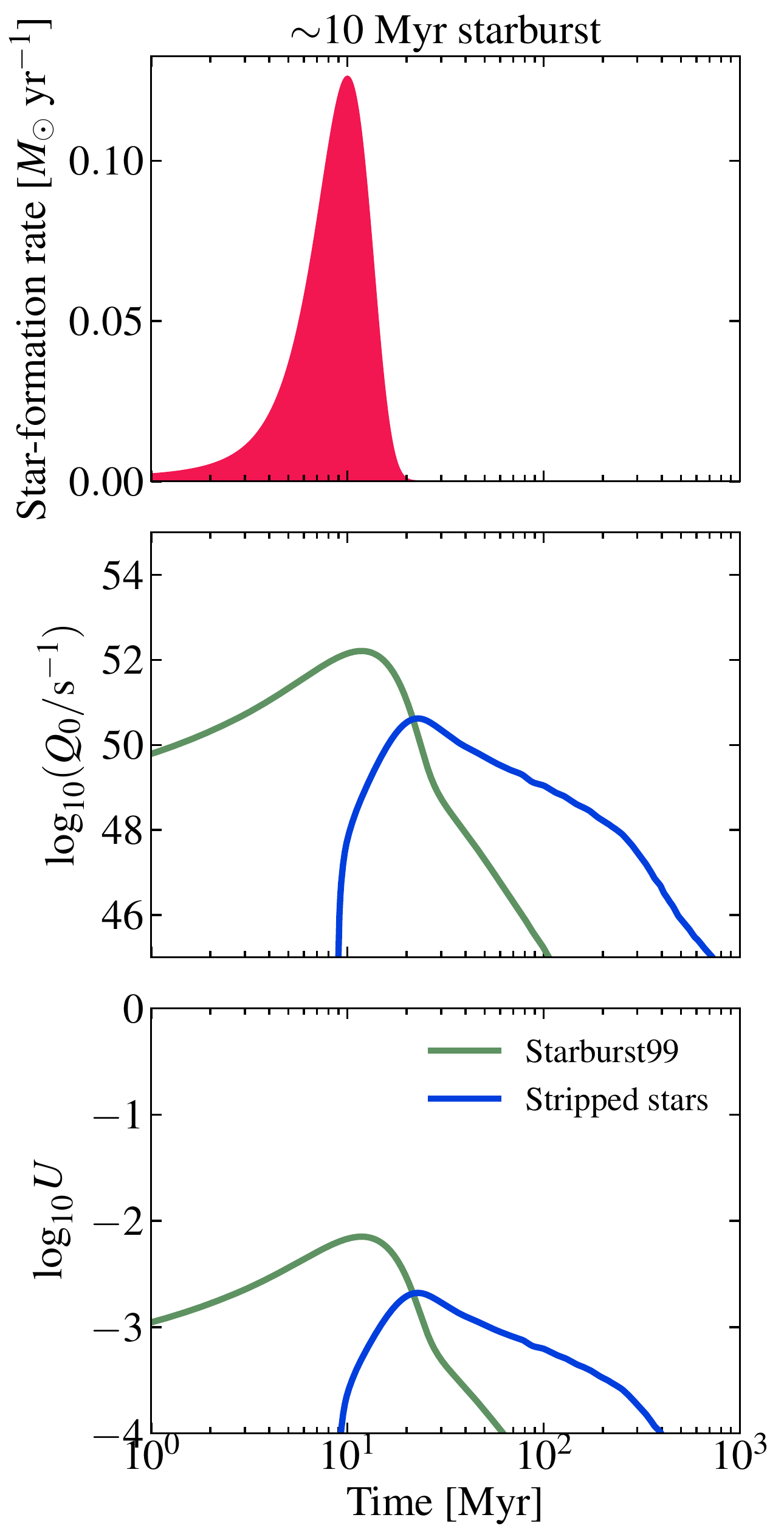}
\includegraphics[width=.33\textwidth]{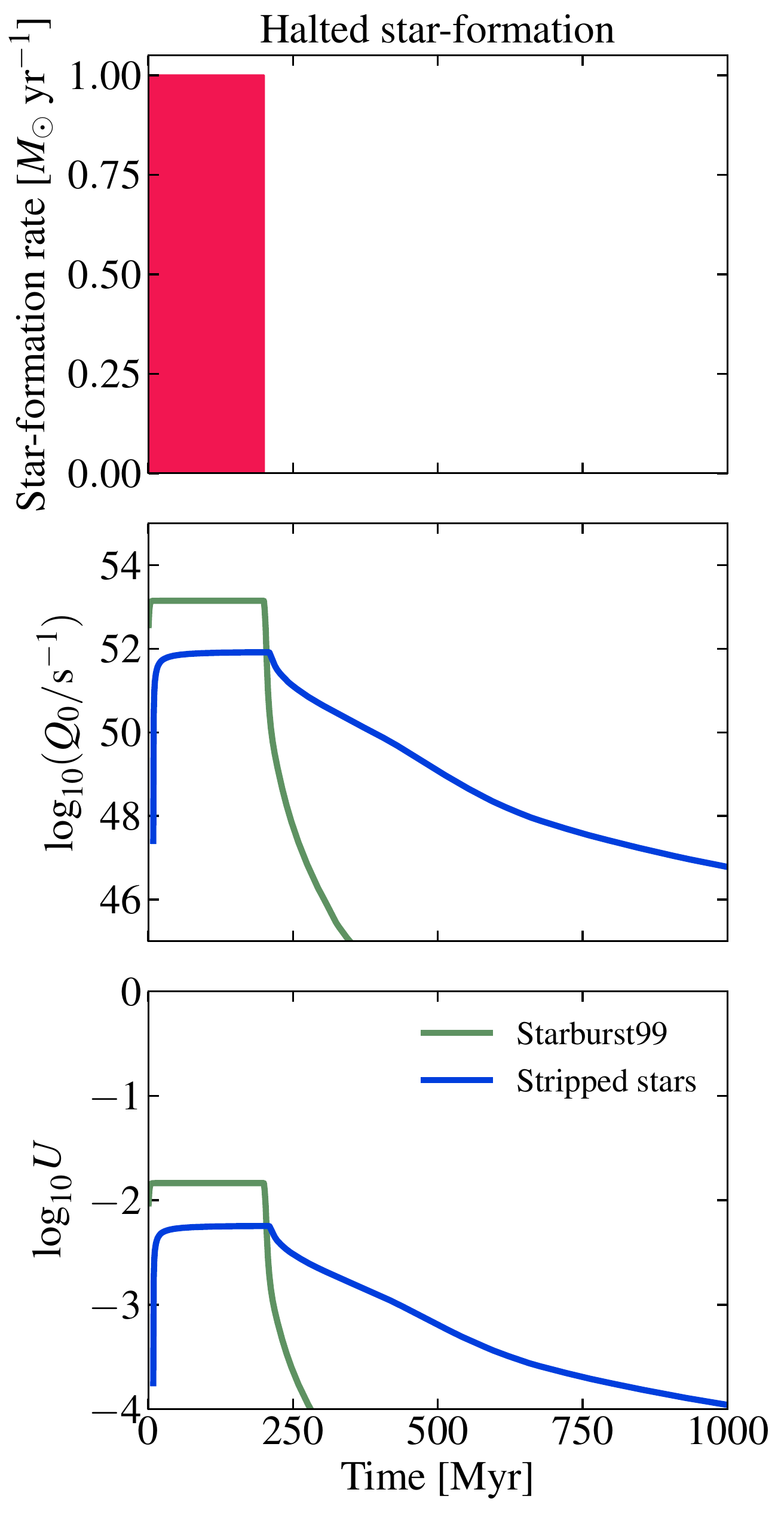} 
\includegraphics[width=.32\textwidth]{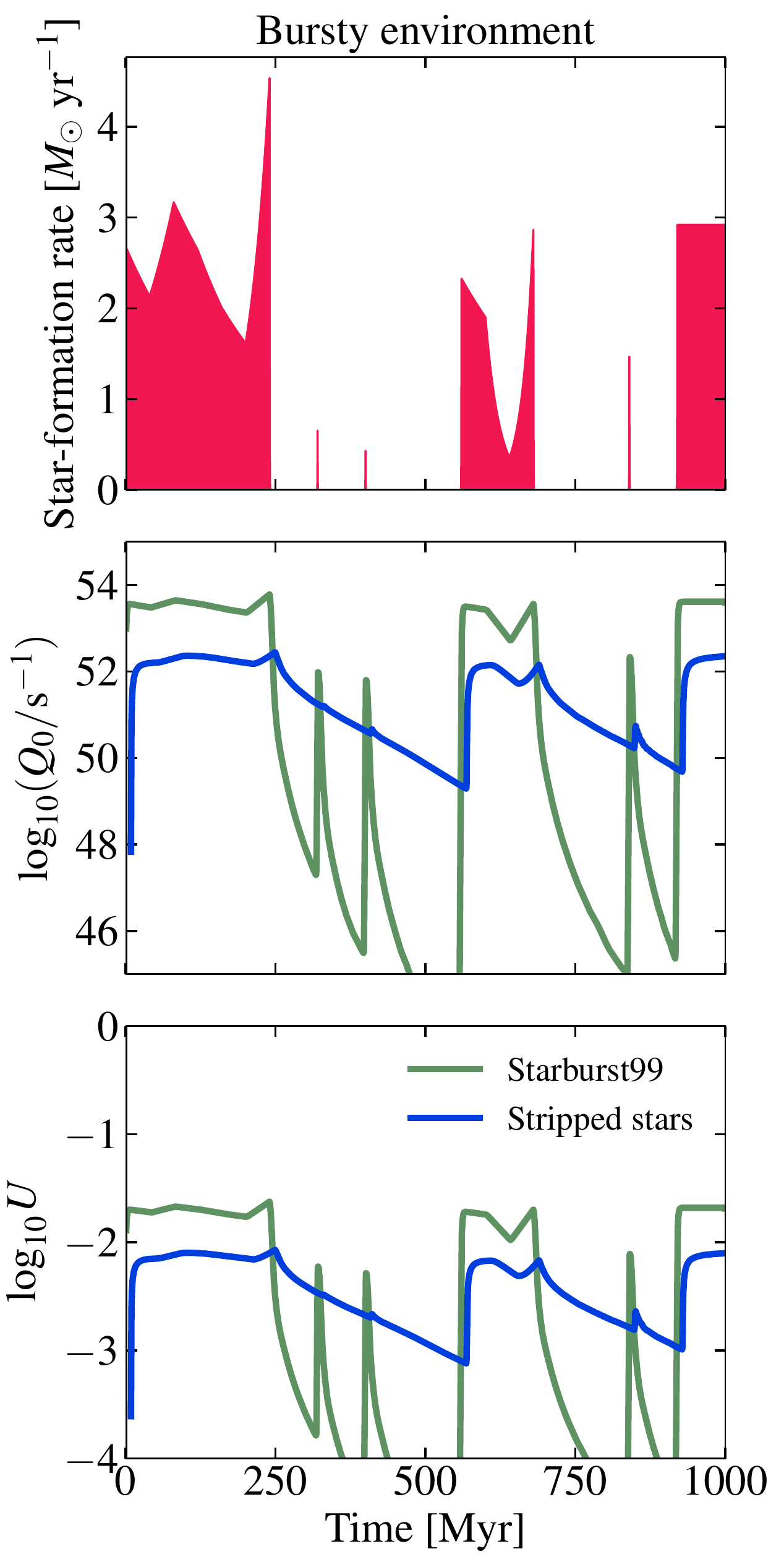} 
\caption{Effect of different star formation histories on the emission rate of \HI-ionizing photons and ionization parameter from a stellar population. We show the contribution from massive stars in green using \code{Starburst99} and the contribution from stripped stars in blue.
From left to right, we show the effect from a Gaussian shaped burst of about 10~Myr, quenched star formation after 200~Myr, and a bursty star-forming environment.
The top panels show the adopted star formation history, the middle panels show the emission rate of \HI-ionizing photons, and the bottom panels show the ionization parameter calculated assuming a gas density of $10^2$ cm$^{-2}$. }
\label{fig:SFH_Q}
\end{figure*}

Here, we consider three additional star formation histories. First, an extended starburst, shaped with a Gaussian profile of width 40~Myr and in total $10^6\Msun$ of stars are created. This could be representative to a super star cluster similar to what are present in high-redshift galaxies and likely similar to 30~Dor in the Large Magellanic Cloud \citep[e.g.,][]{2018A&A...618A..73S}. Second, a stellar population in which stars form at a constant rate for 200~Myr and then star formation is quenched and completely stops. This could be representative to aging stellar populations in which star formation has stopped \citep[e.g., the retired galaxies of][]{2015MNRAS.449..559S}. Third, several, randomly in time distributed, bursts of star formation with fluctuations in order of every 40~Myr and averaging a star formation rate of 1\Msunyr. This could be representative to a clumpy star-forming galaxy in which the emission comes from one or a few bursts at a time \citep[see e.g., the models of][]{2014ApJ...788..121K}. This type of star formation is observed to be common for small galaxies that likely were in majority among galaxies at high redshift and could be the cause of cosmic reionization. The described star formation histories are visualized in the top row of panels in \figref{fig:SFH_Q}. 

We calculate the emission rates of ionizing photons for the described stellar populations by convolving the emission rate of ionizing photons from the co-eval stellar population (described in \secref{sec:ionizing_time}) over time and weighing with the specific star formation history. Since we consider stellar populations that have $10^6$\Msun in stars or more, we do not expect the stochastic effects to be large. We also calculate the ionization parameter for the stellar populations following \eqref{eq:U} and assuming a gas density of $n_{H} = 100$~cm$^{-3}$. With time, the gas density likely decreases, causing slightly smaller ionization parameters than what we calculate. However, a decrease of a factor of ten in density leads to a smaller decrease in the ionization parameter of about a factor of three. We therefore consider this difference to be small. Here, we treat the models from \code{Starburst99} and the contribution from stripped stars separately. 

We show the emission rate of \HI-ionizing photons in the middle row of panels in \figref{fig:SFH_Q}. The figures show that the massive stars dominate the ionizing output when star formation is actively ongoing, while stripped stars provide the majority of ionizing emission shortly after star formation has stopped. This is also visible in the ionization parameter, which is shown in the bottom panels. 

For the burst, the emission rate of \HI-ionizing photons is very similar to what is expected for the co-eval stellar population as seen when comparing the bottom-left panel of \figref{fig:SFH_Q} with panel a of \figref{fig:pop_ionizing_time}. Since we consider the same mass produced in stars, the emission rates not only show the same shapes, but also the same numbers as described in \secref{sec:ionizing_time}. The result is that stripped stars appear to extend the duration of the starburst by providing ionizing emission for a longer time. However, we note that the massive stars in the stellar populations dominate the ionizing emission during the starburst itself. 

For the stellar population in which star formation is quenched, it appears like a combination of the constant star formation population and the co-eval stellar population considered in the main manuscript. First, massive stars completely dominate the ionizing emission during the ongoing star formation. Once the star formation stops, the emission rate of ionizing photons from massive stars rapidly drops, leaving only stripped stars to emit ionizing photons at a significant rate. Their emission rate is lower, but lasts for longer, as seen in the co-eval stellar population. 

In the case of the bursty star formation, the effect is similar; when star formation is ongoing the massive stars dominate the emission of ionizing photons, but once star formation stops their emission rapidly drops and the ionizing emission comes from stripped stars. For the bursty environment that we considered, the stripped stars prevent ionizing emission to ever completely die out, which contrasts with what is expected for massive stars. Stripped stars keep the emission rate of \HI-ionizing photons to never drop below $\sim 10^{51}$ s$^{-1}$ for this particular case. However, we expect this to scale with the frequency of starbursts and the mean star formation rate.

The overall effect of the considered star formation history is similar for the ionization parameter compared to the emission rate of ionizing photons, since the two quantities are closely related (see \eqref{eq:U}). After star formation has stopped, the stripped stars keep the ionization parameter at relatively high values for several hundred Myrs. This can be seen for example in the case of the halted star formation, for which stripped stars cause the ionization parameter to remain larger than $\log_{10}  U  \sim -3$  for another $\sim 300$~Myr after star formation has stopped. For the bursty environment that we considered here, stripped stars never allow the ionization parameter to decrease below $\log_{10} U  \sim -3$, which is in contrast to what is predicted for single stars. 

In summary, stripped stars appear to play an important role as contributors to the emission of ionizing photons in all the three considered star formation histories. Their effect is to extend emission of ionizing photons for a longer period in time. Their emission rate is smaller compared to what is expected for the massive stars formed at the same time. 

Since stripped stars appear to affect the emission rate of ionizing photons significantly for a variety of different stellar populations, we conclude that they cannot be neglected when analyzing data from observed stellar populations. This is also implied from their effect on the ionization parameter.


\section{Evolution during the stripped phase}\label{app:stripped_phase}

Our approach in this study is simple, perhaps more simple than what is expected to reach sufficient accuracy in the results. A concern may be that one spectral model is not sufficient to represent the radiative emission from a stripped star during its entire stripped phase, which is a part of the method we employ and describe in \secref{sec:modeling_population}. Here, we present additional spectral models computed at different times during the stripped phase to show that one spectral model per stripped star is sufficient to represent their emission and estimate their emission rate of ionizing photons. The models that we show here have solar metallicity.



{\small
\begin{sidewaystable*}
\centering
\caption{Table showing how most stellar parameters remain similar throughout the stripped phase. Here, we show three example models with initial masses of $M_{\text{init}} = 8.2$, 12.2, and 18.2\Msun that are stripped as they fill their Roche-lobes during the Hertzsprung gap evolution. The models have solar metallicity.}
\label{tab:params_examples}
\input{params_examples}
\tablefoot{$^{a}$ Computed in \code{MESA}, $^{b}$ Computed in \code{CMFGEN}. The surface temperature $T_{\star}$ is what is referred to as the effective temperature in \code{MESA}. }
\end{sidewaystable*}
}

\subsection{Evolution in the Hertzsprung-Russell diagram during the stripped phase}

\begin{figure*}
\centering
\includegraphics[width=.7\textwidth]{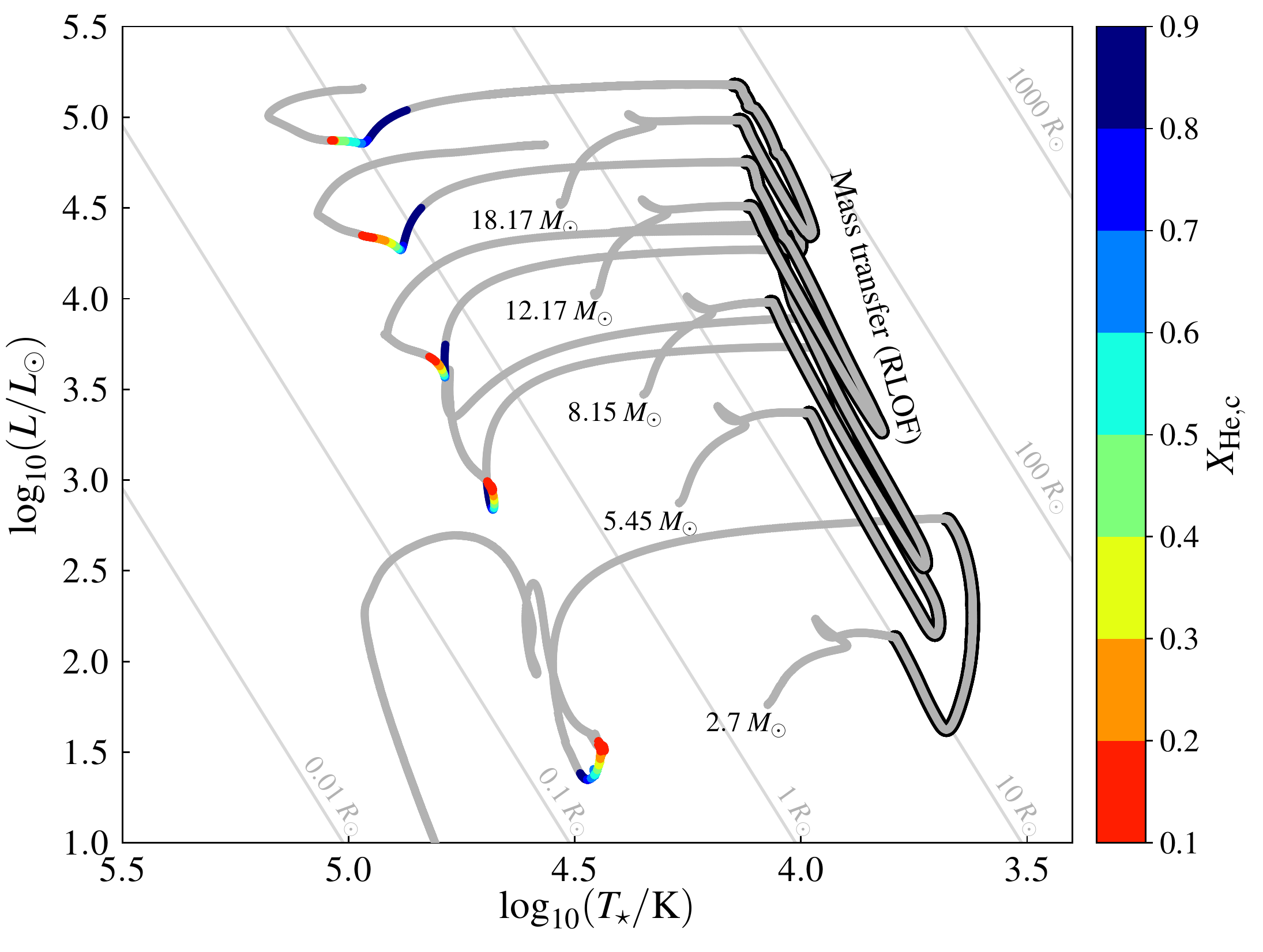}
\caption{Hertzsprung-Russell diagram showing the evolution of five stars that become stripped through mass transfer with a companion star. We show the evolution for models of donor stars with initial masses of 2.7, 5.45, 8.15, 12.17, and 18.17 \Msun, labeled as such. The tracks are gray for all but the central helium burning phase (defined as when the central helium mass fraction is between 0.9 and 0.1), which is colored with the central mass fraction of helium. The colored regions cover only a small part of the HR diagram and therefore indicates that the stripped stars have similar stellar parameters throughout the stripped phase. The models have solar metallicity. (We show here the surface temperature $T_{\star}$, which is referred to as the effective temperature in \code{MESA}.)}
\label{fig:HRD_He}
\end{figure*}

\figref{fig:HRD_He} shows the evolution of five models in the Hertzsprung-Russell diagram computed using \code{MESA} (models presented in \citetalias{2018A&A...615A..78G}). The models are for donor stars in binary systems that initially have masses of 2.7, 5.45, 8.15, 12.17, and 18.17 \Msun. The companion masses are decided such that the initial mass ratio is $q = M_{2, \text{init}}/M_{1, \text{init}} = 0.8$ and the the initial period is set such that interaction is initiated early during the Hertzsprung gap evolution of the most massive star in the system. \figref{fig:HRD_He} shows that the stripped stars contract after Roche-lobe overflow is completed and become hotter than they were at the zero-age main sequence. The color shows the mass fraction of helium in the center of the stars. As the color changes, central helium burning progresses. The figure thus shows that the long-lasting central helium burning occurs in a very small part of the HR diagram, meaning that the luminosity and temperature of the stripped stars remain almost the same throughout the stripped phase.

The evolution after central helium burning is short in relation to the previous helium core burning. However, during this phase the stars expand and sometimes can reach helium giant phases. Even though the parameters change significantly both during this expansion phase and the earlier contraction after mass transfer, these two phases have short durations and only a small fraction of stripped stars would, therefore, reside in those phases. 

\tabref{tab:params_examples} displays the stellar parameters of the three most massive stripped stars shown in \figref{fig:HRD_He}. For nine times during the helium core burning of each star, we show the following parameters computed in \code{MESA}: mass ($M_{\text{strip}}$), surface radius ($R_{\star}$), surface temperature ($T_{\star}$), surface gravity ($log_{10} g_{\star}$), luminosity ($L$), wind mass loss rate ($\dot{M}_{\text{wind}}$), and surface mass fraction of hydrogen and helium ($X_{\text{H, s}}$ and $X_{\text{He, s}}$). These nine times are chosen when the central helium mass fraction is $X_{\text{He, c}} = 0.9$, 0.8, 0.7, 0.6, 0.5, 0.4, 0.3, 0.2, and 0.1, which spans almost the entire stripped phases. Most of the parameters are not varying significantly throughout the stripped phases. The stripped star with a progenitor of initially 8.15\Msun shows least variation in the stellar parameters. The higher mass stripped stars loose the outermost hydrogen layers because of wind mass loss and appear somewhat hotter, smaller, and more compact. 

From the small changes visible in the stellar parameters of the models in \tabref{tab:params_examples}, it is plausible that the changes in the emission rates of ionizing photons also are small. However, to estimate the change of the emission rates of ionizing photons during the stripped phases, modeled spectra give a more accurate result than the black-body spectra that can be created from the parameters computed in \code{MESA}.

\subsection{Atmosphere models for different times during the stripped phase}

To accurately estimate the emission rates of ionizing photons during the stripped phases, we create eight additional atmosphere models for each of the evolutionary models of the stripped stars with progenitors of initially 8.15, 12.17, and 18.17\Msun. These atmosphere models are created such that there is each a model when the central helium mass fraction is $X_{\text{He, c}} = $ 0.9, 0.8, 0.7, 0.6, 0.5, 0.4, 0.3, 0.2, and 0.1, which corresponds to a close to constant temporal separation between the spectral models (see \tabref{tab:params_examples}). The technique for how to create the atmosphere models was described in detail in \citetalias{2017A&A...608A..11G} and \citetalias{2018A&A...615A..78G}.

\subsubsection{Spectral energy distribution}

\begin{figure*}
\centering
\includegraphics[width=.8\textwidth]{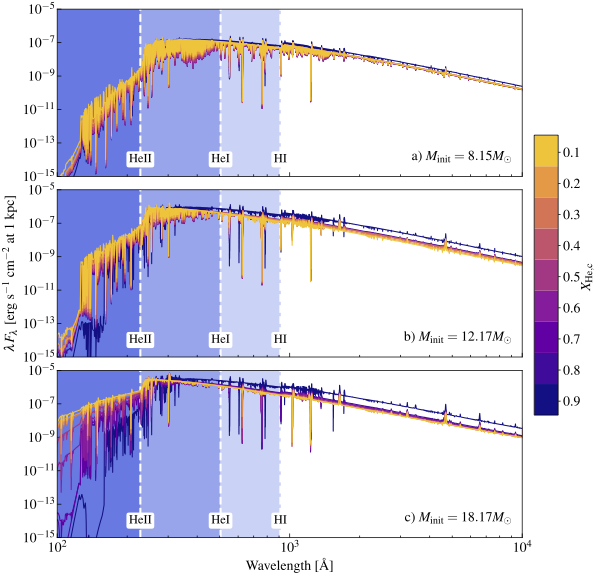}
\caption{Spectral energy distributions modeled at nine different times during the evolution of stripped stars. The displayed models are for stars with initial masses of 8.15 (top), 12.17 (middle), and 18.17 \Msun (bottom). The spectral models are created during the helium core burning at the time when the stripped stars have central helium mass fractions of $X_{\text{He, c}} = 0.9$, 0.8, \dots 0.1, which we highlight using purple, pink, and yellow colors. We use blue colors to mark the parts of the spectra that are \HI-, \HeI-, and \HeII-ionizing. 
The spectral models are very similar during the stripped phases for the three stars, with the exception of the uncertain \HeII-ionizing part of the spectrum.}
\label{fig:SED_He}
\end{figure*}

\noindent In \figref{fig:SED_He}, we show the spectral energy distribution for each of the stripped stars computed at the different times during the stripped phases. The figure shows that the shapes are very similar for most of the wavelength range throughout the stripped phases. The observable UV, optical and infrared spectra are close to identical for all but perhaps the model at $X_{\text{He,c}} = 0.9$, when the stars are still contracting and thus are slightly brighter. The \HI- and \HeI-ionizing parts of the spectral energy distributions are also very similar, while the two most massive stripped stars show that the \HeII-ionizing part of the spectral energy distributions change with time. This is an effect of the outermost hydrogen layers being removed through wind mass loss, causing the star to be hotter. Since the wind mass loss rate from stripped stars is uncertain, the estimated \HeII-ionizing emission is also uncertain. 

\figref{fig:SED_He} shows that all but the uncertain \HeII-ionizing part of the spectral energy distributions seem mostly unaffected by the evolution during the stripped phase. This means that including multiple spectral models during the stripped phases of each star in the evolutionary grid would not significantly improve the estimate of how stripped stars affect the total spectral energy distribution of a stellar population (e.g., \figref{fig:SED_population}).

\subsubsection{Emission rate of ionizing photons}

{\small
\begin{table*}
\centering
\caption{Table showing how the emission rates of \HI-, \HeI-, and \HeII-ionizing vary during the stripped phase for three example models.}
\label{tab:Qs_examples}
\input{Qs_examples}
\tablefoot{$^{a}$ Computed in \code{MESA}, $^{b}$ Computed in \code{CMFGEN}. We use the label BB for the predictions from a blackbody computed using the \code{MESA} parameters.}
\end{table*}
}

\begin{figure*}
\centering
\includegraphics[width=.33\textwidth]{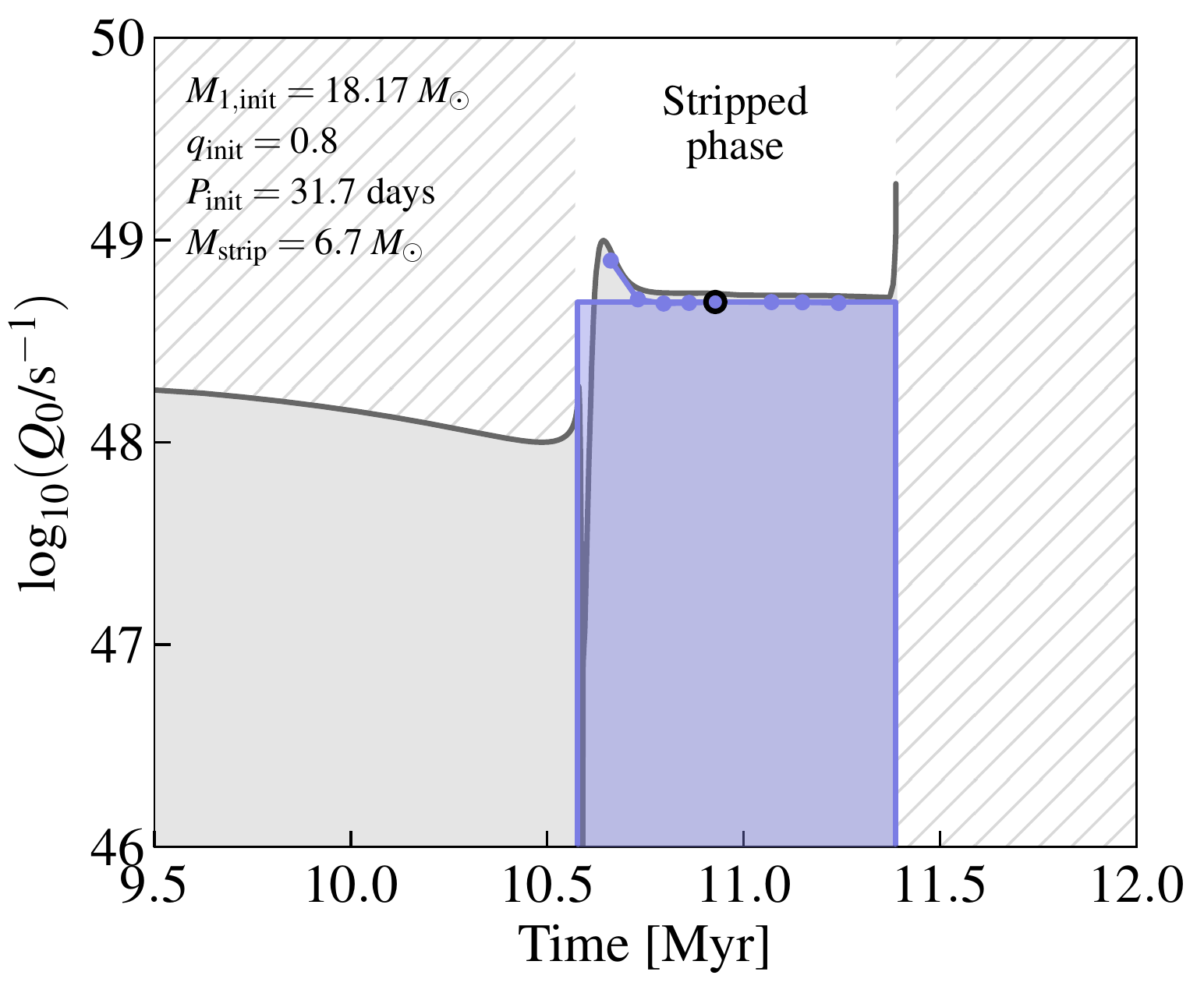}
\includegraphics[width=.33\textwidth]{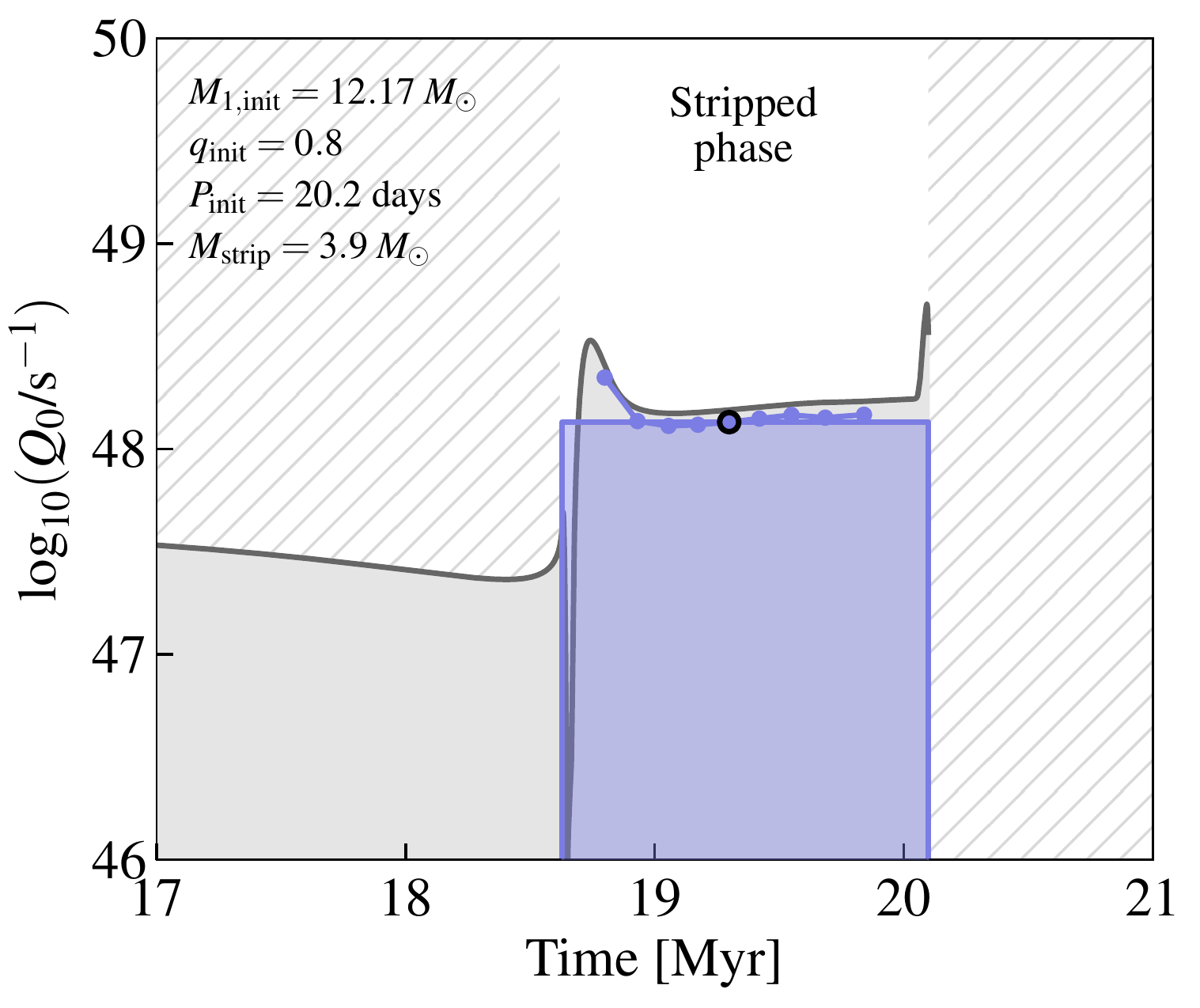}
\includegraphics[width=.33\textwidth]{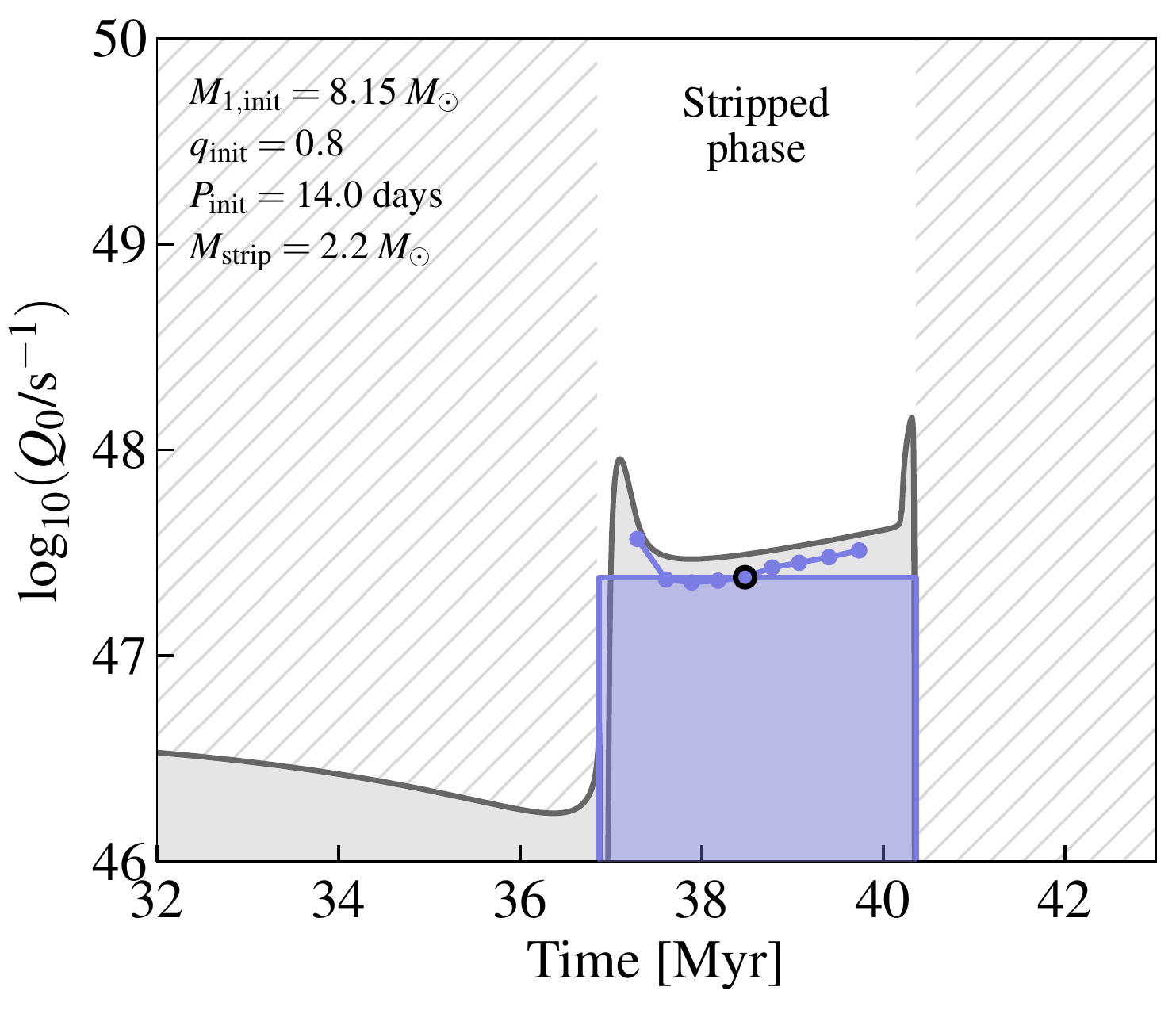}

\includegraphics[width=.33\textwidth]{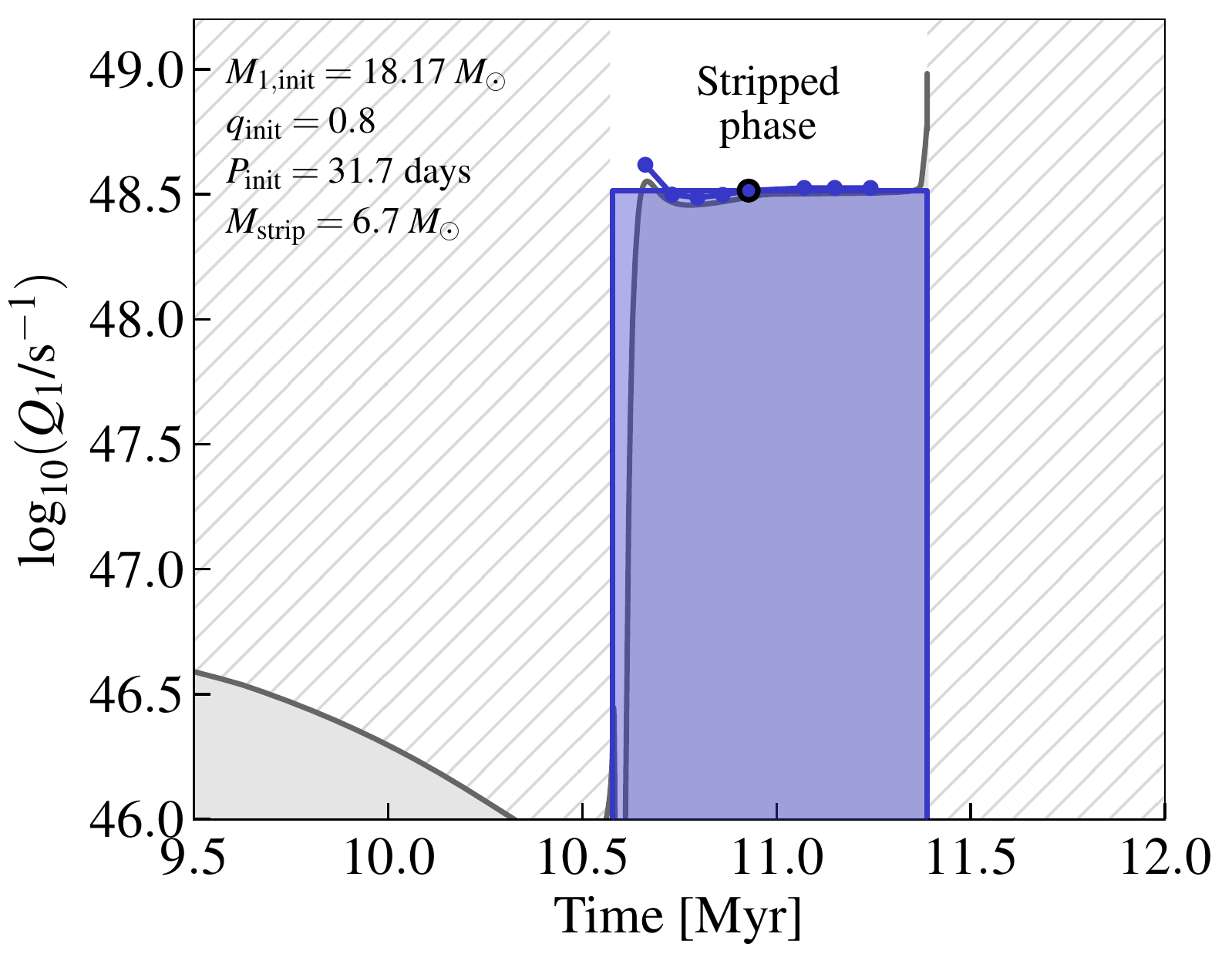}
\includegraphics[width=.33\textwidth]{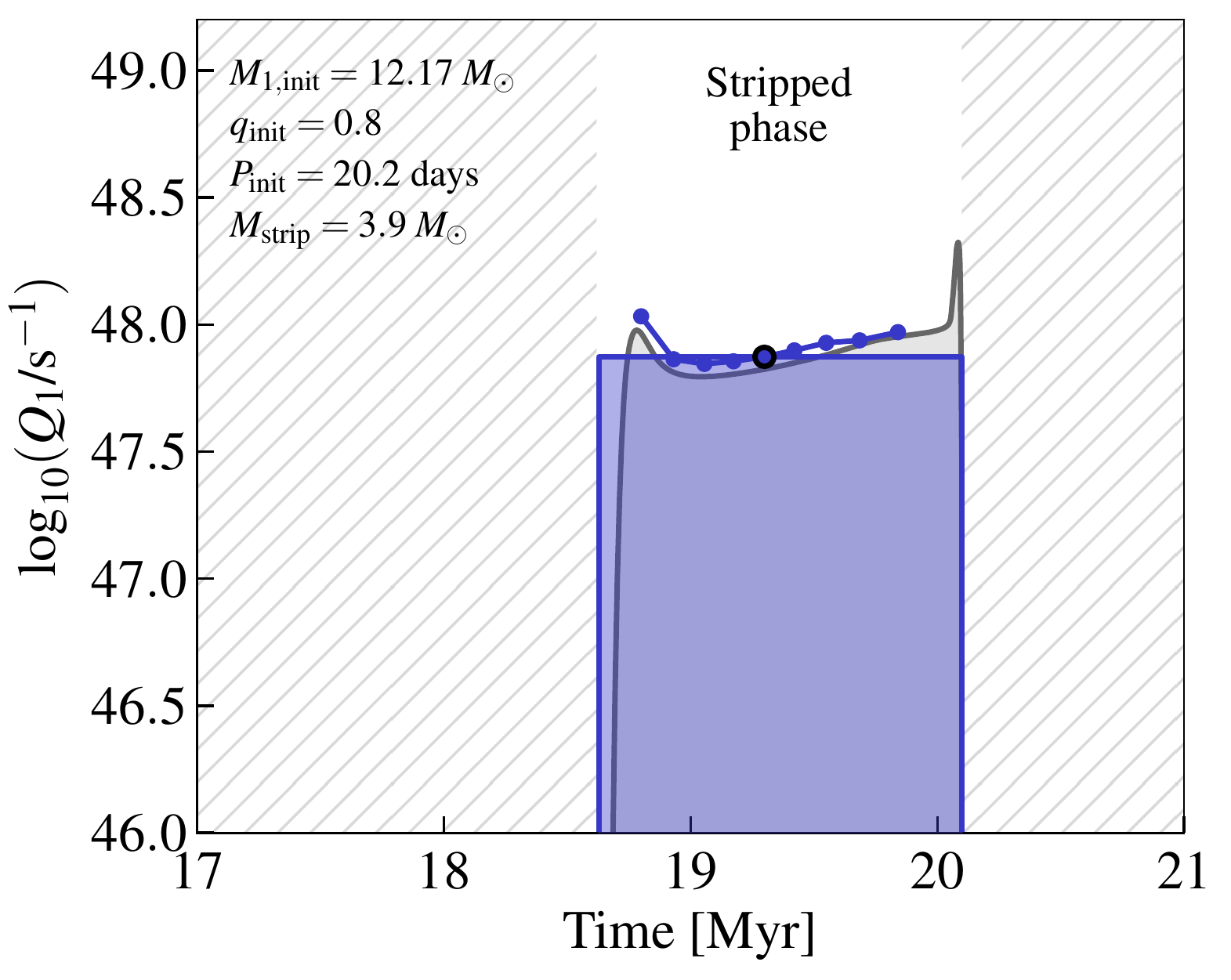}
\includegraphics[width=.33\textwidth]{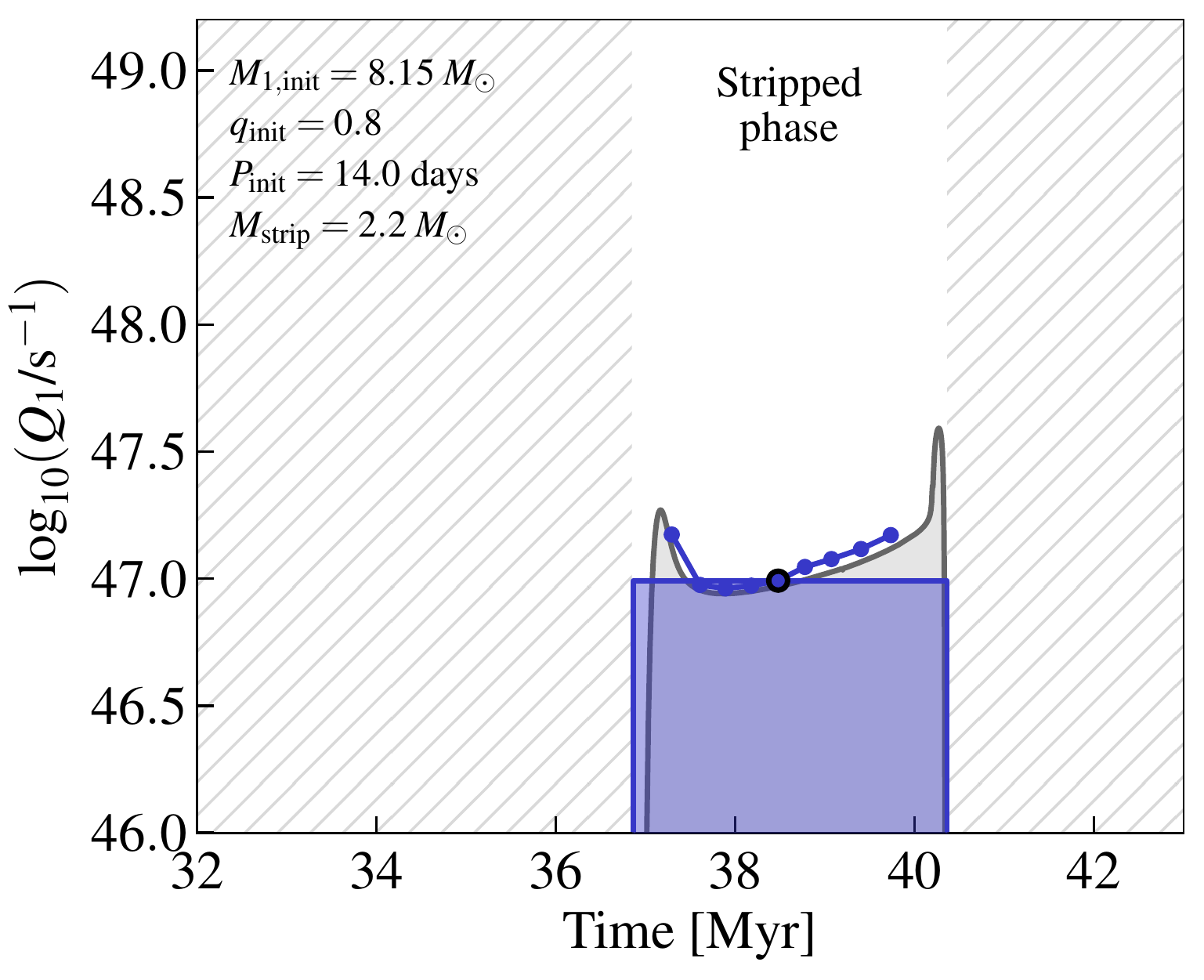}

\includegraphics[width=.33\textwidth]{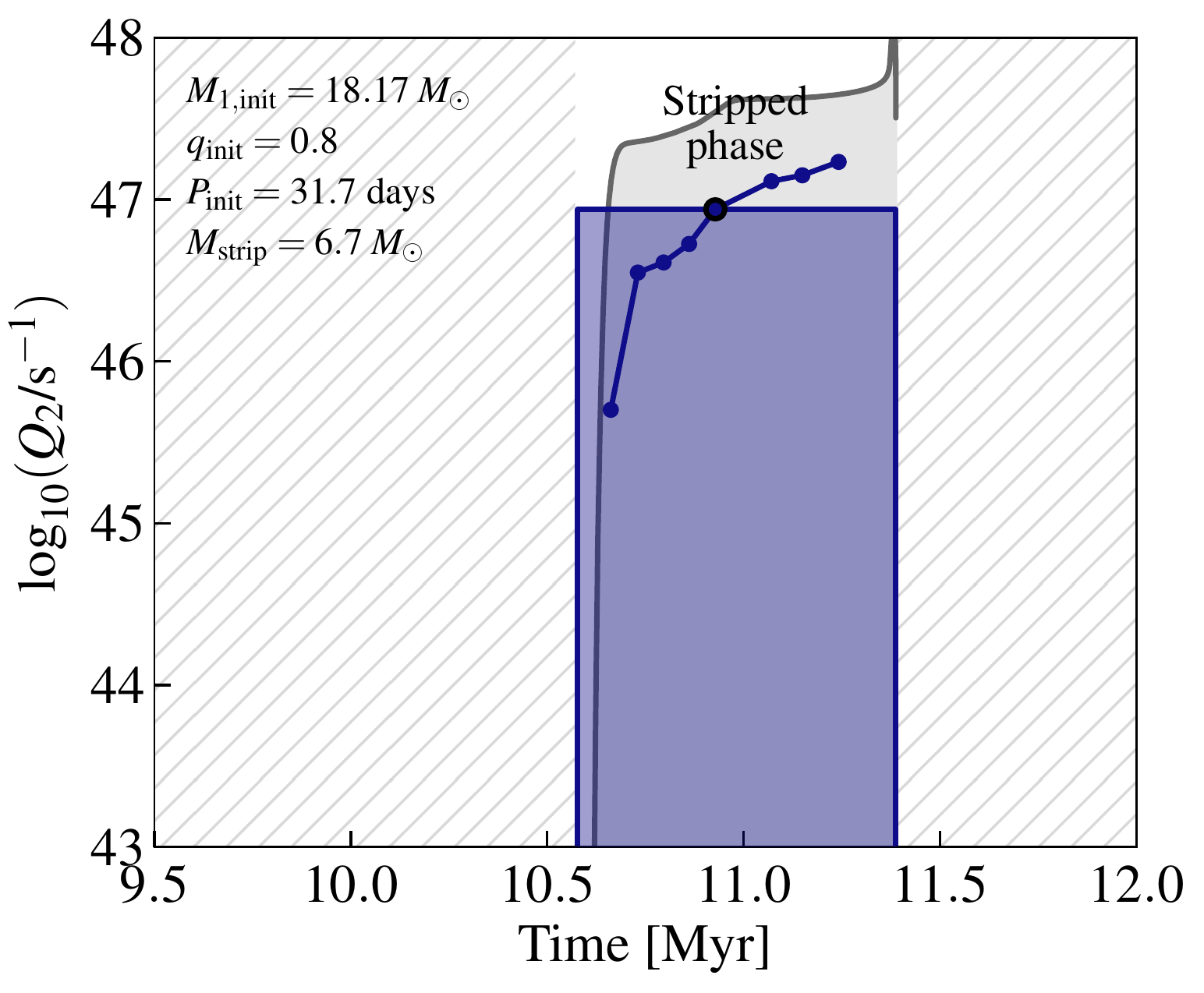}
\includegraphics[width=.33\textwidth]{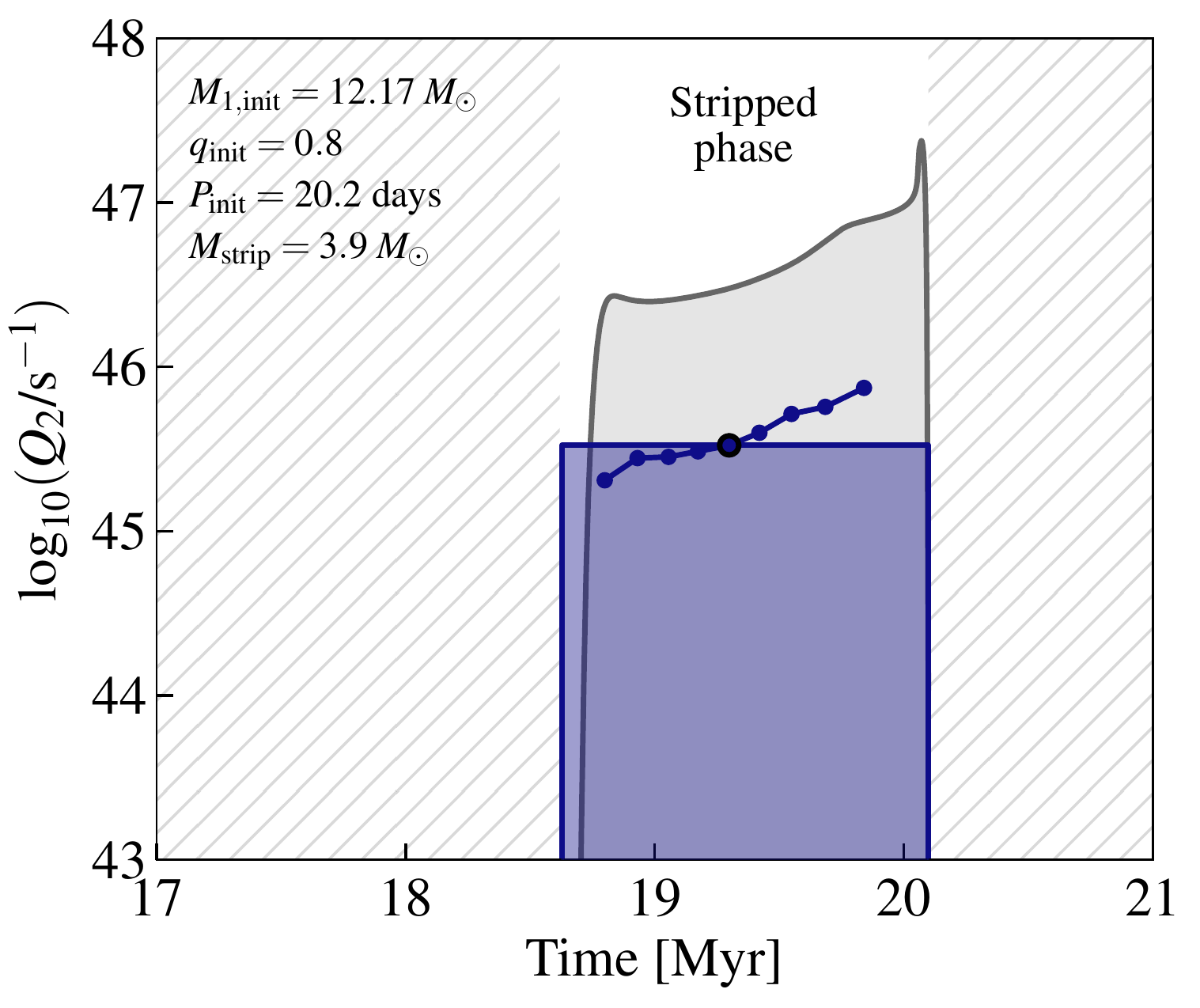}
\includegraphics[width=.33\textwidth]{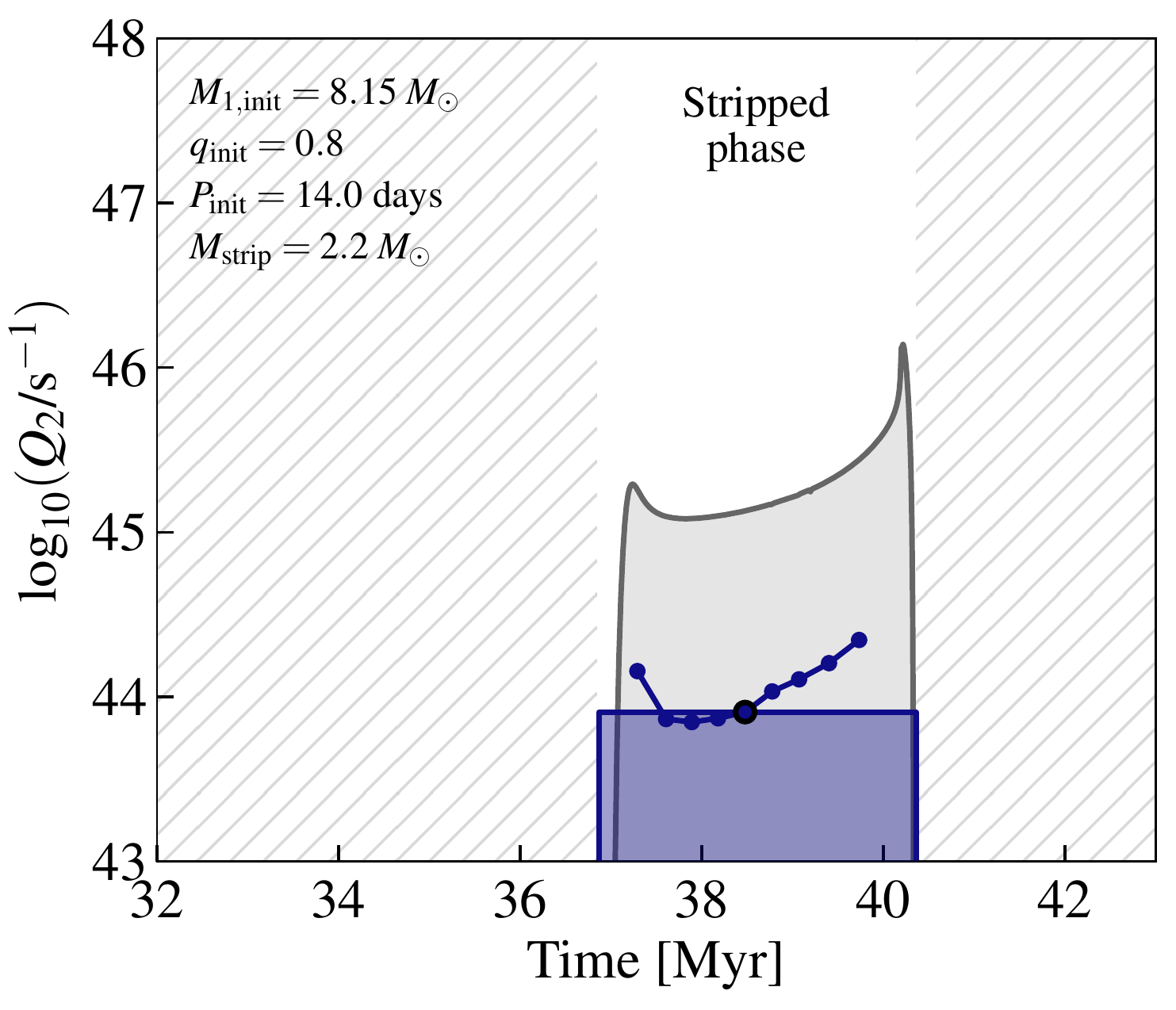}

\caption{Emission rate of ionizing photons as function of time during the stripped phase of three models with initial masses of $M_{\text{init}} = 18.17$ (left column), 12.17 (middle column), and 8.15 \Msun (right column). We show the emission rates of \HI-, \HeI-, and \HeII-ionizing photons in the top, middle and bottom rows respectively. The constant rates that we assume when creating the population synthesis are shown as colored boxes (see \secref{sec:modeling_population}), while the results from atmosphere models that are modeled at different times during the stripped phases are shown as colored dots connected with a line. The prediction from assuming blackbody radiation from the stars is shown in gray. We hatch the time when the star is not stripped.}
\label{fig:Q_M}
\end{figure*}

\noindent We calculate the emission rates of \HI-, \HeI-, and \HeII-ionizing photons at different times during the stripped phases using the additional spectral models. We present the resulting emission rates as a function of time in \figref{fig:Q_M} and give the values in \tabref{tab:Qs_examples}. The variation in the emission rate of \HI-ionizing photons is a factor of $\sim 1.5$ for all the three considered stripped stars. The same is true for the emission rate of \HeI-ionizing photons. The variation in the uncertain emission rate of \HeII-ionizing photons is from a factor of a few up to several ten times. 

We calculate the total number of emitted ionizing photons by the three stripped stars displayed in \figref{fig:Q_M} and in \tabref{tab:Qs_examples} and compare whether multiple spectral models give rise to a significant improvement in the estimate for the total number of emitted ionizing photons. To do this, we first assume that the stripped phase lasts from when the central hydrogen mass fraction has reached $X_{\text{H, c}} < 0.001$ until the end of our computations. The stripped phase starts slightly later, but the difference in time is negligible. We then use either the emission rates as a function of time by using the multiple spectral models for different time bins or just the emission rates for the spectral model computed at $X_{\text{He, c}} = 0.5$. 

The difference between the estimate from just one spectral model and from using a sequence of spectral models is remarkably small. Using just one single spectral model computed at $X_{\text{He, c}} = 0.5$ underestimates the total production of \HI- and  \HeI-ionizing photons with less than 20\% for either of the stars. The emission rate of \HeII-ionizing photons is underestimated by at maximum 40\% for the models we considered here.

In \figref{fig:Q_M} and \tabref{tab:Qs_examples} we also show the evolution of the ionizing emission rates as predicted by blackbody radiation if we would only have known the stellar parameters from \code{MESA}. As we showed in \citetalias{2017A&A...608A..11G}, the difference between the predictions from blackbody and from \code{CMFGEN} is subtle for the emission rates of \HI- and \HeI-ionizing photons, but very significant for the emission rates of \HeII-ionizing photons.

We conclude that modeling a sequence of spectra for each stripped star model will only marginally improve the estimated emission rate of ionizing photons. As \figref{fig:Q_M} and \tabref{tab:Qs_examples} indicate, the stripped stars emit on average ionizing photons at a slightly higher rate than what the model computed half-way through central helium burning suggests. The difference is small and we consider therefore that the simple approach of using just one spectral model per stripped star is sufficient to estimate the emission rate of ionizing photons from stripped stars (as described in \secref{sec:modeling_population}).

\section{Comparison to models of \code{BPASS}}\label{app:BPASS_comparison}

The code \code{BPASS} combines large grids of evolutionary models of interacting binaries and single stars with spectral libraries to simulate a stellar population \citep{2017PASA...34...58E}. \citet{2016MNRAS.456..485S} show that \code{BPASS} predicts higher emission rate of ionizing photons and also a delayed emission, which is very similar to what we find in this study. In the main part of this manuscript (\secref{sec:ionizing_time}), we compare the emission rate of ionizing photons that we predict to come from stripped stars with the emission rate of ionizing photons that \code{BPASS} predicts. Most of our predictions are consistent with the predictions from \code{BPASS}, but there are parts where this is not the case. To address whether the reason is that stripped stars are treated in a way that is very different from ours or whether it actually comes from other types of stars, we go into detail and compare individual evolutionary and spectral models for stripped stars. 

Here, we first compare the evolutionary models of stripped stars included in \code{BPASS} with our evolutionary models, using similar parameters for the initial binary systems. We then compare the spectral energy distribution that \code{BPASS} would choose for the individual stripped stars with the \code{CMFGEN} models that we have computed especially for stripped stars. We use the models from the \code{BPASS} version 2.2.1,  \textit{Tuatara} \citep{2017PASA...34...58E, 2018MNRAS.479...75S}. These models are publicly available on the \code{BPASS} website\footnote{\url{http://bpass.auckland.ac.nz}}. 

We use four models from our model grid with solar metallicity. These have initial masses of 18.17, 12.17, 8.15, and 5.45\Msun. We chose these masses both because they cover a large mass range and thus a diversity of stripped stars and also because these stripped stars are formed roughly at the times we display the integrated spectrum in \figref{fig:SED_population} (11, 20, 50, and 100 Myr after a starburst).

\begin{figure*}
\centering
\includegraphics[width=.35\textwidth]{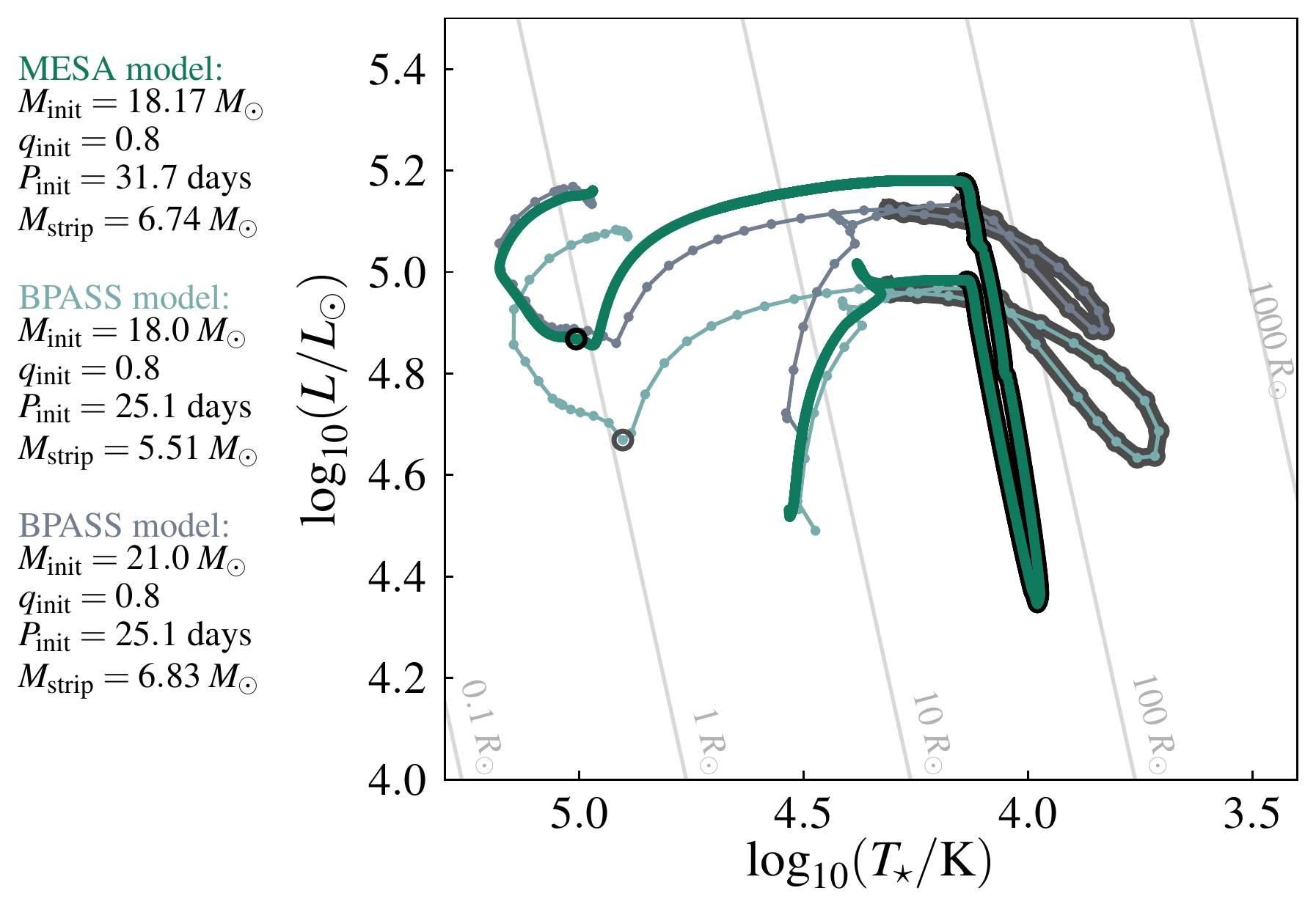}
\hspace{10mm}
\includegraphics[width=.42\textwidth]{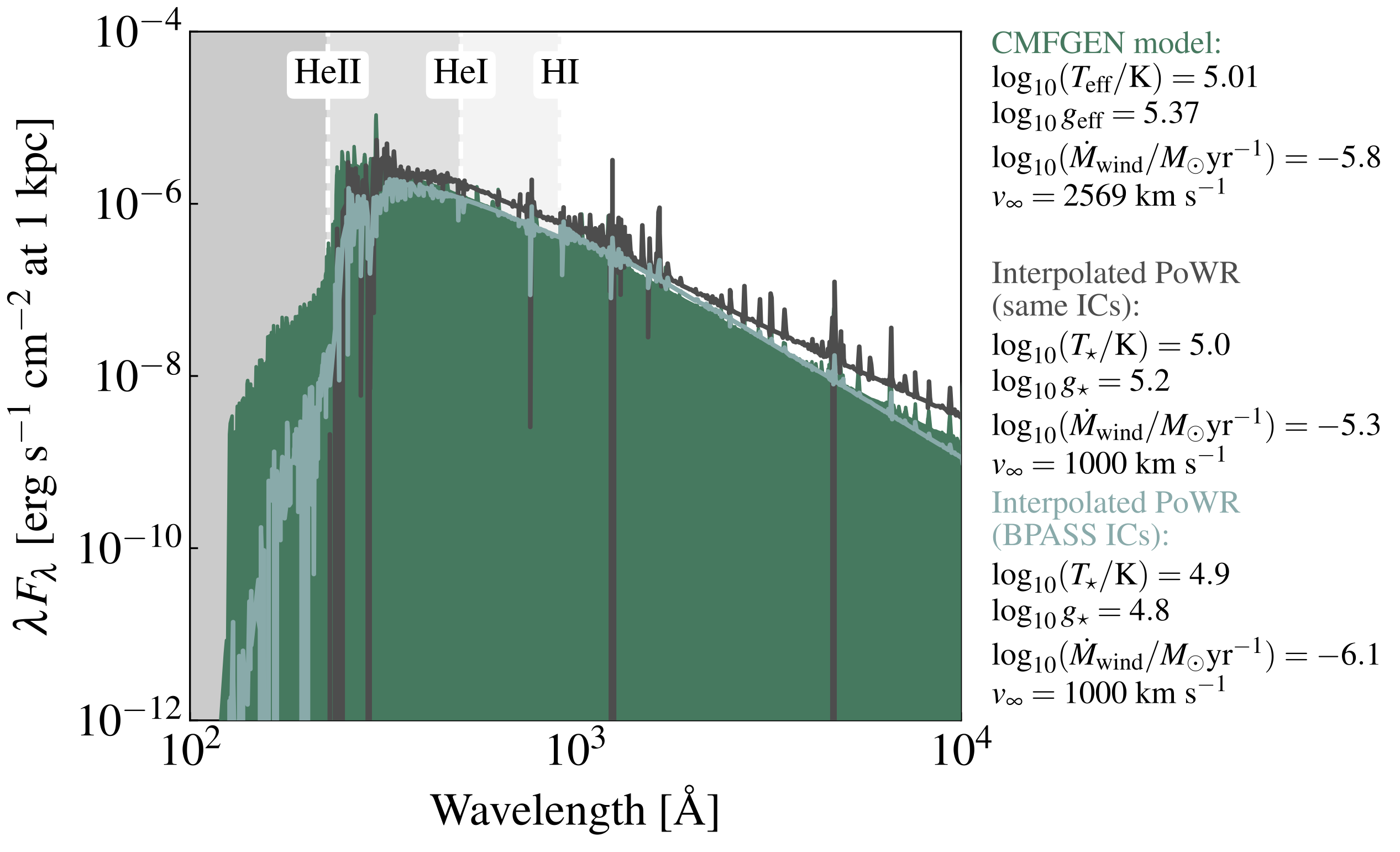}
\includegraphics[width=.35\textwidth]{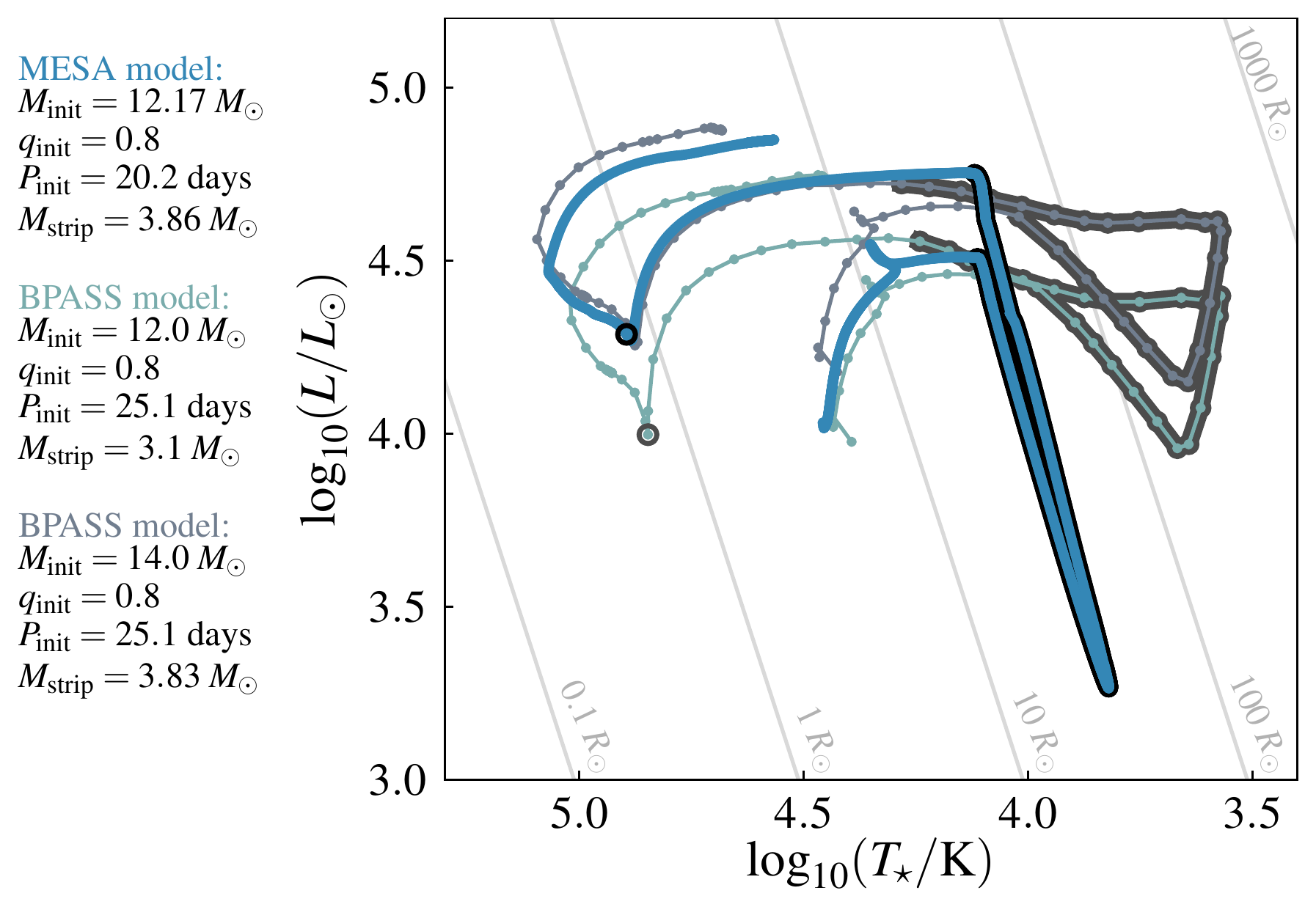}
\hspace{10mm}
\includegraphics[width=.42\textwidth]{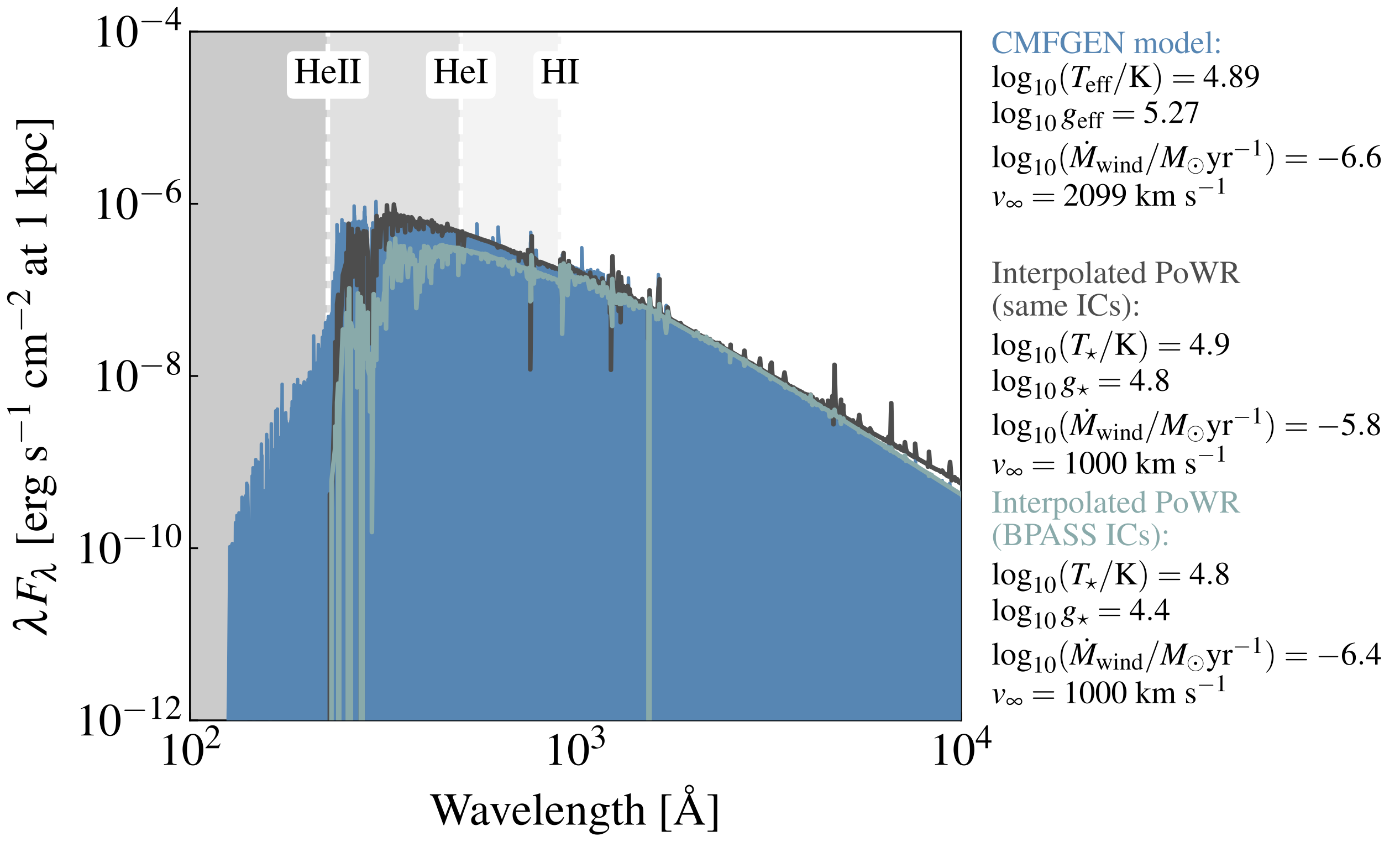}\\
\includegraphics[width=.35\textwidth]{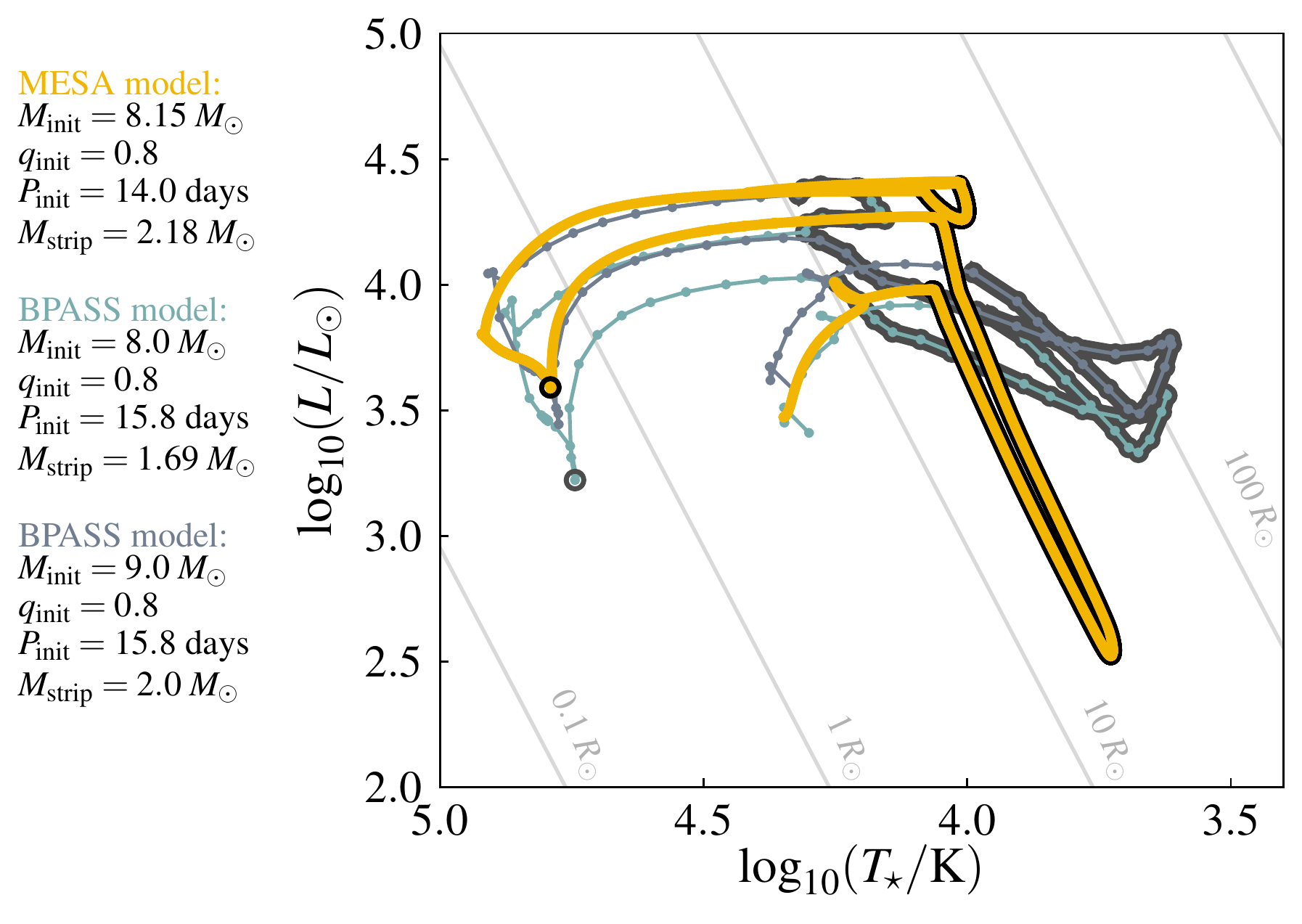}
\hspace{10mm}
\includegraphics[width=.42\textwidth]{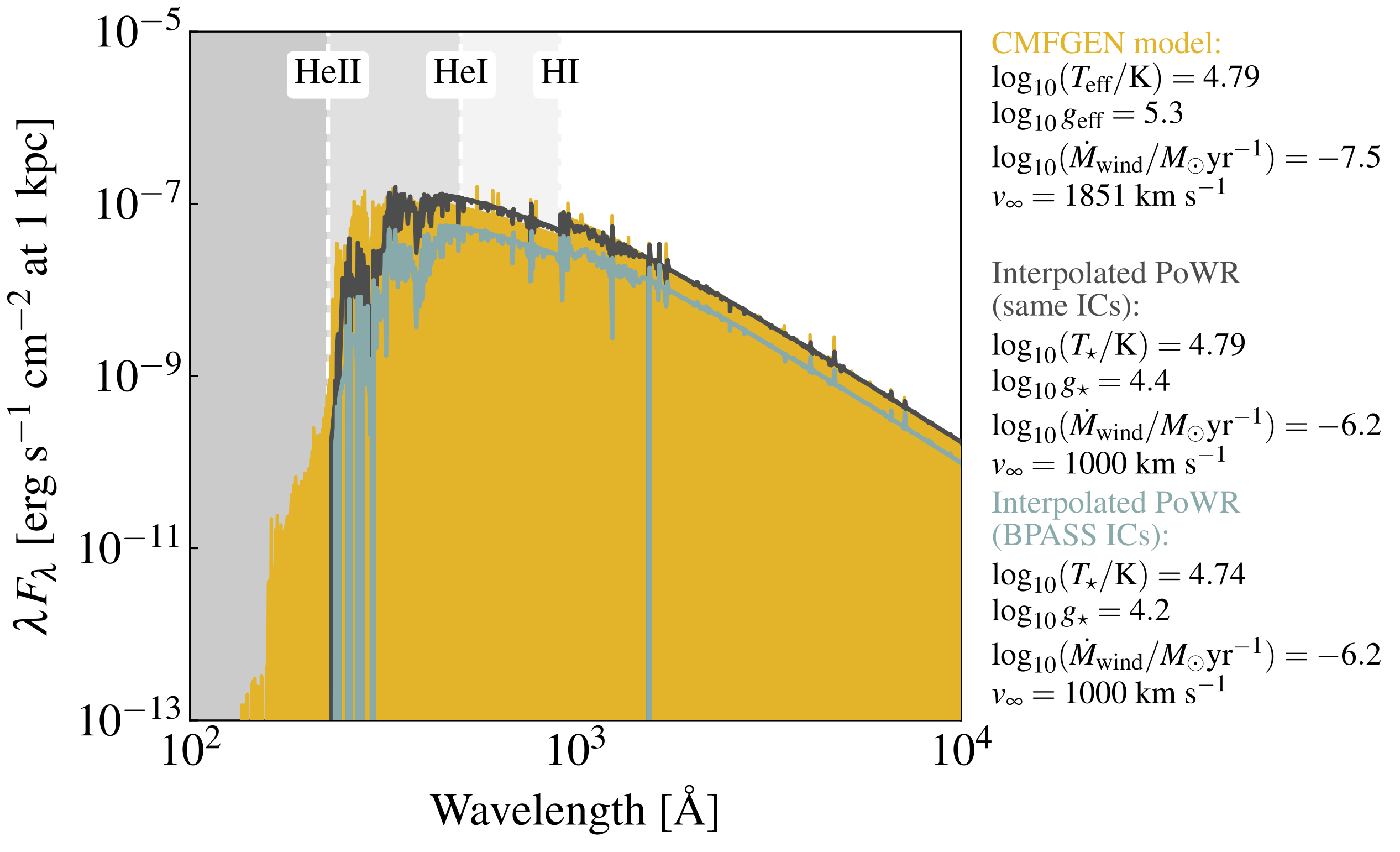}\\
\includegraphics[width=.35\textwidth]{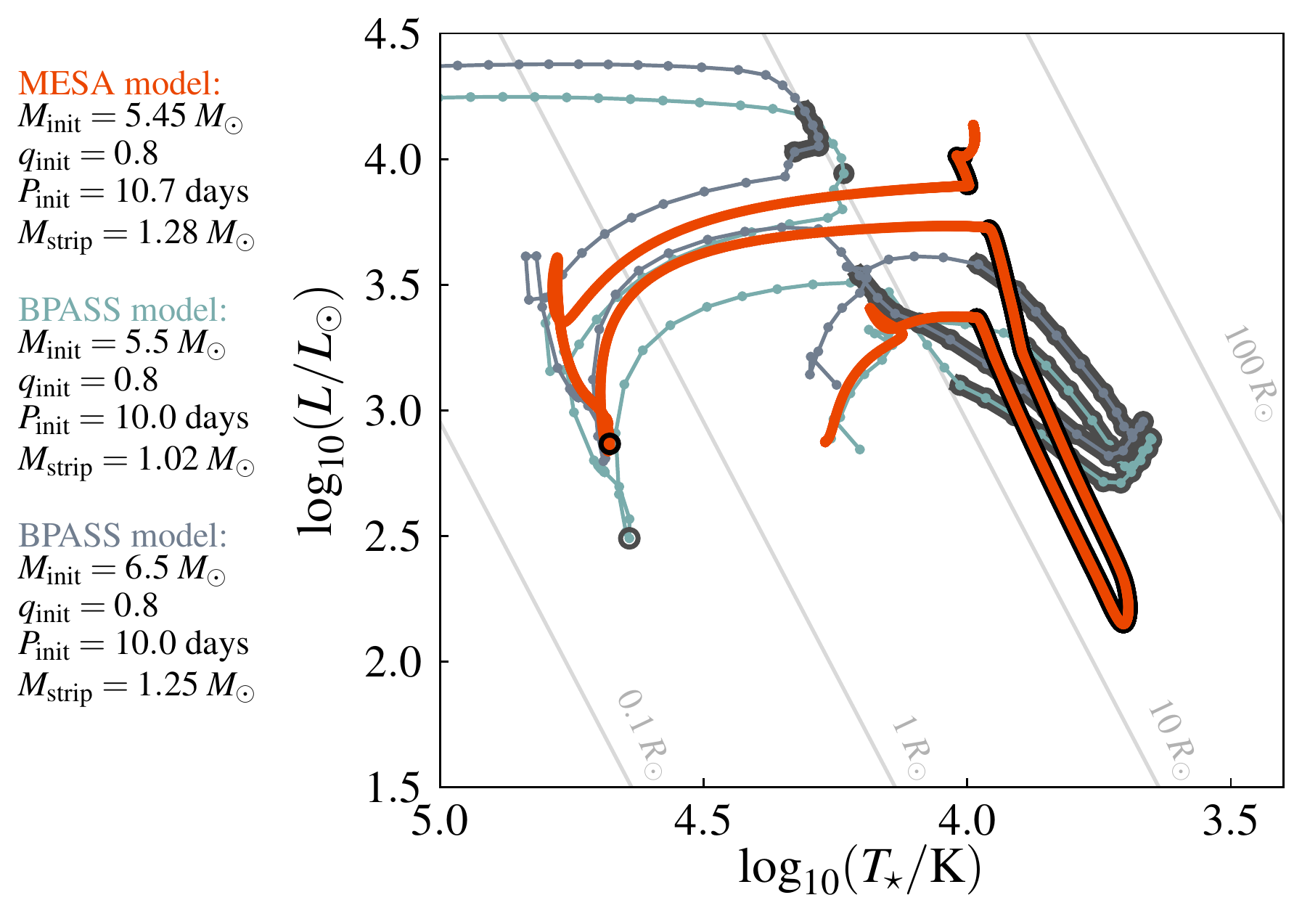}
\hspace{10mm}
\includegraphics[width=.42\textwidth]{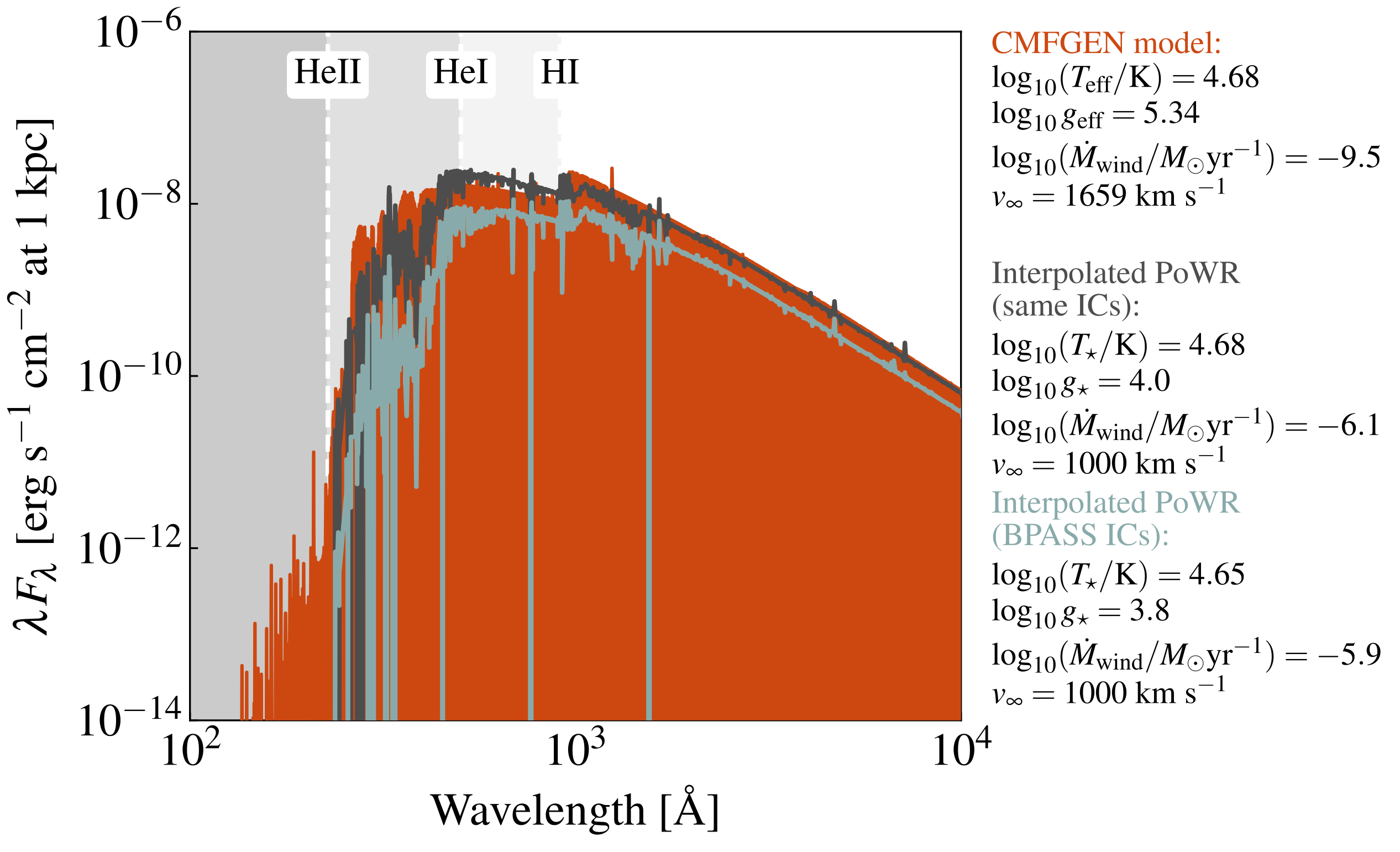}\\
\caption{Comparison between our models and the ones used in \code{BPASS} for the evolution (left) and spectra (right) of stripped stars. We use five example models of stars that lose their envelopes by Roche-lobe overflow initiated early on the Hertzsprung gap and that have initial masses of $M_{\text{init, 1}} = $ 5.5, 8.2, 12.2, and 18.2\Msun and initial mass ratios of $q = M_{\text{init, 2}}/M_{\text{init, 1}} = 0.8$. We show the models we created in orange, yellow, blue, and green, and BPASS models in gray and pale green.}
\label{fig:BPASS_comparison}
\end{figure*}

\subsection{Evolutionary models}

Depending on which assumptions are made for stellar evolution models, the stars evolve slightly or very different. We show the evolutionary tracks of stars that undergo Case~B mass transfer and get stripped in the left column of panels in \figref{fig:BPASS_comparison}. Our models are shown in color  while the models from \code{BPASS} with the most similar initial binary parameters are shown in pale green. As the figure shows, the stripped stars computed in the \code{BPASS} models are fainter than our models for stripped stars. The reason for the difference is likely the lower overshooting that is assumed in \code{BPASS} \citep{2017PASA...34...58E} and the resulting stripped stars therefore have lower masses in \code{BPASS} compared to in our models. 
This means that for the same population initial conditions, \code{BPASS} should predict fewer stripped stars than what our model does since they would reach the lower mass limit for central helium burning at a higher progenitor mass than what our models do. 

To see whether \code{BPASS} can produce stripped stars that are similar to our models, we also show the evolution of a more massive star computed in \code{BPASS} in each the HR diagrams in \figref{fig:BPASS_comparison}. The resulting stripped star in these cases appears in the same region of the HR diagram as our models and likely has very similar properties as our structure models for stripped stars. 

We also note that it is visible in \figref{fig:BPASS_comparison} that mass transfer is treated differently in \code{BPASS} than in \code{MESA}. The shape of the evolutionary tracks during mass transfer is different, and the systems computed in \code{BPASS} tend to tighten in contrast to our models. The reason for this could be that angular momentum is removed from the orbit of the binary system when it cannot be accreted by the accretor star in \code{BPASS}, while we assume that mass is lost with the specific angular momentum of the accretor star itself. 

\subsection{Spectral models}

The code \code{BPASS} uses WR star models from \code{PoWR} to represent the radiative emission from stripped stars with temperatures $T_{\star} \geq 4.45$ and surface hydrogen mass fraction $X_{\text{H, s}} < 0.4$. The coolest subdwarfs are represented by B-star models from the \code{BaSeL} grids. 
The \code{PoWR} models are made in particular for WR stars and therefore are set to have high wind mass loss rates. The models also have a fixed terminal wind speed of $v_{\infty} = 1000\, \kms$. 

To better understand whether the choice of atmosphere model matters for the emission rate of ionizing photons from stripped stars, we compare the spectral model that we computed for a stripped star with what spectral model would have been assumed by \code{BPASS} for the same stripped star parameters (we follow the description for how this is done in \citealt{2017PASA...34...58E}). We show the spectral energy distributions in the right column of \figref{fig:BPASS_comparison}, where our models are shown in color and the model \code{BPASS} would have assumed for the same stellar parameters is shown in dark gray. In pale green, we also show the spectral energy distribution that \code{BPASS} would have assumed for a stripped star with similar initial binary parameters (marked in the corresponding HR diagram).

The figure shows that the spectral energy distributions of the various models are similar for most of the wavelength range. The exception is for the \HeII-ionizing part because the models assumed in \code{BPASS} show a steep drop in emission at the ionization limit of He$^+$. The differences between our models (shown in colors) and the what is expected for stripped stars with similar progenitors (pale green) is also not very large, primarily a difference in brightness can be seen. 

Since the stripped stars in \code{BPASS} are of lower luminosity than in our model, intuitively they should also emit less ionizing radiation. We measure the emission rates of ionizing photons in the spectral models of our models and compare to the emission rates for a stripped star with the same progenitor mass as modeled in \code{BPASS}. We find that the emission rate of \HI-ionizing photons is 30\% lower in the model assumed in \code{BPASS} compared to our model in the case of the stripped star with the progenitor of mass $~\sim18 \Msun$. With lower mass, the difference increases and for the stripped star with a progenitor of $\sim 5.5 \Msun$ the stripped star modeled in  \code{BPASS} emits \HI-ionizing photons at half the rate of our model. For the emission rate of \HeI-ionizing photons, the effect is similar, just slightly larger, spanning from 40\% fainter for higher mass stripped stars in \code{BPASS}, to 75\% fainter for the $5.5\Msun$ progenitor. The emission rate of \HeII-ionizing photons is at highest 6\% of the emission from our $18 \Msun$ model in the \code{BPASS} counterpart. In the other cases we consider, the \HeII-ionizing emission is about seven orders of magnitudes lower compared to the emission from our models. This questions the origin of the \HeII-ionizing emission from \code{BPASS} at later times (see \figref{fig:pop_ionizing_time}). 



In summary, the emission rate of ionizing photons is expected to be lower from the stripped stars in \code{BPASS} than the stripped stars we modeled with the same progenitor stars. Likely because of assumptions that affect how efficiently stripped stars are formed in \code{BPASS}, the predicted total emission rates of ionizing photons is higher in \code{BPASS} than in our model. The emission rate of \HeII-ionizing photons that \code{BPASS} assumes for stripped stars is close to negligible compared to what our models predict. This suggests that the hard ionizing emission originates from different objects in \code{BPASS}.

\end{document}

%% file: Table_parameters_Z0.014.tex
\begin{tabular}{lccccccccc}
\toprule \midrule
\multicolumn{9}{l}{\textbf{Co-eval stellar population} ($10^6 \Msun$, $Z = 0.014$)} \\ 
\midrule
Time & $\log_{10}$ \Qz & $\log_{10}$ \Qo & $\log_{10}$ \Qt & $\log_{10}$ \xii & $\log_{10}$ \xiio & $\log_{10}$ \xiit & $\log_{10} L_{\nu}(1500 \AA)$ & $\log_{10} U$ & $\beta$ \\ 
 $[$Myr$]$ & $[$s$^{-1}]$ & $[$s$^{-1}]$ & $[$s$^{-1}]$ & $[$\ergHz$]$ & $[$\ergHz$]$ & $[$\ergHz$]$ & $[$erg s$^{-1}$ Hz$^{-1}]$ & &  \\ 
Section & \ref{sec:ionizing_time} & \ref{sec:ionizing_time} & \ref{sec:ionizing_time} & \ref{sec:xi_ion} & \ref{sec:xi_ion} & \ref{sec:xi_ion} & \ref{sec:xi_ion} & \ref{sec:U} & \ref{sec:beta} \\ 
\midrule 
$2$ & $52.6$ \, \{-\} \, ($52.6$) & $51.7$ \, \{-\} \,   ($51.7$) & $47.3$ \, \{-\} \,  ($47.3$) & $25.6$ \, ($25.6$) & $24.7$ \, ($24.7$) & $20.3$ \, ($20.3$) & $27.0$ \, ($27.0$) & $-2.0$ \, ($-2.0$) & $-2.97$ \, ($-2.97$)  \\ 
$3$ & $52.3$ \, \{-\} \, ($52.3$) & $51.4$ \, \{-\} \,   ($51.4$) & $41.7$ \, \{-\} \,  ($41.7$) & $25.2$ \, ($25.2$) & $24.3$ \, ($24.3$) & $14.6$ \, ($14.6$) & $27.1$ \, ($27.1$) & $-2.1$ \, ($-2.1$) & $-2.5$ \, ($-2.5$)  \\ 
$5$ & $51.7$ \, \{-\} \, ($51.7$) & $50.3$ \, \{-\} \,   ($50.3$) & $39.9$ \, \{-\} \,  ($39.9$) & $24.9$ \, ($24.9$) & $23.5$ \, ($23.5$) & $13.1$ \, ($13.1$) & $26.8$ \, ($26.8$) & $-2.3$ \, ($-2.3$) & $-2.41$ \, ($-2.41$)  \\ 
$7$ & $50.8$ \, \{-\} \, ($50.8$) & $48.0$ \, \{-\} \,   ($48.0$) & $37.5$ \, \{-\} \,  ($37.5$) & $24.3$ \, ($24.3$) & $21.5$ \, ($21.5$) & $11.0$ \, ($11.0$) & $26.5$ \, ($26.5$) & $-2.6$ \, ($-2.6$) & $-2.41$ \, ($-2.41$)  \\ 
$11$ & $50.9$ \, \{50.9\} \, ($49.8$) & $50.7$ \, \{50.7\} \,   ($46.0$) & $48.7$ \, \{48.7\} \,  ($-$) & $24.8$ \, ($23.6$) & $24.5$ \, ($19.9$) & $22.6$ \, ($-$) & $26.2$ \, ($26.2$) & $-2.6$ \, ($-3.0$) & $-2.32$ \, ($-2.31$)  \\ 
$20$ & $50.3$ \, \{50.3\} \, ($48.6$) & $50.1$ \, \{50.1\} \,   ($43.4$) & $47.7$ \, \{47.7\} \,  ($-$) & $24.5$ \, ($22.8$) & $24.3$ \, ($17.6$) & $21.9$ \, ($-$) & $25.8$ \, ($25.8$) & $-2.8$ \, ($-3.4$) & $-2.21$ \, ($-2.2$)  \\ 
$30$ & $50.0$ \, \{49.9\} \, ($47.9$) & $49.6$ \, \{49.6\} \,   ($41.5$) & $46.7$ \, \{46.7\} \,  ($-$) & $24.3$ \, ($22.3$) & $24.0$ \, ($15.9$) & $21.1$ \, ($-$) & $25.6$ \, ($25.6$) & $-2.9$ \, ($-3.6$) & $-2.02$ \, ($-2.01$)  \\ 
$50$ & $49.5$ \, \{49.5\} \, ($46.7$) & $49.1$ \, \{49.1\} \,   ($39.8$) & $45.5$ \, \{45.5\} \,  ($-$) & $24.2$ \, ($21.4$) & $23.8$ \, ($14.4$) & $20.1$ \, ($-$) & $25.3$ \, ($25.3$) & $-3.0$ \, ($-4.0$) & $-1.83$ \, ($-1.82$)  \\ 
$100$ & $48.9$ \, \{48.9\} \, ($44.9$) & $48.3$ \, \{48.3\} \,   ($37.7$) & $43.2$ \, \{43.2\} \,  ($-$) & $24.0$ \, ($19.9$) & $23.4$ \, ($12.8$) & $18.3$ \, ($-$) & $24.9$ \, ($24.9$) & $-3.2$ \, ($-4.6$) & $-1.51$ \, ($-1.49$)  \\ 
$200$ & $48.2$ \, \{48.2\} \, ($42.8$) & $46.9$ \, \{46.9\} \,   ($-$) & $40.3$ \, \{40.3\} \,  ($-$) & $23.7$ \, ($18.3$) & $22.4$ \, ($-$) & $15.8$ \, ($-$) & $24.5$ \, ($24.5$) & $-3.5$ \, ($-5.3$) & $-1.2$ \, ($-1.17$)  \\ 
$300$ & $47.4$ \, \{47.4\} \, ($41.5$) & $46.1$ \, \{46.1\} \,   ($-$) & $40.7$ \, \{40.7\} \,  ($-$) & $23.3$ \, ($17.4$) & $22.0$ \, ($-$) & $16.6$ \, ($-$) & $24.1$ \, ($24.1$) & $-3.8$ \, ($-5.7$) & $-0.4$ \, ($-0.31$)  \\ 
$500$ & $45.9$ \, \{45.9\} \, ($39.6$) & $43.4$ \, \{43.4\} \,   ($-$) & $37.6$ \, \{37.6\} \,  ($-$) & $22.6$ \, ($16.3$) & $20.1$ \, ($-$) & $14.3$ \, ($-$) & $23.3$ \, ($23.3$) & $-4.3$ \, ($-6.4$) & $1.78$ \, ($2.54$)  \\ 
$800$ & $44.7$ \, \{44.7\} \, ($36.7$) & $41.6$ \, \{41.6\} \,   ($-$) & $34.5$ \, \{34.5\} \,  ($-$) & $22.8$ \, ($15.3$) & $19.7$ \, ($-$) & $12.6$ \, ($-$) & $22.0$ \, ($21.4$) & $-4.6$ \, ($-7.3$) & $4.5$ \, ($9.73$)  \\ 
$1000$ & $44.2$ \, \{44.2\} \, ($35.7$) & $40.8$ \, \{40.8\} \,   ($-$) & $33.8$ \, \{33.8\} \,  ($-$) & $22.5$ \, ($15.2$) & $19.1$ \, ($-$) & $12.1$ \, ($-$) & $21.7$ \, ($20.5$) & $-4.8$ \, ($-7.6$) & $4.8$ \, ($11.97$)  \\ 
\toprule \midrule
\multicolumn{9}{l}{\textbf{Continuous star-formation} ($1 \Msun$/year, $Z = 0.014$)} \\ 
\midrule
Time & $\log_{10}$ \Qz & $\log_{10}$ \Qo & $\log_{10}$ \Qt & $\log_{10}$ \xii & $\log_{10}$ \xiio & $\log_{10}$ \xiit & $\log_{10} L_{\nu}(1500 \AA)$ & $\log_{10} U$ & $\beta$ \\ 
 $[$Myr$]$ & $[$s$^{-1}]$ & $[$s$^{-1}]$ & $[$s$^{-1}]$ & $[$\ergHz$]$ & $[$\ergHz$]$ & $[$\ergHz$]$ & $[$erg s$^{-1}$ Hz$^{-1}]$ & & \\ 
Section & \ref{sec:ionizing_time} & \ref{sec:ionizing_time} & \ref{sec:ionizing_time} & \ref{sec:xi_ion} & \ref{sec:xi_ion} & \ref{sec:xi_ion} & \ref{sec:xi_ion} & \ref{sec:U} & \ref{sec:beta} \\ 
\midrule 
$500$ & $53.1$ \, \{51.9\} \, ($53.1$) & $52.3$ \, \{51.6\} \, ($52.2$) & $49.4$ \, \{49.3\} \, ($48.3$) & $25.1$ \, ($25.1$) & $24.4$ \, ($24.3$) & $21.4$ \, ($20.4$) & $28.0$ \, ($28.0$) & $-1.8$ \, ($-1.9$) & $-2.27$ \, ($-2.27$) \\ 
\bottomrule 
\end{tabular}

%% file: Table_parameters_Z0.006.tex
\begin{tabular}{lccccccccc}
\toprule \midrule
\multicolumn{9}{l}{\textbf{Co-eval stellar population} ($10^6 \Msun$, $Z = 0.006$)} \\ 
\midrule
Time & $\log_{10}$ \Qz & $\log_{10}$ \Qo & $\log_{10}$ \Qt & $\log_{10}$ \xii & $\log_{10}$ \xiio & $\log_{10}$ \xiit & $\log_{10} L_{\nu}(1500 \AA)$ & $\log_{10} U$ & $\beta$ \\ 
 $[$Myr$]$ & $[$s$^{-1}]$ & $[$s$^{-1}]$ & $[$s$^{-1}]$ & $[$\ergHz$]$ & $[$\ergHz$]$ & $[$\ergHz$]$ & $[$erg s$^{-1}$ Hz$^{-1}]$ & &  \\ 
Section & \ref{sec:ionizing_time} & \ref{sec:ionizing_time} & \ref{sec:ionizing_time} & \ref{sec:xi_ion} & \ref{sec:xi_ion} & \ref{sec:xi_ion} & \ref{sec:xi_ion} & \ref{sec:U} & \ref{sec:beta} \\ 
\midrule 
$2$ & $52.6$ \, \{-\} \, ($52.6$) & $51.8$ \, \{-\} \,   ($51.8$) & $47.5$ \, \{-\} \,  ($47.5$) & $25.6$ \, ($25.6$) & $24.8$ \, ($24.8$) & $20.5$ \, ($20.5$) & $27.0$ \, ($27.0$) & $-2.0$ \, ($-2.0$) & $-2.95$ \, ($-2.95$)  \\ 
$3$ & $52.4$ \, \{-\} \, ($52.4$) & $51.5$ \, \{-\} \,   ($51.5$) & $45.5$ \, \{-\} \,  ($45.5$) & $25.3$ \, ($25.3$) & $24.4$ \, ($24.4$) & $18.4$ \, ($18.4$) & $27.1$ \, ($27.1$) & $-2.1$ \, ($-2.1$) & $-2.58$ \, ($-2.58$)  \\ 
$5$ & $51.7$ \, \{-\} \, ($51.7$) & $50.4$ \, \{-\} \,   ($50.4$) & $40.3$ \, \{-\} \,  ($40.3$) & $24.9$ \, ($24.9$) & $23.7$ \, ($23.7$) & $13.5$ \, ($13.5$) & $26.8$ \, ($26.8$) & $-2.3$ \, ($-2.3$) & $-2.54$ \, ($-2.54$)  \\ 
$7$ & $50.7$ \, \{-\} \, ($50.7$) & $47.9$ \, \{-\} \,   ($47.9$) & $37.5$ \, \{-\} \,  ($37.5$) & $24.2$ \, ($24.2$) & $21.4$ \, ($21.4$) & $11.0$ \, ($11.0$) & $26.5$ \, ($26.5$) & $-2.6$ \, ($-2.6$) & $-2.52$ \, ($-2.52$)  \\ 
$11$ & $51.0$ \, \{50.9\} \, ($49.7$) & $50.7$ \, \{50.7\} \,   ($46.0$) & $48.4$ \, \{48.4\} \,  ($-$) & $24.8$ \, ($23.5$) & $24.5$ \, ($19.8$) & $22.3$ \, ($-$) & $26.2$ \, ($26.2$) & $-2.6$ \, ($-3.0$) & $-2.44$ \, ($-2.43$)  \\ 
$20$ & $50.4$ \, \{50.4\} \, ($48.5$) & $50.1$ \, \{50.1\} \,   ($43.4$) & $47.5$ \, \{47.5\} \,  ($-$) & $24.5$ \, ($22.7$) & $24.3$ \, ($17.6$) & $21.7$ \, ($-$) & $25.8$ \, ($25.8$) & $-2.8$ \, ($-3.4$) & $-2.32$ \, ($-2.31$)  \\ 
$30$ & $50.0$ \, \{50.0\} \, ($47.9$) & $49.7$ \, \{49.7\} \,   ($41.6$) & $46.8$ \, \{46.8\} \,  ($-$) & $24.4$ \, ($22.2$) & $24.1$ \, ($16.0$) & $21.1$ \, ($-$) & $25.6$ \, ($25.6$) & $-2.9$ \, ($-3.6$) & $-2.11$ \, ($-2.09$)  \\ 
$50$ & $49.6$ \, \{49.6\} \, ($46.8$) & $49.2$ \, \{49.2\} \,   ($40.0$) & $45.7$ \, \{45.7\} \,  ($-$) & $24.3$ \, ($21.4$) & $23.9$ \, ($14.7$) & $20.4$ \, ($-$) & $25.3$ \, ($25.3$) & $-3.0$ \, ($-4.0$) & $-1.9$ \, ($-1.88$)  \\ 
$100$ & $49.0$ \, \{49.0\} \, ($45.0$) & $48.5$ \, \{48.5\} \,   ($38.1$) & $43.2$ \, \{43.2\} \,  ($-$) & $24.0$ \, ($20.0$) & $23.5$ \, ($13.2$) & $18.2$ \, ($-$) & $24.9$ \, ($24.9$) & $-3.2$ \, ($-4.6$) & $-1.59$ \, ($-1.56$)  \\ 
$200$ & $48.2$ \, \{48.2\} \, ($43.2$) & $47.1$ \, \{47.1\} \,   ($-$) & $41.0$ \, \{41.0\} \,  ($-$) & $23.7$ \, ($18.7$) & $22.6$ \, ($-$) & $16.5$ \, ($-$) & $24.5$ \, ($24.5$) & $-3.5$ \, ($-5.2$) & $-1.31$ \, ($-1.28$)  \\ 
$300$ & $47.5$ \, \{47.5\} \, ($41.9$) & $46.7$ \, \{46.7\} \,   ($-$) & $41.6$ \, \{41.6\} \,  ($-$) & $23.4$ \, ($17.8$) & $22.6$ \, ($-$) & $17.5$ \, ($-$) & $24.1$ \, ($24.1$) & $-3.7$ \, ($-5.6$) & $-0.58$ \, ($-0.49$)  \\ 
$500$ & $46.1$ \, \{46.1\} \, ($40.1$) & $43.8$ \, \{43.8\} \,   ($-$) & $38.2$ \, \{38.2\} \,  ($-$) & $22.6$ \, ($16.7$) & $20.4$ \, ($-$) & $14.8$ \, ($-$) & $23.5$ \, ($23.4$) & $-4.2$ \, ($-6.2$) & $1.28$ \, ($1.98$)  \\ 
$800$ & $44.8$ \, \{44.8\} \, ($37.1$) & $42.0$ \, \{42.0\} \,   ($-$) & $35.9$ \, \{35.9\} \,  ($-$) & $22.7$ \, ($15.2$) & $19.9$ \, ($-$) & $13.7$ \, ($-$) & $22.1$ \, ($21.8$) & $-4.6$ \, ($-7.2$) & $4.35$ \, ($8.46$)  \\ 
$1000$ & $44.3$ \, \{44.3\} \, ($36.0$) & $41.0$ \, \{41.0\} \,   ($-$) & $33.8$ \, \{33.8\} \,  ($-$) & $22.6$ \, ($15.1$) & $19.2$ \, ($-$) & $12.0$ \, ($-$) & $21.7$ \, ($20.9$) & $-4.8$ \, ($-7.5$) & $4.89$ \, ($11.25$)  \\ 
\toprule \midrule
\multicolumn{9}{l}{\textbf{Continuous star-formation} ($1 \Msun$/year, $Z = 0.006$)} \\ 
\midrule
Time & $\log_{10}$ \Qz & $\log_{10}$ \Qo & $\log_{10}$ \Qt & $\log_{10}$ \xii & $\log_{10}$ \xiio & $\log_{10}$ \xiit & $\log_{10} L_{\nu}(1500 \AA)$ & $\log_{10} U$ & $\beta$ \\ 
 $[$Myr$]$ & $[$s$^{-1}]$ & $[$s$^{-1}]$ & $[$s$^{-1}]$ & $[$\ergHz$]$ & $[$\ergHz$]$ & $[$\ergHz$]$ & $[$erg s$^{-1}$ Hz$^{-1}]$ & & \\ 
Section & \ref{sec:ionizing_time} & \ref{sec:ionizing_time} & \ref{sec:ionizing_time} & \ref{sec:xi_ion} & \ref{sec:xi_ion} & \ref{sec:xi_ion} & \ref{sec:xi_ion} & \ref{sec:U} & \ref{sec:beta} \\ 
\midrule 
$500$ & $53.1$ \, \{52.0\} \, ($53.1$) & $52.4$ \, \{51.7\} \, ($52.3$) & $49.5$ \, \{49.2\} \, ($49.1$) & $25.1$ \, ($25.1$) & $24.4$ \, ($24.3$) & $21.5$ \, ($21.1$) & $28.0$ \, ($28.0$) & $-1.8$ \, ($-1.9$) & $-2.34$ \, ($-2.34$) \\ 
\bottomrule 
\end{tabular}

%% file: Table_parameters_Z0.002.tex
\begin{tabular}{lccccccccc}
\toprule \midrule
\multicolumn{9}{l}{\textbf{Co-eval stellar population} ($10^6 \Msun$, $Z = 0.002$)} \\ 
\midrule
Time & $\log_{10}$ \Qz & $\log_{10}$ \Qo & $\log_{10}$ \Qt & $\log_{10}$ \xii & $\log_{10}$ \xiio & $\log_{10}$ \xiit & $\log_{10} L_{\nu}(1500 \AA)$ & $\log_{10} U$ & $\beta$ \\ 
 $[$Myr$]$ & $[$s$^{-1}]$ & $[$s$^{-1}]$ & $[$s$^{-1}]$ & $[$\ergHz$]$ & $[$\ergHz$]$ & $[$\ergHz$]$ & $[$erg s$^{-1}$ Hz$^{-1}]$ & &  \\ 
Section & \ref{sec:ionizing_time} & \ref{sec:ionizing_time} & \ref{sec:ionizing_time} & \ref{sec:xi_ion} & \ref{sec:xi_ion} & \ref{sec:xi_ion} & \ref{sec:xi_ion} & \ref{sec:U} & \ref{sec:beta} \\ 
\midrule 
$2$ & $52.7$ \, \{-\} \, ($52.7$) & $52.2$ \, \{-\} \,   ($52.2$) & $48.9$ \, \{-\} \,  ($48.9$) & $25.8$ \, ($25.8$) & $25.2$ \, ($25.2$) & $21.9$ \, ($21.9$) & $26.9$ \, ($26.9$) & $-2.0$ \, ($-2.0$) & $-3.08$ \, ($-3.08$)  \\ 
$3$ & $52.6$ \, \{-\} \, ($52.6$) & $51.9$ \, \{-\} \,   ($51.9$) & $47.6$ \, \{-\} \,  ($47.6$) & $25.5$ \, ($25.5$) & $24.8$ \, ($24.8$) & $20.5$ \, ($20.5$) & $27.1$ \, ($27.1$) & $-2.0$ \, ($-2.0$) & $-2.74$ \, ($-2.74$)  \\ 
$5$ & $52.0$ \, \{-\} \, ($52.0$) & $50.9$ \, \{-\} \,   ($50.9$) & $45.1$ \, \{-\} \,  ($45.1$) & $25.2$ \, ($25.2$) & $24.1$ \, ($24.1$) & $18.3$ \, ($18.3$) & $26.8$ \, ($26.8$) & $-2.2$ \, ($-2.2$) & $-2.58$ \, ($-2.58$)  \\ 
$7$ & $51.4$ \, \{-\} \, ($51.4$) & $49.5$ \, \{-\} \,   ($49.5$) & $39.8$ \, \{-\} \,  ($39.8$) & $24.7$ \, ($24.7$) & $22.9$ \, ($22.9$) & $13.1$ \, ($13.1$) & $26.7$ \, ($26.7$) & $-2.4$ \, ($-2.4$) & $-2.51$ \, ($-2.51$)  \\ 
$11$ & $51.0$ \, \{50.9\} \, ($50.3$) & $50.7$ \, \{50.7\} \,   ($47.4$) & $48.6$ \, \{48.6\} \,  ($37.1$) & $24.6$ \, ($23.9$) & $24.3$ \, ($21.0$) & $22.3$ \, ($10.7$) & $26.4$ \, ($26.4$) & $-2.6$ \, ($-2.8$) & $-2.43$ \, ($-2.42$)  \\ 
$20$ & $50.5$ \, \{50.4\} \, ($49.1$) & $50.2$ \, \{50.2\} \,   ($45.2$) & $47.2$ \, \{47.2\} \,  ($-$) & $24.5$ \, ($23.2$) & $24.2$ \, ($19.2$) & $21.2$ \, ($-$) & $26.0$ \, ($26.0$) & $-2.7$ \, ($-3.2$) & $-2.34$ \, ($-2.33$)  \\ 
$30$ & $50.1$ \, \{50.1\} \, ($48.5$) & $49.8$ \, \{49.8\} \,   ($43.8$) & $46.5$ \, \{46.5\} \,  ($-$) & $24.4$ \, ($22.7$) & $24.0$ \, ($18.0$) & $20.7$ \, ($-$) & $25.8$ \, ($25.8$) & $-2.8$ \, ($-3.4$) & $-2.28$ \, ($-2.27$)  \\ 
$50$ & $49.7$ \, \{49.7\} \, ($47.7$) & $49.3$ \, \{49.3\} \,   ($41.8$) & $45.2$ \, \{45.2\} \,  ($-$) & $24.2$ \, ($22.2$) & $23.8$ \, ($16.3$) & $19.6$ \, ($-$) & $25.5$ \, ($25.5$) & $-3.0$ \, ($-3.7$) & $-2.1$ \, ($-2.09$)  \\ 
$100$ & $49.1$ \, \{49.1\} \, ($46.5$) & $48.5$ \, \{48.5\} \,   ($40.1$) & $43.2$ \, \{43.2\} \,  ($-$) & $23.9$ \, ($21.3$) & $23.4$ \, ($15.0$) & $18.1$ \, ($-$) & $25.1$ \, ($25.1$) & $-3.2$ \, ($-4.1$) & $-1.84$ \, ($-1.82$)  \\ 
$200$ & $48.3$ \, \{48.3\} \, ($44.9$) & $47.2$ \, \{47.2\} \,   ($38.4$) & $41.4$ \, \{41.4\} \,  ($-$) & $23.5$ \, ($20.2$) & $22.4$ \, ($13.6$) & $16.7$ \, ($-$) & $24.7$ \, ($24.7$) & $-3.5$ \, ($-4.6$) & $-1.54$ \, ($-1.52$)  \\ 
$300$ & $47.6$ \, \{47.6\} \, ($43.9$) & $46.8$ \, \{46.8\} \,   ($36.4$) & $41.7$ \, \{41.7\} \,  ($-$) & $23.1$ \, ($19.5$) & $22.3$ \, ($12.0$) & $17.3$ \, ($-$) & $24.5$ \, ($24.5$) & $-3.7$ \, ($-4.9$) & $-1.33$ \, ($-1.3$)  \\ 
$500$ & $46.2$ \, \{46.2\} \, ($42.9$) & $44.6$ \, \{44.6\} \,   ($-$) & $39.7$ \, \{39.7\} \,  ($-$) & $22.1$ \, ($18.8$) & $20.5$ \, ($-$) & $15.6$ \, ($-$) & $24.1$ \, ($24.1$) & $-4.2$ \, ($-5.3$) & $-0.96$ \, ($-0.93$)  \\ 
$800$ & $45.0$ \, \{45.0\} \, ($41.5$) & $42.2$ \, \{42.2\} \,   ($-$) & $36.1$ \, \{36.1\} \,  ($-$) & $21.3$ \, ($17.9$) & $18.6$ \, ($-$) & $12.5$ \, ($-$) & $23.6$ \, ($23.6$) & $-4.6$ \, ($-5.7$) & $0.1$ \, ($0.19$)  \\ 
$1000$ & $44.3$ \, \{44.3\} \, ($40.8$) & $41.2$ \, \{41.2\} \,   ($-$) & $34.2$ \, \{34.2\} \,  ($-$) & $20.9$ \, ($17.5$) & $17.8$ \, ($-$) & $10.8$ \, ($-$) & $23.4$ \, ($23.4$) & $-4.8$ \, ($-5.9$) & $0.97$ \, ($1.11$)  \\ 
\toprule \midrule
\multicolumn{9}{l}{\textbf{Continuous star-formation} ($1 \Msun$/year, $Z = 0.002$)} \\ 
\midrule
Time & $\log_{10}$ \Qz & $\log_{10}$ \Qo & $\log_{10}$ \Qt & $\log_{10}$ \xii & $\log_{10}$ \xiio & $\log_{10}$ \xiit & $\log_{10} L_{\nu}(1500 \AA)$ & $\log_{10} U$ & $\beta$ \\ 
 $[$Myr$]$ & $[$s$^{-1}]$ & $[$s$^{-1}]$ & $[$s$^{-1}]$ & $[$\ergHz$]$ & $[$\ergHz$]$ & $[$\ergHz$]$ & $[$erg s$^{-1}$ Hz$^{-1}]$ & & \\ 
Section & \ref{sec:ionizing_time} & \ref{sec:ionizing_time} & \ref{sec:ionizing_time} & \ref{sec:xi_ion} & \ref{sec:xi_ion} & \ref{sec:xi_ion} & \ref{sec:xi_ion} & \ref{sec:U} & \ref{sec:beta} \\ 
\midrule 
$500$ & $53.3$ \, \{52.0\} \, ($53.3$) & $52.6$ \, \{51.7\} \, ($52.6$) & $49.4$ \, \{49.1\} \, ($49.1$) & $25.2$ \, ($25.2$) & $24.6$ \, ($24.5$) & $21.3$ \, ($21.0$) & $28.1$ \, ($28.1$) & $-1.8$ \, ($-1.8$) & $-2.35$ \, ($-2.35$) \\ 
\bottomrule 
\end{tabular}

%% file: Table_parameters_Z0.0002.tex
\begin{tabular}{lccccccccc}
\toprule \midrule
\multicolumn{9}{l}{\textbf{Co-eval stellar population} ($10^6 \Msun$, $Z = 0.0002$)} \\ 
\midrule
Time & $\log_{10}$ \Qz & $\log_{10}$ \Qo & $\log_{10}$ \Qt & $\log_{10}$ \xii & $\log_{10}$ \xiio & $\log_{10}$ \xiit & $\log_{10} L_{\nu}(1500 \AA)$ & $\log_{10} U$ & $\beta$ \\ 
 $[$Myr$]$ & $[$s$^{-1}]$ & $[$s$^{-1}]$ & $[$s$^{-1}]$ & $[$\ergHz$]$ & $[$\ergHz$]$ & $[$\ergHz$]$ & $[$erg s$^{-1}$ Hz$^{-1}]$ & &  \\ 
Section & \ref{sec:ionizing_time} & \ref{sec:ionizing_time} & \ref{sec:ionizing_time} & \ref{sec:xi_ion} & \ref{sec:xi_ion} & \ref{sec:xi_ion} & \ref{sec:xi_ion} & \ref{sec:U} & \ref{sec:beta} \\ 
\midrule 
$2$ & $52.8$ \, \{-\} \, ($52.8$) & $52.2$ \, \{-\} \,   ($52.2$) & $48.6$ \, \{-\} \,  ($48.6$) & $25.9$ \, ($25.9$) & $25.3$ \, ($25.3$) & $21.7$ \, ($21.7$) & $26.9$ \, ($26.9$) & $-2.0$ \, ($-2.0$) & $-3.08$ \, ($-3.08$)  \\ 
$3$ & $52.6$ \, \{-\} \, ($52.6$) & $52.0$ \, \{-\} \,   ($52.0$) & $47.8$ \, \{-\} \,  ($47.8$) & $25.5$ \, ($25.5$) & $24.8$ \, ($24.8$) & $20.7$ \, ($20.7$) & $27.1$ \, ($27.1$) & $-2.0$ \, ($-2.0$) & $-2.77$ \, ($-2.77$)  \\ 
$5$ & $52.0$ \, \{-\} \, ($52.0$) & $50.9$ \, \{-\} \,   ($50.9$) & $45.8$ \, \{-\} \,  ($45.8$) & $25.2$ \, ($25.2$) & $24.1$ \, ($24.1$) & $19.0$ \, ($19.0$) & $26.8$ \, ($26.8$) & $-2.2$ \, ($-2.2$) & $-2.61$ \, ($-2.61$)  \\ 
$7$ & $51.4$ \, \{-\} \, ($51.4$) & $49.6$ \, \{-\} \,   ($49.6$) & $40.1$ \, \{-\} \,  ($40.1$) & $24.7$ \, ($24.7$) & $23.0$ \, ($23.0$) & $13.4$ \, ($13.4$) & $26.7$ \, ($26.7$) & $-2.4$ \, ($-2.4$) & $-2.57$ \, ($-2.57$)  \\ 
$11$ & $50.6$ \, \{50.4\} \, ($50.2$) & $49.6$ \, \{49.6\} \,   ($47.3$) & $44.8$ \, \{44.8\} \,  ($37.2$) & $24.2$ \, ($23.8$) & $23.2$ \, ($20.9$) & $18.5$ \, ($10.8$) & $26.4$ \, ($26.4$) & $-2.7$ \, ($-2.8$) & $-2.51$ \, ($-2.5$)  \\ 
$20$ & $50.5$ \, \{50.5\} \, ($49.0$) & $50.0$ \, \{50.0\} \,   ($45.1$) & $45.4$ \, \{45.4\} \,  ($-$) & $24.5$ \, ($23.0$) & $24.0$ \, ($19.1$) & $19.4$ \, ($-$) & $26.0$ \, ($26.0$) & $-2.7$ \, ($-3.2$) & $-2.42$ \, ($-2.39$)  \\ 
$30$ & $50.2$ \, \{50.1\} \, ($48.4$) & $49.7$ \, \{49.7\} \,   ($43.9$) & $44.8$ \, \{44.8\} \,  ($-$) & $24.3$ \, ($22.6$) & $23.8$ \, ($18.1$) & $19.0$ \, ($-$) & $25.8$ \, ($25.8$) & $-2.8$ \, ($-3.4$) & $-2.35$ \, ($-2.32$)  \\ 
$50$ & $49.6$ \, \{49.6\} \, ($47.6$) & $49.0$ \, \{49.0\} \,   ($42.1$) & $44.0$ \, \{44.0\} \,  ($-$) & $24.1$ \, ($22.1$) & $23.5$ \, ($16.6$) & $18.4$ \, ($-$) & $25.5$ \, ($25.5$) & $-3.0$ \, ($-3.7$) & $-2.15$ \, ($-2.13$)  \\ 
$100$ & $49.1$ \, \{49.1\} \, ($46.4$) & $48.2$ \, \{48.2\} \,   ($40.5$) & $42.8$ \, \{42.8\} \,  ($-$) & $23.9$ \, ($21.3$) & $23.1$ \, ($15.3$) & $17.7$ \, ($-$) & $25.1$ \, ($25.1$) & $-3.2$ \, ($-4.1$) & $-1.88$ \, ($-1.85$)  \\ 
$200$ & $48.2$ \, \{48.2\} \, ($45.1$) & $46.8$ \, \{46.8\} \,   ($38.8$) & $41.3$ \, \{41.3\} \,  ($-$) & $23.5$ \, ($20.4$) & $22.1$ \, ($14.1$) & $16.5$ \, ($-$) & $24.7$ \, ($24.7$) & $-3.5$ \, ($-4.5$) & $-1.57$ \, ($-1.53$)  \\ 
$300$ & $47.2$ \, \{47.2\} \, ($44.3$) & $45.4$ \, \{45.4\} \,   ($37.0$) & $40.2$ \, \{40.2\} \,  ($-$) & $22.7$ \, ($19.8$) & $21.0$ \, ($12.6$) & $15.7$ \, ($-$) & $24.5$ \, ($24.5$) & $-3.8$ \, ($-4.8$) & $-1.34$ \, ($-1.3$)  \\ 
$500$ & $46.3$ \, \{46.3\} \, ($43.3$) & $45.0$ \, \{45.0\} \,   ($-$) & $40.2$ \, \{40.2\} \,  ($-$) & $22.1$ \, ($19.2$) & $20.9$ \, ($-$) & $16.1$ \, ($-$) & $24.1$ \, ($24.1$) & $-4.1$ \, ($-5.1$) & $-0.96$ \, ($-0.91$)  \\ 
$800$ & $44.8$ \, \{44.8\} \, ($42.0$) & $42.0$ \, \{42.0\} \,   ($-$) & $35.6$ \, \{35.6\} \,  ($-$) & $21.2$ \, ($18.3$) & $18.3$ \, ($-$) & $11.9$ \, ($-$) & $23.7$ \, ($23.6$) & $-4.6$ \, ($-5.6$) & $0.06$ \, ($0.14$)  \\ 
$1000$ & $41.4$ \, \{-\} \, ($41.4$) & $-$ \, \{-\} \,   ($-$) & $-$ \, \{-\} \,  ($-$) & $18.0$ \, ($18.0$) & $-$ \, ($-$) & $-$ \, ($-$) & $23.4$ \, ($23.4$) & $-5.8$ \, ($-5.8$) & $1.03$ \, ($1.03$)  \\ 
\toprule \midrule
\multicolumn{9}{l}{\textbf{Continuous star-formation} ($1 \Msun$/year, $Z = 0.0002$)} \\ 
\midrule
Time & $\log_{10}$ \Qz & $\log_{10}$ \Qo & $\log_{10}$ \Qt & $\log_{10}$ \xii & $\log_{10}$ \xiio & $\log_{10}$ \xiit & $\log_{10} L_{\nu}(1500 \AA)$ & $\log_{10} U$ & $\beta$ \\ 
 $[$Myr$]$ & $[$s$^{-1}]$ & $[$s$^{-1}]$ & $[$s$^{-1}]$ & $[$\ergHz$]$ & $[$\ergHz$]$ & $[$\ergHz$]$ & $[$erg s$^{-1}$ Hz$^{-1}]$ & & \\ 
Section & \ref{sec:ionizing_time} & \ref{sec:ionizing_time} & \ref{sec:ionizing_time} & \ref{sec:xi_ion} & \ref{sec:xi_ion} & \ref{sec:xi_ion} & \ref{sec:xi_ion} & \ref{sec:U} & \ref{sec:beta} \\ 
\midrule 
$500$ & $53.3$ \, \{52.0\} \, ($53.3$) & $52.7$ \, \{51.4\} \, ($52.6$) & $48.9$ \, \{46.8\} \, ($48.9$) & $25.2$ \, ($25.2$) & $24.6$ \, ($24.6$) & $20.8$ \, ($20.8$) & $28.1$ \, ($28.1$) & $-1.8$ \, ($-1.8$) & $-2.39$ \, ($-2.38$) \\ 
\bottomrule 
\end{tabular}

%% file: params_examples.tex
\begin{tabular}{ccccccccccccc}
\toprule\midrule
$X_{\text{He,c}}$ & Age$^a$ & $M_{\text{strip}}^a$ & $R_{\star}^a$ & $R_{\text{eff}}^b$ & $T_{\star}^a$ & $T_{\text{eff}}^b$ & $\log_{10} g_{\star}^a$ & $\log_{10} g_{\text{eff}}^b$ & $\log_{10} L^a$ & $\log_{10} \dot{M}_{\text{wind}}^a$ & $X_{\text{H,s}}^a$ & $X_{\text{He,s}}^a$ \\ 
 & $[$Myr$]$ & $[M_{\odot}]$ & $[R_{\odot}]$ & $[R_{\odot}]$ & $[$kK$]$ & $[$kK$]$ & $[$cm s$^{-2}]$ & $[$cm s$^{-2}]$ & $[L_{\odot}]$ & $[M_{\odot}\text{ yr}^{-1}]$ &  &  \\ 
\midrule
\multicolumn{13}{l}{$M_{\text{init}} = 8.15 M_{\odot}$} \\ 
\midrule
$0.9$& $37.3$ & $2.22$ & $0.67$ & $0.67$ & $61.1$ & $61.0$ & $5.13$ & $5.13$ & $3.75$ & $-7.3$ & $0.24$ & $0.746$\\ 
$0.8$& $37.6$ & $2.21$ & $0.55$ & $0.55$ & $61.2$ & $61.1$ & $5.3$ & $5.3$ & $3.58$ & $-7.53$ & $0.245$ & $0.741$\\ 
$0.7$& $37.9$ & $2.2$ & $0.54$ & $0.54$ & $61.2$ & $61.1$ & $5.31$ & $5.31$ & $3.57$ & $-7.55$ & $0.243$ & $0.743$\\ 
$0.6$& $38.2$ & $2.19$ & $0.54$ & $0.54$ & $61.4$ & $61.3$ & $5.31$ & $5.31$ & $3.57$ & $-7.54$ & $0.239$ & $0.747$\\ 
$0.5$& $38.5$ & $2.18$ & $0.55$ & $0.55$ & $61.7$ & $61.6$ & $5.3$ & $5.3$ & $3.59$ & $-7.52$ & $0.238$ & $0.748$\\ 
$0.4$& $38.8$ & $2.17$ & $0.55$ & $0.55$ & $62.2$ & $62.0$ & $5.3$ & $5.29$ & $3.61$ & $-7.47$ & $0.219$ & $0.767$\\ 
$0.3$& $39.1$ & $2.16$ & $0.55$ & $0.55$ & $62.9$ & $62.8$ & $5.29$ & $5.29$ & $3.63$ & $-7.45$ & $0.22$ & $0.766$\\ 
$0.2$& $39.4$ & $2.15$ & $0.54$ & $0.55$ & $64.2$ & $64.0$ & $5.3$ & $5.3$ & $3.65$ & $-7.4$ & $0.212$ & $0.774$\\ 
$0.1$& $39.7$ & $2.14$ & $0.52$ & $0.52$ & $66.4$ & $66.4$ & $5.33$ & $5.33$ & $3.68$ & $-7.37$ & $0.21$ & $0.776$\\ 
\midrule
\multicolumn{13}{l}{$M_{\text{init}} = 12.17 M_{\odot}$} \\ 
\midrule
$0.9$& $18.8$ & $3.99$ & $1.24$ & $1.25$ & $69.1$ & $68.8$ & $4.85$ & $4.84$ & $4.5$ & $-6.34$ & $0.247$ & $0.739$\\ 
$0.8$& $18.9$ & $3.95$ & $0.81$ & $0.82$ & $75.8$ & $75.6$ & $5.21$ & $5.21$ & $4.29$ & $-6.6$ & $0.236$ & $0.75$\\ 
$0.7$& $19.1$ & $3.92$ & $0.77$ & $0.78$ & $76.7$ & $76.5$ & $5.26$ & $5.25$ & $4.27$ & $-6.63$ & $0.226$ & $0.76$\\ 
$0.6$& $19.2$ & $3.9$ & $0.76$ & $0.76$ & $77.5$ & $77.3$ & $5.27$ & $5.26$ & $4.27$ & $-6.62$ & $0.218$ & $0.768$\\ 
$0.5$& $19.3$ & $3.86$ & $0.75$ & $0.76$ & $78.5$ & $78.3$ & $5.27$ & $5.27$ & $4.29$ & $-6.59$ & $0.213$ & $0.773$\\ 
$0.4$& $19.4$ & $3.83$ & $0.74$ & $0.74$ & $80.1$ & $80.0$ & $5.29$ & $5.28$ & $4.3$ & $-6.56$ & $0.201$ & $0.785$\\ 
$0.3$& $19.5$ & $3.8$ & $0.7$ & $0.7$ & $82.8$ & $82.7$ & $5.32$ & $5.32$ & $4.32$ & $-6.52$ & $0.179$ & $0.807$\\ 
$0.2$& $19.7$ & $3.75$ & $0.63$ & $0.63$ & $88.1$ & $87.9$ & $5.41$ & $5.41$ & $4.34$ & $-6.44$ & $0.106$ & $0.88$\\ 
$0.1$& $19.8$ & $3.68$ & $0.57$ & $0.57$ & $93.5$ & $93.4$ & $5.49$ & $5.49$ & $4.35$ & $-6.33$ & $7.735\times 10^{-4}$ & $0.986$\\ 
\midrule
\multicolumn{13}{l}{$M_{\text{init}} = 18.17 M_{\odot}$} \\ 
\midrule
$0.9$& $10.7$ & $7.16$ & $1.99$ & $2.02$ & $74.4$ & $73.9$ & $4.69$ & $4.68$ & $5.04$ & $-5.63$ & $0.232$ & $0.755$\\ 
$0.8$& $10.7$ & $7.04$ & $1.11$ & $1.11$ & $90.7$ & $90.4$ & $5.2$ & $5.19$ & $4.87$ & $-5.83$ & $0.213$ & $0.774$\\ 
$0.7$& $10.8$ & $6.95$ & $1.02$ & $1.03$ & $93.4$ & $93.2$ & $5.26$ & $5.25$ & $4.86$ & $-5.84$ & $0.195$ & $0.791$\\ 
$0.6$& $10.9$ & $6.85$ & $0.97$ & $0.97$ & $96.2$ & $95.9$ & $5.3$ & $5.29$ & $4.86$ & $-5.82$ & $0.176$ & $0.811$\\ 
$0.5$& $10.9$ & $6.74$ & $0.88$ & $0.88$ & $101.6$ & $101.3$ & $5.38$ & $5.38$ & $4.87$ & $-5.76$ & $0.117$ & $0.87$\\ 
$0.4$& $11.0$ & $6.61$ & $0.79$ & -- & $107.1$ & -- & $5.46$ & -- & $4.87$ & $-5.66$ & $1.12\times 10^{-3}$ & $0.985$\\ 
$0.3$& $11.1$ & $6.44$ & $0.79$ & $0.8$ & $107.2$ & $106.5$ & $5.45$ & $5.44$ & $4.87$ & $-5.66$ & $3.556\times 10^{-9}$ & $0.986$\\ 
$0.2$& $11.1$ & $6.27$ & $0.78$ & $0.79$ & $107.8$ & $107.0$ & $5.45$ & $5.43$ & $4.87$ & $-5.66$ & $4.273\times 10^{-15}$ & $0.986$\\ 
$0.1$& $11.2$ & $6.07$ & $0.76$ & $0.77$ & $109.4$ & $108.6$ & $5.46$ & $5.44$ & $4.87$ & $-5.66$ & $8.791\times 10^{-18}$ & $0.986$\\ 
\bottomrule
\end{tabular}

%% file: Qs_examples.tex
\begin{tabular}{ccccccccc}
\toprule\midrule
$X_{\text{He,c}}$ & Age$^a$ & $M_{\text{strip}}^a$ & $\log_{10} Q_{\text{0}}^b$ & $\log_{10} Q_{\text{0,BB}}^a$ & $\log_{10} Q_{\text{1}}^b$ & $\log_{10} Q_{\text{1,BB}}^a$ & $\log_{10} Q_{\text{2}}^b$ & $\log_{10} Q_{\text{2,BB}}^a$ \\ 
 & [Myr] & [$M_{\odot}$] & $[$s$^{-1}]$ & $[$s$^{-1}]$ & $[$s$^{-1}]$ & $[$s$^{-1}]$ & $[$s$^{-1}]$ & $[$s$^{-1}]$ \\ 
\midrule
\multicolumn{9}{l}{$M_{\text{init}} = 8.15 M_{\odot}$} \\ 
\midrule
$0.9$ & $37.3$ & $2.22$ & $47.57$ & $47.65$ & $47.17$ & $47.12$ & $44.16$ & $45.26$\\ 
$0.8$ & $37.6$ & $2.21$ & $47.37$ & $47.48$ & $46.98$ & $46.95$ & $43.86$ & $45.1$\\ 
$0.7$ & $37.9$ & $2.2$ & $47.36$ & $47.47$ & $46.96$ & $46.94$ & $43.85$ & $45.08$\\ 
$0.6$ & $38.2$ & $2.19$ & $47.36$ & $47.48$ & $46.97$ & $46.95$ & $43.87$ & $45.1$\\ 
$0.5$ & $38.5$ & $2.18$ & $47.38$ & $47.49$ & $46.99$ & $46.97$ & $43.91$ & $45.13$\\ 
$0.4$ & $38.8$ & $2.17$ & $47.43$ & $47.51$ & $47.05$ & $47.0$ & $44.03$ & $45.17$\\ 
$0.3$ & $39.1$ & $2.16$ & $47.45$ & $47.53$ & $47.08$ & $47.03$ & $44.11$ & $45.23$\\ 
$0.2$ & $39.4$ & $2.15$ & $47.48$ & $47.56$ & $47.12$ & $47.07$ & $44.2$ & $45.31$\\ 
$0.1$ & $39.7$ & $2.14$ & $47.51$ & $47.59$ & $47.17$ & $47.12$ & $44.35$ & $45.44$\\ 
\midrule
\multicolumn{9}{l}{$M_{\text{init}} = 12.17 M_{\odot}$} \\ 
\midrule
$0.9$ & $18.8$ & $3.99$ & $48.35$ & $48.41$ & $48.03$ & $47.97$ & $45.31$ & $46.37$\\ 
$0.8$ & $18.9$ & $3.95$ & $48.14$ & $48.2$ & $47.86$ & $47.81$ & $45.44$ & $46.4$\\ 
$0.7$ & $19.1$ & $3.92$ & $48.11$ & $48.17$ & $47.85$ & $47.79$ & $45.45$ & $46.41$\\ 
$0.6$ & $19.2$ & $3.9$ & $48.12$ & $48.18$ & $47.86$ & $47.8$ & $45.49$ & $46.43$\\ 
$0.5$ & $19.3$ & $3.86$ & $48.13$ & $48.19$ & $47.87$ & $47.82$ & $45.52$ & $46.48$\\ 
$0.4$ & $19.4$ & $3.83$ & $48.15$ & $48.2$ & $47.9$ & $47.85$ & $45.6$ & $46.54$\\ 
$0.3$ & $19.5$ & $3.8$ & $48.17$ & $48.22$ & $47.93$ & $47.88$ & $45.71$ & $46.63$\\ 
$0.2$ & $19.7$ & $3.75$ & $48.15$ & $48.23$ & $47.94$ & $47.92$ & $45.76$ & $46.77$\\ 
$0.1$ & $19.8$ & $3.68$ & $48.17$ & $48.23$ & $47.97$ & $47.95$ & $45.87$ & $46.89$\\ 
\midrule
\multicolumn{9}{l}{$M_{\text{init}} = 18.17 M_{\odot}$} \\ 
\midrule
$0.9$ & $10.7$ & $7.16$ & $48.9$ & $48.95$ & $48.62$ & $48.55$ & $45.7$ & $47.1$\\ 
$0.8$ & $10.7$ & $7.04$ & $48.71$ & $48.76$ & $48.5$ & $48.47$ & $46.55$ & $47.36$\\ 
$0.7$ & $10.8$ & $6.95$ & $48.69$ & $48.74$ & $48.49$ & $48.46$ & $46.61$ & $47.39$\\ 
$0.6$ & $10.9$ & $6.85$ & $48.69$ & $48.74$ & $48.5$ & $48.47$ & $46.73$ & $47.45$\\ 
$0.5$ & $10.9$ & $6.74$ & $48.69$ & $48.74$ & $48.51$ & $48.49$ & $46.94$ & $47.54$\\ 
$0.4$ & $11.0$ & $6.61$ & -- & $48.73$ & -- & $48.5$ & -- & $47.62$\\ 
$0.3$ & $11.1$ & $6.44$ & $48.69$ & $48.73$ & $48.53$ & $48.5$ & $47.11$ & $47.62$\\ 
$0.2$ & $11.1$ & $6.27$ & $48.69$ & $48.73$ & $48.53$ & $48.5$ & $47.15$ & $47.63$\\ 
$0.1$ & $11.2$ & $6.07$ & $48.69$ & $48.72$ & $48.52$ & $48.5$ & $47.23$ & $47.65$\\ 
\bottomrule
\end{tabular}